\documentclass[review]{elsarticle}
\usepackage[utf8]{inputenc}
\usepackage{hyperref,amsmath,amssymb,amsbsy,bbold,color,graphics,graphicx,bigints,relsize,cancel}
\usepackage{lineno,hyperref,appendix}

\newcommand{\dotp}{
\mathop{\mathchoice{\vcenter{\hbox{\LARGE$\cdot$}}}
{\vcenter{\hbox{\LARGE$\cdot$}}}
{\vcenter{\hbox{\normalsize$\cdot$}}}
{\vcenter{\hbox{\small$\cdot$}}}}}

\modulolinenumbers[5]
\pdfoutput=1
\allowdisplaybreaks

\journal{Journal of \LaTeX\ Templates}

\begin{document}

\begin{frontmatter}

\title{Rotations in classical mechanics using geometric algebra}

\author{Sylvain D. Brechet\fnref{myfootnote}}
\address{Institute of Physics, Station 3, Ecole Polytechnique F\'ed\'erale de Lausanne\,-\,EPFL, CH-1015 Lausanne,\,Switzerland}

\ead{sylvain.brechet@epfl.ch}

\begin{abstract}

\noindent In geometric algebra, the rotation of a vector is described using rotors. Rotors are phasors where the imaginary number has been replaced by a oriented plane element of unit area called a unit bivector. The algebra in three dimensional space relating vectors and bivectors is the Pauli algebra. Multivectors consisting of linear combinations of scalars and bivectors are isomorphic to quaternions. The rotational dynamics can be expressed entirely in the plane of rotation using bivectors. In particular, the Poisson formula providing the time derivative of the unit vectors of a moving frame are recast in terms of the angular velocity bivector and applied to cylindrical and spherical frames. The rotational dynamics of a point particle and a rigid body are fully determined by the time evolution of rotors. The mapping of the angular velocity bivector onto the angular momentum bivector is the inertia map. In the principal axis frame of the rigid body, the inertia map is characterised by symmetric coefficients representing the moments of inertia. The Huygens-Steiner theorem, the kinetic energy of a rigid body and the Euler equations are expressed in terms of bivector components. This formalism is applied to study the rotational dynamics of a gyroscope.  
	
\end{abstract}

\end{frontmatter}


\tableofcontents

\section{Introduction}

\noindent Rotational motion is a key component of the classical mechanics of rigid bodies. Textbooks describing rotations in linear algebra~\cite{Goldstein:2001,Pettofrezzo:1978,Atanasiu:2020,Strang:2009}, usually write vectors in components with respect to a given frame and represent rotations as matrices. The approach followed in geometric algebra is simpler, geometrically more meaningful and insightful, as we shall show in the article.

To understand the main idea of geometric algebra, we consider the geometric interpretation of complex numbers. The complex plane $\mathbb{C}$ is isomorphic to $\mathbb{R}^2$. The rotation of any complex number $z\in\mathbb{C}$ by an angle $\theta$ in this plane is performed by multiplying this number by the complex phasor $e^{i\theta} = \cos\theta + i\,\sin\theta$. The imaginary number $i$ is not merely a useful algebraic concept. It has also a geometric meaning as an oriented plane element of unit norm or modulus, called a unit bivector in geometric algebra (GA) $\mathbb{G}_3$. To establish this geometric interpretation, we write the geometric product of two vectors $\boldsymbol{a}$ and $\boldsymbol{b}$ as the sum of the symmetric inner product $\boldsymbol{a}\cdot\boldsymbol{b}$, that is identical to the scalar product, and the antisymmetric outer product $\boldsymbol{a}\wedge\boldsymbol{b}$, that represents an oriented plane element and is the dual of the cross product $\boldsymbol{a}\times\boldsymbol{b}$,
\begin{equation}\label{geometric product}
\boldsymbol{a}\,\boldsymbol{b} = \boldsymbol{a}\cdot\boldsymbol{b} + \boldsymbol{a}\wedge\boldsymbol{b}
\end{equation}
In a plane spanned by the orthonormal units vectors $\boldsymbol{\hat{e}}_1$ and $\boldsymbol{\hat{e}}_2$, using the orthogonality of the vector and the antisymmetry of their outer product and the symmetry of their inner product,
\begin{equation}\label{orthongonality}
\begin{split}
&\boldsymbol{\hat{e}}_1\cdot\boldsymbol{\hat{e}}_2 = 0 \qquad\qquad\quad\ \,\text{and}\qquad \boldsymbol{\hat{e}}_1\,\boldsymbol{\hat{e}}_2 = \boldsymbol{\hat{e}}_1\wedge\boldsymbol{\hat{e}}_2 = -\,\boldsymbol{\hat{e}}_2\wedge\boldsymbol{\hat{e}}_1 = -\,\boldsymbol{\hat{e}}_2\,\boldsymbol{\hat{e}}_1\\
&\boldsymbol{\hat{e}}_1\,\boldsymbol{\hat{e}}_1 = \boldsymbol{\hat{e}}_1\cdot\boldsymbol{\hat{e}}_1 = 1 \qquad\text{and}\qquad \boldsymbol{\hat{e}}_2\,\boldsymbol{\hat{e}}_2 = \boldsymbol{\hat{e}}_2\cdot\boldsymbol{\hat{e}}_2 = 1
\end{split}
\end{equation}
the square of the their geometric product reads,
\begin{equation}\label{geometric product}
\left(\boldsymbol{\hat{e}}_1\,\boldsymbol{\hat{e}}_2\right)^2 = \boldsymbol{\hat{e}}_1\,\boldsymbol{\hat{e}}_2\,\boldsymbol{\hat{e}}_1\,\boldsymbol{\hat{e}}_2 = -\,\boldsymbol{\hat{e}}_1\,\boldsymbol{\hat{e}}_1\,\boldsymbol{\hat{e}}_2\,\boldsymbol{\hat{e}}_2 = -\,1
\end{equation}
which demonstrates that the bivector $\boldsymbol{\hat{e}}_1\,\boldsymbol{\hat{e}}_2$ behaves geometrically as an imaginary number. Similarly, in the planes spanned by the units vectors $\boldsymbol{\hat{e}}_2$ and $\boldsymbol{\hat{e}}_3$ and by the unit vectors $\boldsymbol{\hat{e}}_3$ and $\boldsymbol{\hat{e}}_1$, the bivectors $\boldsymbol{\hat{e}}_1\,\boldsymbol{\hat{e}}_2$ and $\boldsymbol{\hat{e}}_3\,\boldsymbol{\hat{e}}_1$ behave like two other imaginary numbers. Thus, the subalgebra of the bivectors of $\mathbb{G}^3$ is isomorphic to the quaternions $\mathbb{H}$. As we shall show in this article, the rotation of any vector can be described using compositions of the three phasors $e^{\,\boldsymbol{\hat{e}}_1\,\boldsymbol{\hat{e}}_2}$, $e^{\,\boldsymbol{\hat{e}}_2\,\boldsymbol{\hat{e}}_3}$, and $e^{\,\boldsymbol{\hat{e}}_3\,\boldsymbol{\hat{e}}_1}$ called rotors in geometric algebra.

This geometric structure can be established by showing that every rotation is the result of the composition of two reflections. Each reflection of a vector over a plane can be written in a beautiful yet simple form using the geometric product of this vector with the vector orthogonal to the plane. By iterating the process twice, rotors appear.

Rotations in classical mechanics are usually described in terms of the angular velocity $\boldsymbol{\omega}$ and the angular momentum $\boldsymbol{\ell}$. These quantities are pseudovectors that were introduced in order to describe classical mechanics in a vector space. The right hand rule used to define the angular momentum pseudovector $\boldsymbol{\ell}_O$ evaluated at the origin $O$ as a cross product of the position vector $\boldsymbol{r}$ and the momentum vector $\boldsymbol{p}$, i.e. $\boldsymbol{\ell}_O = \boldsymbol{r}\times\boldsymbol{p}$, is a pure mathematical convention. In order not to break the rotational symmetry, the angular pseudovector $\boldsymbol{\ell}_O$ has to be chosen along the axis orthogonal to the plane of motion. Then, all that is left is the chirality : history chose the right hand ! However, when an object rotates in a plane, why should a vector orthogonal to the plane of rotation be introduced in the first place ? Would it not be more natural to introduce an oriented plane in the rotation plane ? It turns out that this oriented plane element is the angular momentum bivector $\boldsymbol{L}_O$ evaluated at the origin $O$. It is the outer product of the position vector $\boldsymbol{r}$ and the momentum vector $\boldsymbol{p}$, i.e. $\boldsymbol{L}_O = \boldsymbol{r}\wedge\boldsymbol{p}$. The pseudovector $\boldsymbol{\ell}_O$ is the dual of the bivector $\boldsymbol{L}_O$, i.e $\boldsymbol{\ell}_O = \boldsymbol{L}_O^{\ast}$. Similarly, the angular velocity pseudovector $\boldsymbol{\omega}$ is the dual of the angular velocity bivector $\boldsymbol{\Omega}$, i.e $\boldsymbol{\omega} = \boldsymbol{\Omega}^{\ast}$. The rotation of rigid bodies can thus shall be written in terms of these bivectors, as we shall see in this publication.

This article is structured as follows. In the first part, we begin with a geometric description of projections, rejection, reflections and rotation in geometric algebra. Then, in Sec.~\ref{Rotors, quaternions and the Pauli algebra}, rotors are written in terms of quaternions and their algebra is shown to be isomorphic to the Pauli algebra. In the second part, we describe the rotation of point particles. In Sec.~\ref{Rotating frame and angular velocity}, the time evolution of the vector frame is expressed in terms of the angular velocity bivector. The rotation of cylindrical and spherical frames is described in Sec.~\ref{Rotating cylindrical frame} and~\ref{Rotating spherical frame}. The rotational dynamics a point particles in detailed in Sec.~\ref{Point particle motion}. In the third part, we describe the intrinsic rotational dynamics of rigid bodies. The rotation of rigid bodies is discussed in Sec.~\ref{Rigid body motion}. In Sec.~\ref{Inertia and principal body frame}, inertia is expressed as a mapping of unit bivectors of the principal body frame. The Huygens-Steiner theorem is established in Sec.~\ref{Huygens-Steiner theorem}. The angular momentum and the kinetic energy are discussed in Sec.~\ref{Angular momentum of a rigid body} and~\ref{Kinetic energy of a rigid body}. The Euler equations are established in Sec.~\ref{Euler equations section} and the rotational dynamics of a spinning disk is described in Sec.~\ref{Symmetric spinning disk}. The foundations of geometric algebra are presented in Appendix~\ref{Geometric algebra}. The duality operation in geometric algebra (GA) is defined in Appendix~\ref{Duality in geometric algebra}. Important geometric identities in geometric algebra are established in Appendix~\ref{Algebraic identities in geometric algebra}.


\section{Projection, rejection and reflection}
\label{Projection, rejection and reflection}

\begin{figure}[!ht]
\begin{center}
\includegraphics[scale=0.85]{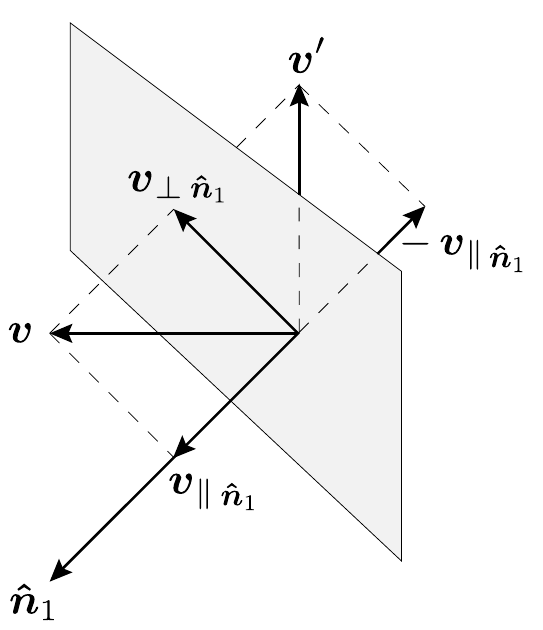}
\end{center}
\caption{Reflection of a vector $\boldsymbol{v}$ over the plane orthogonal to a unit vector $\boldsymbol{\hat{n}}_1$.}\label{Fig: Reflection}
\end{figure}

\noindent From a geometric perspective, a rotation can be described as the composition of two reflections over two different planes. In order to show this in the framework of geometric algebra, we begin by studying reflections. We consider an arbitrary vector $\boldsymbol{v}$ and an arbitrary unit vector $\boldsymbol{\hat{n}}_1$. The geometric product~\eqref{geometric product space u v} of the vectors $\boldsymbol{\hat{n}}_1$ and $\boldsymbol{v}$ reads,
\begin{equation}\label{reflection geometric product}
\boldsymbol{\hat{n}}_1\,\boldsymbol{v} = \boldsymbol{\hat{n}}_1\cdot\boldsymbol{v} + \boldsymbol{\hat{n}}_1\wedge\boldsymbol{v}
\end{equation}
Since $\boldsymbol{\hat{n}}_1$ is a unit vector, the vector $\boldsymbol{v}$ can be resolved into a parallel and a perpendicular part to $\boldsymbol{\hat{n}}_1$ using the geometric product~\eqref{reflection geometric product},
\begin{equation}\label{reflection v}
\boldsymbol{v} = \boldsymbol{\hat{n}}_1^2\,\boldsymbol{v} = \boldsymbol{\hat{n}}_1\left(\boldsymbol{\hat{n}}_1\boldsymbol{v}\right) = \boldsymbol{\hat{n}}_1\left(\boldsymbol{\hat{n}}_1\cdot\boldsymbol{v}\right) + \boldsymbol{\hat{n}}_1\left(\boldsymbol{\hat{n}}_1\wedge\boldsymbol{v}\right)
\end{equation}
Using the geometric product~\eqref{geometric product space v B} of the vector $\boldsymbol{\hat{n}}_1$ with the bivector $\boldsymbol{\hat{n}}_1\wedge\boldsymbol{v}$, the second part of vector $\boldsymbol{v}$ can be recast as,
\begin{equation}\label{reflection v antisymetric term}
\boldsymbol{\hat{n}}_1\left(\boldsymbol{\hat{n}}_1\wedge\boldsymbol{v}\right) = \boldsymbol{\hat{n}}_1\wedge\boldsymbol{\hat{n}}_1\wedge\boldsymbol{v} + \boldsymbol{\hat{n}}_1\cdot\left(\boldsymbol{\hat{n}}_1\wedge\boldsymbol{v}\right) = \boldsymbol{\hat{n}}_1\cdot\left(\boldsymbol{\hat{n}}_1\wedge\boldsymbol{v}\right)
\end{equation}
In view of relations~\eqref{reflection v} and~\eqref{reflection v antisymetric term}, the vector $\boldsymbol{v}$ is resolved into parallel and perpendicular parts (Fig.~\ref{Fig: Reflection}),
\begin{equation}\label{reflection v para perp}
\boldsymbol{v} = \boldsymbol{v}_{\parallel\,\boldsymbol{\hat{n}}_1} + \boldsymbol{v}_{\perp\,\boldsymbol{\hat{n}}_1}
\end{equation}
according to,
\begin{equation}\label{v para perp}
\begin{split}
&\boldsymbol{v}_{\parallel\,\boldsymbol{\hat{n}}_1} = \left(\boldsymbol{\hat{n}}_1\cdot\boldsymbol{v}\right)\boldsymbol{\hat{n}}_1\\
&\boldsymbol{v}_{\perp\,\boldsymbol{\hat{n}}_1} = \boldsymbol{\hat{n}}_1\cdot\left(\boldsymbol{\hat{n}}_1\wedge\boldsymbol{v}\right)
\end{split}
\end{equation}

\noindent Using identity~\eqref{id dot cross}, the geometric interpretation of the perpendicular part becomes clear,
\begin{equation}\label{v perp interpretation}
\boldsymbol{v}_{\perp\,\boldsymbol{\hat{n}}_1} = \boldsymbol{\hat{n}}_1\cdot\left(\boldsymbol{\hat{n}}_1\wedge\boldsymbol{v}\right) = \left(\boldsymbol{\hat{n}}_1\cdot\boldsymbol{\hat{n}}_1\right)\boldsymbol{v} -\,\left(\boldsymbol{\hat{n}}_1\cdot\boldsymbol{v}\right)\boldsymbol{\hat{n}}_1 = \boldsymbol{v} -\,\boldsymbol{v}_{\parallel\,\boldsymbol{\hat{n}}_1}
\end{equation}
The parallel part of a vector $\boldsymbol{v}$ is the result of the projection onto the unit vector $\boldsymbol{\hat{n}}_1$. In view of relation~\eqref{v para perp}, taking into account that a unit vector is its own inverse, i.e. $\boldsymbol{\hat{n}} = \boldsymbol{\hat{n}}^{-1}$, the projection of the vector $\boldsymbol{v}$ onto a unit vector $\boldsymbol{\hat{n}}$ is an automorphism $\mathsf{P}_{\boldsymbol{\hat{n}}} : \mathbb{R}^{3} \to \mathbb{R}^{3}$ defined as (Fig.~\ref{Fig: Reflection}),
\begin{equation}\label{projection automorphism}
\mathsf{P}_{\boldsymbol{\hat{n}}}\left(\boldsymbol{v}\right) = \left(\boldsymbol{\hat{n}}\cdot\boldsymbol{v}\right)\boldsymbol{\hat{n}} = \frac{\boldsymbol{\hat{n}}\cdot\boldsymbol{v}}{\boldsymbol{\hat{n}}}
\end{equation}
The perpendicular part of a vector $\boldsymbol{v}$ is the result of the rejection onto the unit vector $\boldsymbol{\hat{n}}_1$, which is a projection on the orthogonal complement of the subspace $\mathsf{P}_{\boldsymbol{\hat{n}}_1}$ in the plane spanned by $\boldsymbol{\hat{n}}_1$ and $\boldsymbol{v}$. In view of relations~\eqref{reflection v antisymetric term} and~\eqref{v para perp}, taking into account that a unit vector is its own inverse, i.e. $\boldsymbol{\hat{n}} = \boldsymbol{\hat{n}}^{-1}$, the rejection of the vector $\boldsymbol{v}$ onto the unit vector $\boldsymbol{\hat{n}}$ is an automorphism $\mathsf{\bar{P}}_{\boldsymbol{\hat{n}}} : \mathbb{R}^{3} \to \mathbb{R}^{3}$ defined as (Fig.~\ref{Fig: Reflection}),
\begin{equation}\label{rejection automorphism}
\mathsf{\bar{P}}_{\boldsymbol{\hat{n}}}\left(\boldsymbol{v}\right) = \boldsymbol{\hat{n}}\cdot\left(\boldsymbol{\hat{n}}\wedge\boldsymbol{v}\right) = \boldsymbol{\hat{n}}\left(\boldsymbol{\hat{n}}\wedge\boldsymbol{v}\right) = \frac{\boldsymbol{\hat{n}}\wedge\boldsymbol{v}}{\boldsymbol{\hat{n}}}
\end{equation}
\begin{figure}[!ht]
\begin{center}
\includegraphics[scale=0.85]{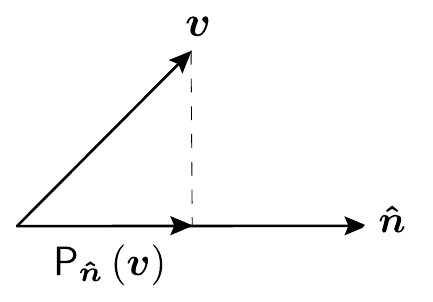}
\hspace{0.75cm}
\includegraphics[scale=0.85]{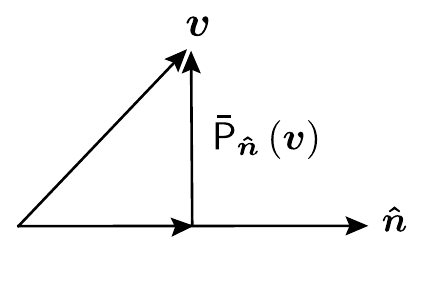}
\end{center}
\caption{Projection $\mathsf{P}_{\boldsymbol{\hat{n}}}\left(\boldsymbol{v}\right)$ and rejection $\mathsf{\bar{P}}_{\boldsymbol{\hat{n}}}\left(\boldsymbol{v}\right)$ of a vector $\boldsymbol{v}$ on the unit vector $\boldsymbol{\hat{n}}_1$.}\label{Fig: Projection and Rejection}
\end{figure}

\noindent In view of relations~\eqref{reflection v para perp},~\eqref{projection automorphism} and~\eqref{rejection automorphism}, the complementarity between the subspaces $\mathsf{P}_{\boldsymbol{\hat{n}}_1}$ and $\mathsf{\bar{P}}_{\boldsymbol{\hat{n}}_1}$ is written as,
\begin{equation}\label{reflection complementarity}
\boldsymbol{v} = \mathsf{P}_{\boldsymbol{\hat{n}}_1}\left(\boldsymbol{v}\right) + \mathsf{\bar{P}}_{\boldsymbol{\hat{n}}_1}\left(\boldsymbol{v}\right)
\end{equation}
The reflection of a vector $\boldsymbol{v}$ over the plan orthogonal to the unit vector $\boldsymbol{\hat{n}}_1$ yields the vector $\boldsymbol{v}^{\prime}$ according to (Fig.~\ref{Fig: Reflection}),
\begin{equation}\label{v reflected}
\boldsymbol{v}^{\prime} = -\,\mathsf{P}_{\boldsymbol{\hat{n}}_1}\left(\boldsymbol{v}\right) + \mathsf{\bar{P}}_{\boldsymbol{\hat{n}}_1}\left(\boldsymbol{v}\right) = -\,\boldsymbol{v}_{\parallel\,\boldsymbol{\hat{n}}_1} + \boldsymbol{v}_{\perp\,\boldsymbol{\hat{n}}_1}
\end{equation}
where the projection $\mathsf{P}_{\boldsymbol{\hat{n}}_1}\left(\boldsymbol{v}\right)$ changes sign and the rejection $\mathsf{\bar{P}}_{\boldsymbol{\hat{n}}_1}\left(\boldsymbol{v}\right)$ remains unchanged. Using the parallel and perpendicular parts~\eqref{v para perp} of the vector $\boldsymbol{v}$ and the antisymmetry~\eqref{inner product v B antisymmetric} of the inner product of the vector $\boldsymbol{\hat{n}}_1$ and the bivector $\boldsymbol{\hat{n}}_1\wedge\boldsymbol{v}$, the reflected vector~\eqref{v reflected} is recast as,
\begin{equation}\label{v reflected bis}
\boldsymbol{v}^{\prime} = -\,\left(\boldsymbol{\hat{n}}_1\cdot\boldsymbol{v}\right)\boldsymbol{\hat{n}}_1 + \boldsymbol{\hat{n}}_1\cdot\left(\boldsymbol{\hat{n}}_1\wedge\boldsymbol{v}\right) = -\,\left(\boldsymbol{\hat{n}}_1\cdot\boldsymbol{v}\right)\boldsymbol{\hat{n}}_1 -\,\left(\boldsymbol{\hat{n}}_1\wedge\boldsymbol{v}\right)\cdot\boldsymbol{\hat{n}}_1
\end{equation}
Using the geometric product~\eqref{geometric product space B v} of the bivector $\boldsymbol{\hat{n}}_1\wedge\boldsymbol{v}$ with the vector $\boldsymbol{\hat{n}}_1$, the second part of vector $\boldsymbol{v}^{\prime}$ can be recast as,
\begin{equation}\label{reflection v antisymetric term bis}
\left(\boldsymbol{\hat{n}}_1\wedge\boldsymbol{v}\right)\cdot\boldsymbol{\hat{n}}_1 = \left(\boldsymbol{\hat{n}}_1\wedge\boldsymbol{v}\right)\boldsymbol{\hat{n}}_1 -\,\boldsymbol{\hat{n}}_1\wedge\boldsymbol{v}\wedge\boldsymbol{\hat{n}}_1 = \left(\boldsymbol{\hat{n}}_1\wedge\boldsymbol{v}\right)\boldsymbol{\hat{n}}_1
\end{equation}
Taking into account the geometric product~\eqref{reflection geometric product} and the identity~\eqref{reflection v antisymetric term bis}, the reflected vector~\eqref{v reflected bis} is recast as,
\begin{equation}\label{v reflected ter}
\boldsymbol{v}^{\prime} = -\,\left(\boldsymbol{\hat{n}}_1\cdot\boldsymbol{v} + \boldsymbol{\hat{n}}_1\wedge\boldsymbol{v}\right)\boldsymbol{\hat{n}}_1 = -\,\boldsymbol{\hat{n}}_1\,\boldsymbol{v}\,\boldsymbol{\hat{n}}_1
\end{equation}
Thus, we conclude that the reflection of a vector $\boldsymbol{v}$ over a plane orthogonal to the unit vector $\boldsymbol{\hat{n}}_1$ is an automorphism $\mathsf{F}_{\boldsymbol{\hat{n}}_1} : \mathbb{R}^{3} \to \mathbb{R}^{3}$ defined as,
\begin{equation}\label{reflection automorphism}
\mathsf{F}_{\boldsymbol{\hat{n}}_1}\left(\boldsymbol{v}\right) = -\,\boldsymbol{\hat{n}}_1\,\boldsymbol{v}\,\boldsymbol{\hat{n}}_1
\end{equation}
%


\section{Rotation}
\label{Rotation}

\begin{figure}[!ht]
\begin{center}
\includegraphics[scale=0.85]{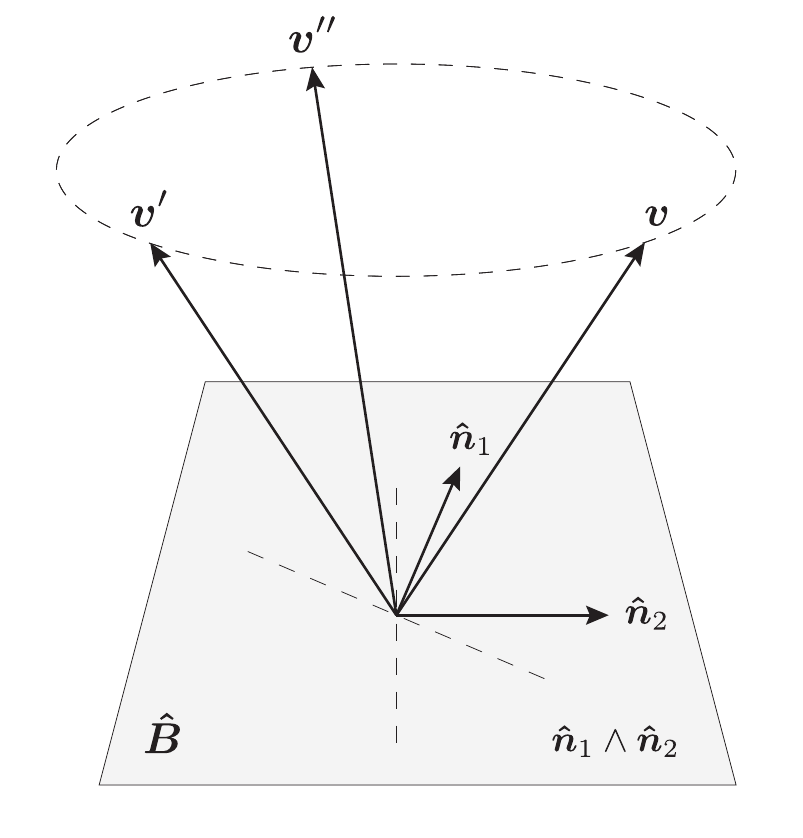}
\end{center}
\caption{Vector $\boldsymbol{v}^{\prime}$ is the reflection of vector $\boldsymbol{v}$ over the plane orthogonal to vector $\boldsymbol{\hat{n}}_1$ and vector $\boldsymbol{v}^{\prime\prime}$ is the reflection of vector $\boldsymbol{v}^{\prime}$ over the plane orthogonal to vector $\boldsymbol{\hat{n}}_2$. Vector $\boldsymbol{v}^{\prime\prime}$ is the rotation of vector $\boldsymbol{v}$ along the bivector $\boldsymbol{\hat{n}}_1\wedge\boldsymbol{\hat{n}}_2$ or the unit bivector $\boldsymbol{\hat{B}}$.}\label{Fig: Rotation with reflections}
\end{figure}

\noindent The rotation of a vector $\boldsymbol{v}$ in a plane spanned by the unit vectors $\boldsymbol{\hat{n}}_1$ and $\boldsymbol{\hat{n}}_2$ is obtained by performing a reflection over the plane orthogonal to the unit vector $\boldsymbol{\hat{n}}_1$ followed by a reflection over the plane orthogonal to the unit vector $\boldsymbol{\hat{n}}_2$. According to relations~\eqref{v reflected bis} and~\eqref{reflection automorphism}, the composition of the reflection $\mathsf{F}_{\boldsymbol{\hat{n}}_1}\left(\boldsymbol{v}\right)$ of a vector $\boldsymbol{v}$ over a plane orthogonal to the unit vector $\boldsymbol{\hat{n}}_1$ that yields the vector $\boldsymbol{v}^{\prime}$ and a reflection $\mathsf{F}_{\boldsymbol{\hat{n}}_1}\left(\boldsymbol{v}\right)$ over a plane orthogonal to the unit vector $\boldsymbol{\hat{n}}_1$ yields the vector $\boldsymbol{v}^{\prime\prime}$ (Fig:~\ref{Fig: Rotation with reflections}),
\begin{equation}\label{double reflection}
\begin{split}
&\boldsymbol{v}^{\prime\prime} = \mathsf{F}_{\boldsymbol{\hat{n}}_2}\circ\mathsf{F}_{\boldsymbol{\hat{n}}_1}\left(\boldsymbol{v}\right) = \mathsf{F}_{\boldsymbol{\hat{n}}_2}\,\Big(\mathsf{F}_{\boldsymbol{\hat{n}}_1}\left(\boldsymbol{v}\right)\!\Big) = \mathsf{F}_{\boldsymbol{\hat{n}}_2}\left(\boldsymbol{v}^{\prime}\right)\\
&\phantom{\boldsymbol{v}^{\prime\prime}} = -\,\mathsf{F}_{\boldsymbol{\hat{n}}_2}\left(\boldsymbol{\hat{n}}_1\,\boldsymbol{v}\,\boldsymbol{\hat{n}}_1\right) = \boldsymbol{\hat{n}}_2\,\boldsymbol{\hat{n}}_1\,\boldsymbol{v}\,\boldsymbol{\hat{n}}_1\,\boldsymbol{\hat{n}}_2
\end{split}
\end{equation}

\begin{figure}[!ht]
\begin{center}
\includegraphics[scale=0.85]{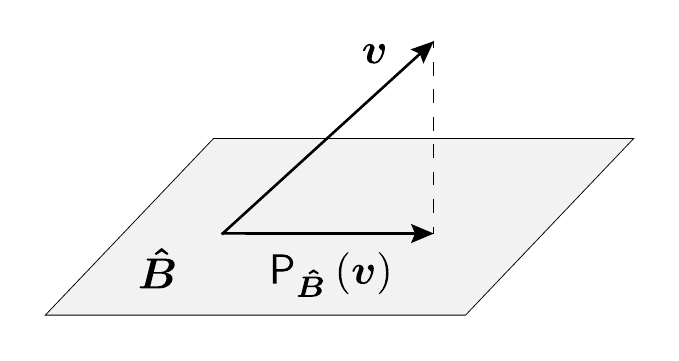}
\hspace{0.25cm}
\includegraphics[scale=0.85]{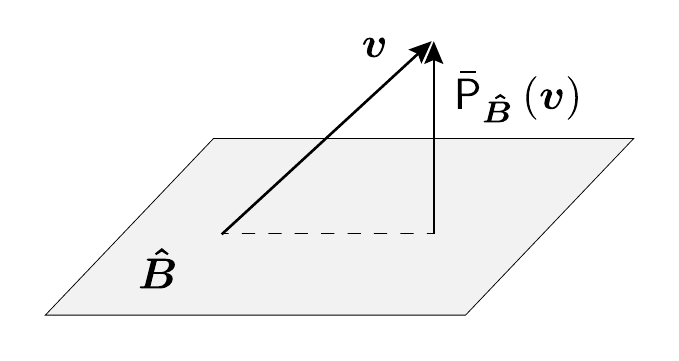}
\end{center}
\caption{Projection $\mathsf{P}_{\boldsymbol{\hat{B}}}\left(\boldsymbol{v}\right)$ and rejection $\mathsf{\bar{P}}_{\boldsymbol{\hat{B}}}\left(\boldsymbol{v}\right)$ of a vector $\boldsymbol{v}$ on the plane defined by the bivector vector $\boldsymbol{\hat{B}}$.}\label{Fig: Projection and Rejection plane}
\end{figure}

\noindent The key geometric object to describe rotations is the multivector rotor $R$ and its reverse $R^{\dag}$ that are given by,
\begin{equation}\label{rotor}
R = \boldsymbol{\hat{n}}_2\,\boldsymbol{\hat{n}}_1 \qquad\text{and}\qquad R^{\dag} = \left(\boldsymbol{\hat{n}}_2\,\boldsymbol{\hat{n}}_1\right)^{\dag} = \boldsymbol{\hat{n}}_1^{\dag} \boldsymbol{\hat{n}}_2^{\dag} = \boldsymbol{\hat{n}}_1\,\boldsymbol{\hat{n}}_2
\end{equation}
Thus, the rotation~\eqref{double reflection} of the vector $\boldsymbol{v}$ that yields the vector $\boldsymbol{v}^{\prime\prime}$ is written in terms of the rotor $R$ and its reverse $R^{\dag}$ as,
\begin{equation}\label{rotation rotor}
\boldsymbol{v}^{\prime\prime} = R\,\boldsymbol{v}\,R^{\dag}
\end{equation}
The rotor and its reverse $R$ and its reverse $R^{\dag}$ satisfy the orthogonality relation,
\begin{equation}\label{orthogonality rotor}
R\,R^{\dag} = \boldsymbol{\hat{n}}_2\,\boldsymbol{\hat{n}}_1\,\boldsymbol{\hat{n}}_1\,\boldsymbol{\hat{n}}_2 = 1
\end{equation}
Thus, the rotor is a multivector of unit modulus,
\begin{equation}\label{modulus rotor}
\vert R \vert^2 = R\,R^{\dag} = 1
\end{equation}
The inner and outer products of the vectors $\boldsymbol{\hat{n}}_2$ and $\boldsymbol{\hat{n}}_1$ can be written in terms of the angle $\phi$ between the vectors,
\begin{equation}\label{inner outer reflect}
\boldsymbol{\hat{n}}_2\cdot\boldsymbol{\hat{n}}_1 = \boldsymbol{\hat{n}}_1\cdot\boldsymbol{\hat{n}}_2 = \cos\phi \qquad\text{and}\qquad \boldsymbol{\hat{n}}_2\wedge\boldsymbol{\hat{n}}_1 = -\,\boldsymbol{\hat{n}}_1\wedge\boldsymbol{\hat{n}}_2 = -\,\sin\phi\,\boldsymbol{\hat{B}}
\end{equation}
where $\boldsymbol{\hat{B}}$ is the unit bivector in the rotation plane oriented in the rotation direction from vector $\boldsymbol{\hat{n}}_1$ to vector $\boldsymbol{\hat{n}}_2$ (Fig:~\ref{Fig: Rotation with reflections}). According to these products~\eqref{inner outer reflect}, the rotor and its reverse~\eqref{rotor} are recast as,
\begin{equation}\label{rotor angle}
\begin{split}
&R = \boldsymbol{\hat{n}}_2\cdot\boldsymbol{\hat{n}}_1 + \boldsymbol{\hat{n}}_2\wedge\boldsymbol{\hat{n}}_1 = \cos\phi -\,\sin\phi\,\boldsymbol{\hat{B}}\\
&R^{\dag} = \boldsymbol{\hat{n}}_1\cdot\boldsymbol{\hat{n}}_2 + \boldsymbol{\hat{n}}_1\wedge\boldsymbol{\hat{n}}_2 = \cos\phi + \sin\phi\,\boldsymbol{\hat{B}}
\end{split}
\end{equation}
which shows that the rotor and its reverse are even multivectors of unit modulus which are linear combination of a scalar and a bivector. Moreover, the unit bivector $\boldsymbol{\hat{B}}$ satisfies the relation,
\begin{equation}\label{unit bivector squared}
\boldsymbol{\hat{B}}^2 = \boldsymbol{\hat{B}}\,\boldsymbol{\hat{B}} = -\,\boldsymbol{\hat{B}}\,\boldsymbol{\hat{B}}^{\dag} = -\,\vert \boldsymbol{\hat{B}} \vert^2 = -\,1
\end{equation}
which means that the unit bivector $\boldsymbol{\hat{B}}$ in the rotation plane is isomorphic to the imaginary number $i$ in the complex plane $\mathbb{C}$. This isomorphism yields the Euler formula in the plane of rotation where $i$ is replaced by $\boldsymbol{\hat{B}}$,
\begin{equation}\label{Euler formula}
e^{\boldsymbol{\hat{B}}\phi} = \cos\phi + \sin\phi\,\boldsymbol{\hat{B}} \qquad\text{and}\qquad e^{-\,\boldsymbol{\hat{B}}\phi} = \cos\phi -\,\sin\phi\,\boldsymbol{\hat{B}}
\end{equation}
In view of the Euler formula~\eqref{Euler formula}, the rotor and its reverse~\eqref{rotor angle} are recast as,
\begin{equation}\label{rotor angle bis}
R = e^{-\,\boldsymbol{\hat{B}}\phi} \qquad\text{and}\qquad R^{\dag} = e^{\boldsymbol{\hat{B}}\phi}
\end{equation}
and the rotation~\eqref{rotation rotor} is recast as,
\begin{equation}\label{rotation rotor phi}
\boldsymbol{v}^{\prime\prime} = e^{-\,\boldsymbol{\hat{B}}\phi}\,\boldsymbol{v}\,e^{\boldsymbol{\hat{B}}\phi}
\end{equation}
The rotation~\eqref{rotation rotor phi} of the vector $\boldsymbol{v}$ is expressed in terms of the angle $\phi$ between the vectors $\boldsymbol{\hat{n}}_1$ and $\boldsymbol{\hat{n}}_2$. It has to be expressed now in terms of the angle $\theta$ between the projections of the vectors $\boldsymbol{v}$ and $\boldsymbol{v}^{\prime\prime}$ in the rotation plane. The projection of the vector $\boldsymbol{v}$ on the rotation plane is a projection on the unit bivector $\boldsymbol{\hat{B}}$. Using the geometric product~\eqref{geometric product space B v} of the bivector $\boldsymbol{\hat{B}}$ with the vector $\boldsymbol{v}$, the vector $\boldsymbol{v}$ is resolved into parallel and perpendicular parts as,
\begin{equation}\label{v projection and rejection plane}
\boldsymbol{v} = \boldsymbol{\hat{B}}^{-1}\,\boldsymbol{\hat{B}}\,\boldsymbol{v} = \boldsymbol{\hat{B}}^{-1}\left(\boldsymbol{\hat{B}}\cdot\boldsymbol{v} + \boldsymbol{\hat{B}}\wedge\boldsymbol{v}\right) = \frac{\boldsymbol{\hat{B}}\cdot\boldsymbol{v}}{\boldsymbol{\hat{B}}} + \frac{\boldsymbol{\hat{B}}\wedge\boldsymbol{v}}{\boldsymbol{\hat{B}}}
\end{equation}
where $\boldsymbol{\hat{B}}^{-1} = \boldsymbol{\hat{B}}^{\dag}$. The projection of the vector $\boldsymbol{v}$ onto the unit bivector $\boldsymbol{\hat{B}}$ is an automorphism $\mathsf{P}_{\boldsymbol{\hat{B}}} : \mathbb{R}^{3} \to \mathbb{R}^{3}$ defined as (Fig.~\ref{Fig: Projection and Rejection plane}),
\begin{equation}\label{projection automorphism plane}
\mathsf{P}_{\boldsymbol{\hat{B}}}\left(\boldsymbol{v}\right) = \boldsymbol{\hat{B}}^{\dag}\cdot\left(\boldsymbol{\hat{B}}\cdot\boldsymbol{v}\right) = \frac{\boldsymbol{\hat{B}}\cdot\boldsymbol{v}}{\boldsymbol{\hat{B}}}
\end{equation}
The rejection of the vector $\boldsymbol{v}$ onto the unit bivector $\boldsymbol{\hat{B}}$ is an automorphism $\mathsf{\bar{P}}_{\boldsymbol{\hat{B}}} : \mathbb{R}^{3} \to \mathbb{R}^{3}$ defined as (Fig.~\ref{Fig: Projection and Rejection plane}),
\begin{equation}\label{rejection automorphism plane}
\mathsf{\bar{P}}_{\boldsymbol{\hat{B}}}\left(\boldsymbol{v}\right) = \boldsymbol{\hat{B}}^{\dag}\cdot\left(\boldsymbol{\hat{B}}\wedge\boldsymbol{v}\right) = \frac{\boldsymbol{\hat{B}}\wedge\boldsymbol{v}}{\boldsymbol{\hat{B}}}
\end{equation}

\begin{figure}[!ht]
\begin{center}
\includegraphics[scale=0.85]{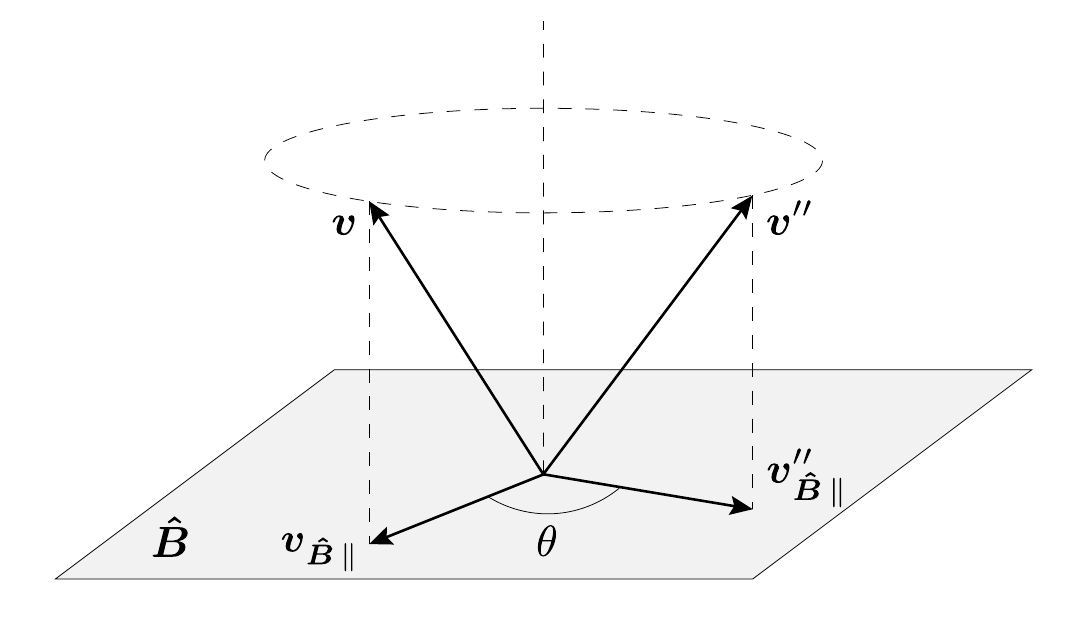}
\end{center}
\caption{Rotation from a projected vector $\boldsymbol{v}_{\parallel\,\boldsymbol{\hat{B}}}$ by an angle $\theta$ to the projected vector $\boldsymbol{v}^{\prime\prime}_{\parallel\,\boldsymbol{\hat{B}}}$.}\label{Fig: Projected rotation}
\end{figure}

\noindent In view of relations~\eqref{v projection and rejection plane},~\eqref{projection automorphism plane} and~\eqref{rejection automorphism plane}, the complementarity between the subspaces $\mathsf{P}_{\boldsymbol{\hat{B}}}$ and $\mathsf{\bar{P}}_{\boldsymbol{\hat{B}}}$ is written as,
\begin{equation}\label{reflection complementarity plane}
\boldsymbol{v} = \mathsf{P}_{\boldsymbol{\hat{B}}}\left(\boldsymbol{v}\right) + \mathsf{\bar{P}}_{\boldsymbol{\hat{B}}}\left(\boldsymbol{v}\right)
\end{equation}
In view of relations~\eqref{rotor angle}, the projection of the rotor $R$ and its reverse $R^{\dag}$ onto the unit bivector $\boldsymbol{\hat{B}}$ is the identity,
\begin{equation}\label{rotor angle projection 0}
\begin{split}
&\mathsf{P}_{\boldsymbol{\hat{B}}}\left(R\right) = \boldsymbol{\hat{B}}^{\dag}\cdot\,\Big(\boldsymbol{\hat{B}}\left(\cos\phi -\,\sin\phi\,\boldsymbol{\hat{B}}\right)\!\Big) = \cos\phi -\,\sin\phi\,\boldsymbol{\hat{B}} = R\\
&\mathsf{P}_{\boldsymbol{\hat{B}}}\left(R^{\dag}\right) = \boldsymbol{\hat{B}}^{\dag}\cdot\,\Big(\boldsymbol{\hat{B}}\left(\cos\phi + \sin\phi\,\boldsymbol{\hat{B}}\right)\!\Big) = \cos\phi + \sin\phi\,\boldsymbol{\hat{B}} = R^{\dag}
\end{split}
\end{equation}
Taking into account relations~\eqref{rotor angle bis} and~\eqref{rotor angle projection 0}, we obtain the following projection identities,
\begin{equation}\label{rotor angle projection}
\mathsf{P}_{\boldsymbol{\hat{B}}}\left(e^{-\,\boldsymbol{\hat{B}}\phi}\right) = e^{-\,\boldsymbol{\hat{B}}\phi} \qquad\text{and}\qquad \mathsf{P}_{\boldsymbol{\hat{B}}}\left(e^{\boldsymbol{\hat{B}}\phi}\right) = e^{\boldsymbol{\hat{B}}\phi}
\end{equation}
Since a projection is a linear map, the projection of the rotation~\eqref{rotation rotor phi} onto the rotation plane is written as,
\begin{equation}\label{rotation rotor phi projection 0}
\mathsf{P}_{\boldsymbol{\hat{B}}}\left(\boldsymbol{v}^{\prime\prime}\right) = \mathsf{P}_{\boldsymbol{\hat{B}}}\left(e^{-\,\boldsymbol{\hat{B}}\phi}\,\boldsymbol{v}\,e^{\boldsymbol{\hat{B}}\phi}\right) = \mathsf{P}_{\boldsymbol{\hat{B}}}\left(e^{-\,\boldsymbol{\hat{B}}\phi}\right)\mathsf{P}_{\boldsymbol{\hat{B}}}\left(\boldsymbol{v}\right)\mathsf{P}_{\boldsymbol{\hat{B}}}\left(e^{\boldsymbol{\hat{B}}\phi}\right)
\end{equation}
In view of the identities~\eqref{rotor angle projection}, the projection~\eqref{rotation rotor phi projection 0} becomes,
\begin{equation}\label{rotation rotor phi projection}
\mathsf{P}_{\boldsymbol{\hat{B}}}\left(\boldsymbol{v}^{\prime\prime}\right) = e^{-\,\boldsymbol{\hat{B}}\phi}\,\mathsf{P}_{\boldsymbol{\hat{B}}}\left(\boldsymbol{v}\right)e^{\boldsymbol{\hat{B}}\phi}
\end{equation}
Using the projection~\eqref{projection automorphism plane} and the Euler formula~\eqref{Euler formula}, the projection~\eqref{rotation rotor phi projection} becomes,
\begin{equation}\label{rotation rotor phi projection bis}
\boldsymbol{\hat{B}}^{\dag}\cdot\left(\boldsymbol{\hat{B}}\cdot\boldsymbol{v}^{\prime\prime}\right) = \left(\cos\phi -\,\sin\phi\,\boldsymbol{\hat{B}}\right)\boldsymbol{\hat{B}}^{\dag}\cdot\left(\boldsymbol{\hat{B}}\cdot\boldsymbol{v}\right)\left(\cos\phi + \sin\phi\,\boldsymbol{\hat{B}}\right)
\end{equation}
Using the fact that the unit bivector commutes with its reverse, i.e. $\boldsymbol{\hat{B}}\,\boldsymbol{\hat{B}}^{\dag} = \boldsymbol{\hat{B}}^{\dag}\,\boldsymbol{\hat{B}}$, and the antisymmetry of the inner product~\eqref{inner product v B antisymmetric} of the bivector $\boldsymbol{\hat{B}}$ with the vector $\boldsymbol{\hat{B}}\cdot\boldsymbol{v}$, we obtain the following identity,
\begin{equation}\label{identity projection B v}
\boldsymbol{\hat{B}}\left(\boldsymbol{\hat{B}}\cdot\boldsymbol{v}\right) = \boldsymbol{\hat{B}}\cdot\left(\boldsymbol{\hat{B}}\cdot\boldsymbol{v}\right) = -\,\left(\boldsymbol{\hat{B}}\cdot\boldsymbol{v}\right)\cdot\boldsymbol{\hat{B}} = -\,\left(\boldsymbol{\hat{B}}\cdot\boldsymbol{v}\right)\boldsymbol{\hat{B}}
\end{equation}
Using the identity~\eqref{identity projection B v}, the projection~\eqref{rotation rotor phi projection bis} is recast as (Fig.~\ref{Fig: Projected rotation}),
\begin{equation}\label{rotation rotor phi projection ter}
\boldsymbol{\hat{B}}^{\dag}\cdot\left(\boldsymbol{\hat{B}}\cdot\boldsymbol{v}^{\prime\prime}\right) = \boldsymbol{\hat{B}}^{\dag}\cdot\left(\boldsymbol{\hat{B}}\cdot\boldsymbol{v}\right)\left(\cos\phi + \sin\phi\,\boldsymbol{\hat{B}}\right)^2
\end{equation}
Using the projection~\eqref{projection automorphism plane} and the Euler formula~\eqref{Euler formula}, the projection~\eqref{rotation rotor phi projection ter} becomes,
\begin{equation}\label{rotation rotor phi projection quad}
\mathsf{P}_{\boldsymbol{\hat{B}}}\left(\boldsymbol{v}^{\prime\prime}\right) = \mathsf{P}_{\boldsymbol{\hat{B}}}\left(\boldsymbol{v}\right)e^{2\boldsymbol{\hat{B}}\phi}
\end{equation}
The projection of the vectors $\boldsymbol{v}$ and $\boldsymbol{v}^{\prime\prime}$ on the rotation plane are defined as,
\begin{equation}\label{projection rotation plane}
\boldsymbol{v}_{\parallel\,\boldsymbol{\hat{B}}} = \mathsf{P}_{\boldsymbol{\hat{B}}}\left(\boldsymbol{v}\right) 
\qquad\text{and}\qquad
\boldsymbol{v}^{\prime\prime}_{\parallel\,\boldsymbol{\hat{B}}} = \mathsf{P}_{\boldsymbol{\hat{B}}}\left(\boldsymbol{v}^{\prime\prime}\right)
\end{equation}
Thus, the projection~\eqref{rotation rotor phi projection quad} is recast as,
\begin{equation}\label{rotation rotor phi projection pent}
\boldsymbol{v}^{\prime\prime}_{\parallel\,\boldsymbol{\hat{B}}} = \boldsymbol{v}_{\parallel\,\boldsymbol{\hat{B}}}\,e^{2\boldsymbol{\hat{B}}\phi}
\end{equation}
The geometric product of the vectors $\boldsymbol{v}_{\parallel\,\boldsymbol{\hat{B}}}$ and $\boldsymbol{v}^{\prime\prime}_{\parallel\,\boldsymbol{\hat{B}}}$ is written as (Fig.~\ref{Fig: Projected rotation}),
\begin{equation}\label{rotation rotor phi projection hex}
\boldsymbol{v}_{\parallel\,\boldsymbol{\hat{B}}}\,\boldsymbol{v}^{\prime\prime}_{\parallel\,\boldsymbol{\hat{B}}} = \boldsymbol{v}_{\parallel\,\boldsymbol{\hat{B}}}\cdot\boldsymbol{v}^{\prime\prime}_{\parallel\,\boldsymbol{\hat{B}}} + \boldsymbol{v}_{\parallel\,\boldsymbol{\hat{B}}}\wedge\boldsymbol{v}^{\prime\prime}_{\parallel\,\boldsymbol{\hat{B}}}
\end{equation}

\noindent The inner and outer products of the vectors $\boldsymbol{v}_{\parallel\,\boldsymbol{\hat{B}}}$ and $\boldsymbol{v}^{\prime\prime}_{\parallel\,\boldsymbol{\hat{B}}}$ can be written in terms of the angle $\theta$ between the vectors in the rotation plane,
\begin{equation}\label{inner outer rotate}
\begin{split}
&\boldsymbol{v}_{\parallel\,\boldsymbol{\hat{B}}}\cdot\boldsymbol{v}^{\prime\prime}_{\parallel\,\boldsymbol{\hat{B}}} = \vert\boldsymbol{v}_{\parallel\,\boldsymbol{\hat{B}}}\vert\,\vert\boldsymbol{v}^{\prime\prime}_{\parallel\,\boldsymbol{\hat{B}}}\vert\cos\theta\\ &\boldsymbol{v}_{\parallel\,\boldsymbol{\hat{B}}}\wedge\boldsymbol{v}^{\prime\prime}_{\parallel\,\boldsymbol{\hat{B}}} = \vert\boldsymbol{v}_{\parallel\,\boldsymbol{\hat{B}}}\vert\,\vert\boldsymbol{v}^{\prime\prime}_{\parallel\,\boldsymbol{\hat{B}}}\vert\sin\theta\,\boldsymbol{\hat{B}}
\end{split}
\end{equation}
where the same unit bivector $\boldsymbol{\hat{B}}$ rotates $\boldsymbol{v}_{\parallel\,\boldsymbol{\hat{B}}}\cdot\boldsymbol{v}_{\parallel\,\boldsymbol{\hat{B}}}$ onto $\boldsymbol{v}_{\parallel\,\boldsymbol{\hat{B}}}\cdot\boldsymbol{v}^{\prime\prime}_{\parallel\,\boldsymbol{\hat{B}}}$ and $\boldsymbol{\hat{n}}_1$ onto $\boldsymbol{v}$. In view of the products~\eqref{inner outer rotate} and the Euler formula~\eqref{Euler formula} for the angle $\theta$, the geometric product~\eqref{rotation rotor phi projection hex} is recast as,
\begin{equation}\label{rotation rotor phi projection hep}
\boldsymbol{v}_{\parallel\,\boldsymbol{\hat{B}}}\,\boldsymbol{v}^{\prime\prime}_{\parallel\,\boldsymbol{\hat{B}}} = \vert\boldsymbol{v}_{\parallel\,\boldsymbol{\hat{B}}}\vert\,\vert\boldsymbol{v}^{\prime\prime}_{\parallel\,\boldsymbol{\hat{B}}}\vert\left(\cos\theta + \sin\theta\,\boldsymbol{\hat{B}}\right) = \vert\boldsymbol{v}_{\parallel\,\boldsymbol{\hat{B}}}\vert\,\vert\boldsymbol{v}^{\prime\prime}_{\parallel\,\boldsymbol{\hat{B}}}\vert\,e^{\boldsymbol{\hat{B}}\theta}
\end{equation}
Since the vector $\boldsymbol{v}^{\prime\prime}_{\parallel\,\boldsymbol{\hat{B}}}$ is obtained by rotating the vector $\boldsymbol{v}_{\parallel\,\boldsymbol{\hat{B}}}$ in the rotation plane, their moduli are equal,
\begin{equation}\label{rotation moduli}
\vert\boldsymbol{v}^{\prime\prime}_{\parallel\,\boldsymbol{\hat{B}}}\vert = \vert\boldsymbol{v}_{\parallel\,\boldsymbol{\hat{B}}}\vert
\end{equation}
Multiplying relation~\eqref{rotation rotor phi projection hep} on the left by the vector $\boldsymbol{v}_{\parallel\,\boldsymbol{\hat{B}}}$ and taking into account the identity~\eqref{rotation moduli} yields,
\begin{equation}\label{rotation rotor phi projection oct}
\boldsymbol{v}_{\parallel\,\boldsymbol{\hat{B}}}^2\,\boldsymbol{v}^{\prime\prime}_{\parallel\,\boldsymbol{\hat{B}}} = \vert\boldsymbol{v}_{\parallel\,\boldsymbol{\hat{B}}}\vert^2\boldsymbol{v}^{\prime\prime}_{\parallel\,\boldsymbol{\hat{B}}} = \vert\boldsymbol{v}_{\parallel\,\boldsymbol{\hat{B}}}\vert^2\,\boldsymbol{v}_{\parallel\,\boldsymbol{\hat{B}}}\,e^{\boldsymbol{\hat{B}}\theta}
\end{equation}
which implies that,
\begin{equation}\label{rotation rotor theta projection}
\boldsymbol{v}^{\prime\prime}_{\parallel\,\boldsymbol{\hat{B}}} = \boldsymbol{v}_{\parallel\,\boldsymbol{\hat{B}}}\,e^{\boldsymbol{\hat{B}}\theta}
\end{equation}
Comparing the rotations~\eqref{rotation rotor phi projection pent} and~\eqref{rotation rotor theta projection}, we conclude that the angle $\theta$ between the vectors $\boldsymbol{v}_{\parallel\,\boldsymbol{\hat{B}}}$ and $\boldsymbol{v}^{\prime\prime}_{\parallel\,\boldsymbol{\hat{B}}}$ is the double of the angle $\phi$ between the units vectors $\boldsymbol{\hat{n}}_1$ and $\boldsymbol{\hat{n}}_2$,
\begin{equation}\label{rotation angles theta phi}
\theta = 2\,\phi
\end{equation}
Thus, the rotation~\eqref{rotation rotor phi} of the vector $\boldsymbol{v}$ is recast in terms of the angle $\theta$ as,
\begin{equation}\label{rotation rotor phi bis}
\boldsymbol{v}^{\prime\prime} = e^{-\,\boldsymbol{\hat{B}}\theta/2}\,\boldsymbol{v}\,e^{\boldsymbol{\hat{B}}\theta/2}
\end{equation}
The rotor and its reverse~\eqref{rotor angle bis} can be recast in terms of the angle $\theta$,
\begin{equation}\label{rotor angle theta}
R = e^{-\,\boldsymbol{\hat{B}}\theta/2} \qquad\text{and}\qquad R^{\dag} = e^{\boldsymbol{\hat{B}}\theta/2}
\end{equation}
In view of relation~\eqref{rotation rotor phi bis}, we conclude that the rotation of a vector $\boldsymbol{v}$ by an angle $\theta$ in a plane along the unit bivector $\boldsymbol{\hat{B}}$ is an automorphism $\mathsf{R}_{\boldsymbol{\hat{B}}\,\theta} : \mathbb{R}^{3} \to \mathbb{R}^{3}$ defined as,
\begin{equation}\label{rotation automorphism}
\mathsf{R}_{\boldsymbol{\hat{B}}\,\theta}\left(\boldsymbol{v}\right) = e^{-\,\boldsymbol{\hat{B}}\theta/2}\,\boldsymbol{v}\,e^{\boldsymbol{\hat{B}}\theta/2}
\end{equation}
According to the orthogonality condition~\eqref{modulus rotor}, the transformation law for the geometric product of two vectors $\boldsymbol{u}\,\boldsymbol{v}$ under rotation by an angle $\theta$ in a plane along the unit bivector $\boldsymbol{\hat{B}}$ is an automorphism,
\begin{equation}\label{rotation geometric product}
\mathsf{R}_{\boldsymbol{\hat{B}}\,\theta}\left(\boldsymbol{u}\,\boldsymbol{v}\right) = R\,\boldsymbol{u}\,\boldsymbol{v}\,R^{\dag} = R\,\boldsymbol{u}\,R^{\dag}\,R\,\boldsymbol{v}\,R^{\dag} = \mathsf{R}_{\boldsymbol{\hat{B}}\,\theta}\left(\boldsymbol{u}\right)\mathsf{R}_{\boldsymbol{\hat{B}}\,\theta}\left(\boldsymbol{v}\right)
\end{equation}
According to the orthogonality condition~\eqref{modulus rotor}, the transformation law for the inner product of two vectors $\boldsymbol{u}\cdot\boldsymbol{v}$ under rotation by an angle $\theta$ in a plane along the unit bivector $\boldsymbol{\hat{B}}$ is an innermorphism,
\begin{equation}\label{rotation inner product}
\mathsf{R}_{\boldsymbol{\hat{B}}\,\theta}\left(\boldsymbol{u}\cdot\boldsymbol{v}\right) = R\,\boldsymbol{u}\cdot\boldsymbol{v}\,R^{\dag} = \left(R\,\boldsymbol{u}\,R^{\dag}\right)\cdot\left(R\,\boldsymbol{v}\,R^{\dag}\right) = \mathsf{R}_{\boldsymbol{\hat{B}}\,\theta}\left(\boldsymbol{u}\right)\cdot\mathsf{R}_{\boldsymbol{\hat{B}}\,\theta}\left(\boldsymbol{v}\right)
\end{equation}
In view of the transformation laws~\eqref{rotation geometric product} and~\eqref{rotation inner product}, the transformation law for the outer product of two vectors $\boldsymbol{u}\wedge\boldsymbol{v}$ under rotation by an angle $\theta$ in a plane along the unit bivector $\boldsymbol{\hat{B}}$ is an outermorphism,
\begin{equation}\label{rotation outer product}
\mathsf{R}_{\boldsymbol{\hat{B}}\,\theta}\left(\boldsymbol{u}\wedge\boldsymbol{v}\right) = R\,\boldsymbol{u}\wedge\boldsymbol{v}\,R^{\dag} = \left(R\,\boldsymbol{u}\,R^{\dag}\right)\wedge\left(R\,\boldsymbol{v}\,R^{\dag}\right) = \mathsf{R}_{\boldsymbol{\hat{B}}\,\theta}\left(\boldsymbol{u}\right)\wedge\mathsf{R}_{\boldsymbol{\hat{B}}\,\theta}\left(\boldsymbol{v}\right)
\end{equation}
In view of the rotor~\eqref{rotor angle theta} and the transformation law~\eqref{rotation outer product}, the transformation law for a bivector $\boldsymbol{A} = \boldsymbol{u}\wedge\boldsymbol{v}$ under rotation by an angle $\theta$ in a plane along the unit bivector $\boldsymbol{\hat{B}}$ is an automorphism,
\begin{equation}\label{rotation bivector}
\mathsf{R}_{\boldsymbol{\hat{B}}\,\theta}\left(\boldsymbol{A}\right) = R\,\boldsymbol{A}\,R^{\dag} = e^{-\,\boldsymbol{\hat{B}}\theta/2}\,\boldsymbol{A}\,e^{\boldsymbol{\hat{B}}\theta/2}
\end{equation}
Using the transformation laws for vectors~\eqref{rotation automorphism} and bivectors~\eqref{rotation bivector}, the transformation law for a multivector $M = s + \boldsymbol{v} + \boldsymbol{A} + s^{\prime}\,I$~\eqref{multivector} is an automorphism $\mathsf{R}_{\boldsymbol{\hat{B}}\,\theta} : \mathbb{G}^{3} \to \mathbb{G}^{3}$ defined as,
\begin{equation}\label{rotation multivector}
\mathsf{R}_{\boldsymbol{\hat{B}}\,\theta}\left(M\right) = R\,M\,R^{\dag} = e^{-\,\boldsymbol{\hat{B}}\theta/2}\,M\,e^{\boldsymbol{\hat{B}}\theta/2}
\end{equation}
The composition of two rotations of a vector $\boldsymbol{v}$ can be written as,
\begin{equation}\label{composition rotations}
\begin{split}
&\mathsf{R}_{\boldsymbol{\hat{B}}_2\,\theta_2}\circ\mathsf{R}_{\boldsymbol{\hat{B}}_1\,\theta_1}\left(\boldsymbol{v}\right) = \mathsf{R}_{\boldsymbol{\hat{B}}_2\,\theta_2}\,\Big(\mathsf{R}_{\boldsymbol{\hat{B}}_1\,\theta_1}\left(\boldsymbol{v}\right)\!\Big) = \mathsf{R}_{\boldsymbol{\hat{B}}_2\,\theta_2}\,\Big(e^{-\,\boldsymbol{\hat{B}}_1\theta_1/2}\,\boldsymbol{v}\,e^{\boldsymbol{\hat{B}}_1\theta_1/2}\Big)\\
&\phantom{\mathsf{R}_{\boldsymbol{\hat{B}}_2\,\theta_2}\circ\mathsf{R}_{\boldsymbol{\hat{B}}_1\,\theta_1}\left(\boldsymbol{v}\right)} = e^{-\,\boldsymbol{\hat{B}}_2\theta_2/2}\,e^{-\,\boldsymbol{\hat{B}}_1\theta_1/2}\,\boldsymbol{v}\,e^{\boldsymbol{\hat{B}}_1\theta_1/2}\,e^{\boldsymbol{\hat{B}}_2\theta_2/2}
\end{split}
\end{equation}
In view of the rotors~\eqref{rotor angle theta} the composition of rotations~\eqref{composition rotations} is a rotation,
\begin{equation}\label{composition rotations bis}
\mathsf{R}_{\boldsymbol{\hat{B}}\,\theta}\left(\boldsymbol{v}\right) = \mathsf{R}_{\boldsymbol{\hat{B}}_2\,\theta_2}\circ\mathsf{R}_{\boldsymbol{\hat{B}}_1\,\theta_1}\left(\boldsymbol{v}\right) = R_2\,R_1\,\boldsymbol{v}\,R_1^{\dag}\,R_2^{\dag} = R\,\boldsymbol{v}\,R^{\dag}
\end{equation}
where the rotor characterising the composition of the two rotations is,
\begin{equation}\label{composition rotations rotor}
R = R_2\,R_1 = e^{-\left(\boldsymbol{\hat{B}}_1\theta_1 + \boldsymbol{\hat{B}}_2\theta_2\right)/2} = e^{-\,\boldsymbol{\hat{B}}\theta/2}
\end{equation}
and the bivector characterising the composition of the two rotations is,
\begin{equation}\label{composition rotations bivector}
\frac{1}{2}\left(\boldsymbol{\hat{B}}\theta\right) = \frac{1}{2}\left(\boldsymbol{\hat{B}}_1\theta_1 + \boldsymbol{\hat{B}}_2\theta_2\right)
\end{equation}
The composition of two rotations reveals that a rotation $\mathsf{R}_{\boldsymbol{\hat{B}}\,\theta}$ is an element of a Lie group called the rotation group. It also shows that a rotor $R$ is an element of another Lie group called the spin group, and the bivector $\boldsymbol{\hat{B}}\theta/2$ is an element of the spin algebra. There is a fundamental difference between a vector and a rotor under rotation. Unlike the rotation of a vector that is subject to a double-sided transformation law~\eqref{rotation automorphism}, rotor are subject to a single-sided transformation law~\eqref{composition rotations rotor}. In order to see how the consequence of this difference, we consider a rotation by an angle $\theta = 2\pi$. In view of relation~\eqref{rotation automorphism}, a $360^{\circ}$ rotation of the vector $\boldsymbol{v}$ leaves it unchanged,
\begin{equation}\label{rotation vector 2pi}
\mathsf{R}_{\boldsymbol{\hat{B}}\,2\pi}\left(\boldsymbol{v}\right) = e^{-\,\boldsymbol{\hat{B}}\pi}\,\boldsymbol{v}\,e^{\boldsymbol{\hat{B}}\pi} = \left(-\,1\right)\boldsymbol{v}\left(-\,1\right) = \boldsymbol{v}
\end{equation}
as expected. In view of relation~\eqref{composition rotations rotor}, a $360^{\circ}$ rotation of the rotor $R_1$ characterised by the rotor $R_2 = e^{-\,\boldsymbol{\hat{B}}\pi}$ changes the sign of the rotor,
\begin{equation}\label{rotation rotor 2pi}
R = e^{-\,\boldsymbol{\hat{B}}\pi}\,R_1 = \left(-\,1\right)R_1 = -\,R_1
\end{equation}
which implies that a $720^{\circ}$ rotation of the rotor $R_1$ leaves it unchanged. Rotors transform under rotation as spinors that describe the state of fermions in quantum mechanics. The rotation group is isomorphic to the special orthogonal group $\text{SO}\left(3\right)$ and the spin group $\text{Spin}\left(3\right)$ is isomorphic to the special unitary group $\text{SU}\left(2\right)$, i.e $\text{Spin}\left(3\right) \cong \text{SU}\left(2\right)$. The difference between a vector and a rotor under rotation is due to the fact that the spin group is the double cover of the rotation group exactly like the special unitary group that is the double cover of the special orthogonal group, i.e. $\text{SU}\left(2\right)/\mathbb{Z}_2 \cong \text{SO}\left(3\right)$.


\section{Rotors, quaternions and the Pauli algebra}
\label{Rotors, quaternions and the Pauli algebra}

\noindent A unit bivector $\boldsymbol{\hat{B}}$ in a rotation plane can be written in an orthonormal frame $\left\{\boldsymbol{\hat{e}}_1,\boldsymbol{\hat{e}}_2,\boldsymbol{\hat{e}}_3\right\}$ as,
\begin{equation}\label{bivector orthonormal basis}
\boldsymbol{\hat{B}} = B_{12}\,\boldsymbol{\hat{e}}_1\wedge\boldsymbol{\hat{e}}_2 + B_{23}\,\boldsymbol{\hat{e}}_2\wedge\boldsymbol{\hat{e}}_3 + B_{31}\,\boldsymbol{\hat{e}}_3\wedge\boldsymbol{\hat{e}}_1 = B_{12}\,\boldsymbol{\hat{e}}_1\,\boldsymbol{\hat{e}}_2 + B_{23}\,\boldsymbol{\hat{e}}_2\,\boldsymbol{\hat{e}}_3 + B_{31}\,\boldsymbol{\hat{e}}_3\,\boldsymbol{\hat{e}}_1
\end{equation}
where the normalisation condition is,
\begin{equation}\label{bivector unitary}
B_{12}^2 + B_{23}^2 + B_{31}^2 = 1
\end{equation}
The basis unit bivectors satisfy the following conditions,
\begin{equation}\label{bivector basis condition}
\begin{split}
&\left(\boldsymbol{\hat{e}}_1\,\boldsymbol{\hat{e}}_2\right)^2 = \boldsymbol{\hat{e}}_1\,\boldsymbol{\hat{e}}_2\,\boldsymbol{\hat{e}}_1\,\boldsymbol{\hat{e}}_2 = -\,\boldsymbol{\hat{e}}_1\,\boldsymbol{\hat{e}}_1\,\boldsymbol{\hat{e}}_2\,\boldsymbol{\hat{e}}_2 = -\,1\\
&\left(\boldsymbol{\hat{e}}_2\,\boldsymbol{\hat{e}}_3\right)^2 = \boldsymbol{\hat{e}}_2\,\boldsymbol{\hat{e}}_3\,\boldsymbol{\hat{e}}_2\,\boldsymbol{\hat{e}}_3 = -\,\boldsymbol{\hat{e}}_2\,\boldsymbol{\hat{e}}_2\,\boldsymbol{\hat{e}}_3\,\boldsymbol{\hat{e}}_3 = -\,1\\
&\left(\boldsymbol{\hat{e}}_3\,\boldsymbol{\hat{e}}_1\right)^2 = \boldsymbol{\hat{e}}_3\,\boldsymbol{\hat{e}}_1\,\boldsymbol{\hat{e}}_3\,\boldsymbol{\hat{e}}_1 = -\,\boldsymbol{\hat{e}}_3\,\boldsymbol{\hat{e}}_3\,\boldsymbol{\hat{e}}_1\,\boldsymbol{\hat{e}}_1 = -\,1\\
&\left(\boldsymbol{\hat{e}}_1\,\boldsymbol{\hat{e}}_2\right)\left(\boldsymbol{\hat{e}}_2\,\boldsymbol{\hat{e}}_3\right)\left(\boldsymbol{\hat{e}}_3\,\boldsymbol{\hat{e}}_1\right) = 1
\end{split}
\end{equation}
In view of relations~\eqref{rotor angle} and~\eqref{bivector orthonormal basis}, a rotor $R$ generating a rotation of angle $\theta$ in a rotation plane along the unit bivector $\boldsymbol{\hat{B}}$ can be expressed in the orthonormal bivector basis as,
\begin{equation}\label{rotor orthornormal basis}
R = \cos\left(\frac{\theta}{2}\right) -\,B_{12}\sin\left(\frac{\theta}{2}\right)\boldsymbol{\hat{e}}_1\,\boldsymbol{\hat{e}}_2 -\,B_{23}\sin\left(\frac{\theta}{2}\right)\boldsymbol{\hat{e}}_2\,\boldsymbol{\hat{e}}_3 -\,B_{31}\sin\left(\frac{\theta}{2}\right)\boldsymbol{\hat{e}}_3\,\boldsymbol{\hat{e}}_1
\end{equation}
The bivector identities~\eqref{bivector basis condition} are isomorphic to the formula for the quaternions with the identifications,
\begin{equation}\label{bivectors quaternions}
i \leftrightarrow -\,\boldsymbol{\hat{e}}_1\,\boldsymbol{\hat{e}}_2 \qquad\text{and}\qquad
j \leftrightarrow -\,\boldsymbol{\hat{e}}_2\,\boldsymbol{\hat{e}}_3 \qquad\text{and}\qquad
k \leftrightarrow -\,\boldsymbol{\hat{e}}_3\,\boldsymbol{\hat{e}}_1
\end{equation}
Indeed, in view of the identifications~\eqref{bivectors quaternions}, bivector identities~\eqref{bivector basis condition} yield formula for the quaternions,
\begin{equation}\label{formula quaternions}
i^2 = j^2 = k^2 = ijk = -\,1
\end{equation}
With the identification~\eqref{bivectors quaternions}, the rotor~\eqref{rotor orthornormal basis} can be written as a quaternion,
\begin{equation}\label{rotor quaternion}
R = a + b\,i + c\,j + d\,k
\end{equation}
where the real coefficients are,
\begin{equation}\label{rotor quaternion coefficients}
\begin{split}
&a = \cos\left(\frac{\theta}{2}\right) \quad\quad\quad\ \,\text{and}\quad b = -\,B_{12}\sin\left(\frac{\theta}{2}\right)\\
&c = -\,B_{23}\sin\left(\frac{\theta}{2}\right) \quad\text{and}\quad d = -\,B_{31}\sin\left(\frac{\theta}{2}\right)
\end{split}
\end{equation}
and satisfy the normalisation condition,
\begin{equation}\label{normalisation condition}
a^2 + b^2 + c^2 + d^2 = 1
\end{equation}
This identification shows that the spin group $\text{Spin}\left(3\right)$ is isomorphic to the hyperunitary group $\text{Sp}\left(1\right)$ consisting of the unit quaternions $\mathbb{H}$, i.e. $\text{Spin}\left(3\right) \cong \text{Sp}\left(1\right)$. After showing that rotors are isomorphic to unit quaternions, we now show how rotors or unit bivectors are related to the Pauli algebra. The totally antisymmetric Levi-Civita symbols are defined as,
\begin{equation}\label{Levi Civita}
\varepsilon_{\sigma\left(123\right)} = \left(-\,1\right)^{\text{sgn}\left(\sigma\right)}\varepsilon_{123}
\end{equation}
where $\sigma\left(123\right)$ is a permutation of the indices $123$ and $\text{sgn}\left(\sigma\right)$ is its signature. In view of the pseudoscalar $I = \boldsymbol{\hat{e}}_1\,\boldsymbol{\hat{e}}_2\,\boldsymbol{\hat{e}}_3$ and the Levi-Civita symbol~\eqref{Levi Civita}, the unit bivectors can be recast as,
\begin{equation}\label{bivectors algebra}
\begin{split}
&\boldsymbol{\hat{e}}_1\,\boldsymbol{\hat{e}}_2 = \boldsymbol{\hat{e}}_1\,\boldsymbol{\hat{e}}_2\,\boldsymbol{\hat{e}}_3\,\boldsymbol{\hat{e}}_3 = I\,\boldsymbol{\hat{e}}_3 = I\,\varepsilon_{123}\,\boldsymbol{\hat{e}}_3\\
&\boldsymbol{\hat{e}}_2\,\boldsymbol{\hat{e}}_3 = \boldsymbol{\hat{e}}_2\,\boldsymbol{\hat{e}}_3\,\boldsymbol{\hat{e}}_1\,\boldsymbol{\hat{e}}_1 = I\,\boldsymbol{\hat{e}}_1 = I\,\varepsilon_{231}\,\boldsymbol{\hat{e}}_1\\
&\boldsymbol{\hat{e}}_3\,\boldsymbol{\hat{e}}_1 = \boldsymbol{\hat{e}}_3\,\boldsymbol{\hat{e}}_1\,\boldsymbol{\hat{e}}_2\,\boldsymbol{\hat{e}}_2 = I\,\boldsymbol{\hat{e}}_2 = I\,\varepsilon_{312}\,\boldsymbol{\hat{e}}_2
\end{split}
\end{equation}
where $\varepsilon_{123} = \varepsilon_{231} = \varepsilon_{312} = 1$. Thus, the outer product of the unit vectors $\boldsymbol{\hat{e}}_i$ and $\boldsymbol{\hat{e}}_j$ with $i,j,k =1,2,3$ is written as,
\begin{equation}\label{outer product unit vectors}
\boldsymbol{\hat{e}}_i\wedge\boldsymbol{\hat{e}}_j = I\,\varepsilon_{ijk}\,\boldsymbol{\hat{e}}_k
\end{equation}
The geometric product of the unit vectors $\boldsymbol{\hat{e}}_i$ and $\boldsymbol{\hat{e}}_j$ reads,
\begin{equation}\label{geometric product unit}
\boldsymbol{\hat{e}}_i\,\boldsymbol{\hat{e}}_j = \boldsymbol{\hat{e}}_i\cdot\boldsymbol{\hat{e}}_j + \boldsymbol{\hat{e}}_i\wedge\boldsymbol{\hat{e}}_j
\end{equation}
where the orthonormality condition is given by,
\begin{equation}\label{orthonormlity unit}
\boldsymbol{\hat{e}}_i\cdot\boldsymbol{\hat{e}}_j = \delta_{ij}
\end{equation}
In view of the outer product~\eqref{outer product unit vectors} and the inner product~\eqref{orthonormlity unit} of unit vectors, the geometric product~\eqref{geometric product unit} yields the Pauli algebra,
\begin{equation}\label{Pauli algebra}
\boldsymbol{\hat{e}}_i\,\boldsymbol{\hat{e}}_j = \delta_{ij} + I\,\varepsilon_{ijk}\,\boldsymbol{\hat{e}}_k
\end{equation}
Therefore, we conclude that the Pauli algebra, that is usually introduced to characterise the spin in quantum mechanics, has a purely geometric interpretation as the underlying algebra of physical space.


\section{Rotating frame and angular velocity}
\label{Rotating frame and angular velocity}

\begin{figure}[!ht]
\begin{center}
\includegraphics[scale=0.85]{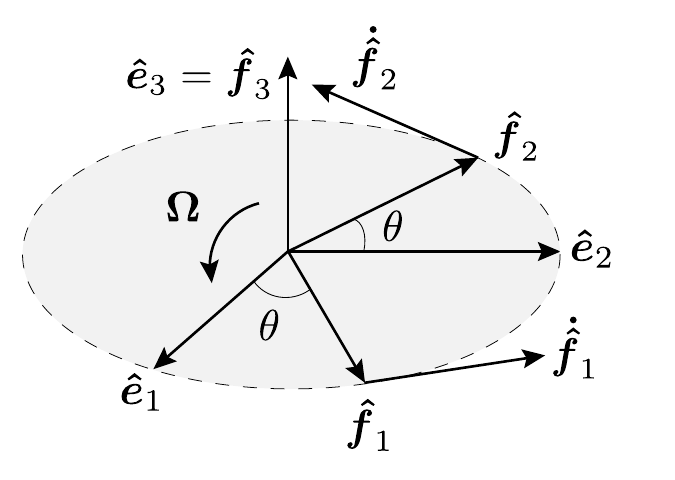}
\end{center}
\caption{Orthonormal frame $\left\{\boldsymbol{\hat{f}}_1,\boldsymbol{\hat{f}}_2,\boldsymbol{\hat{f}}_3\right\}$ rotating with an angular velocity $\boldsymbol{\Omega}$ with respect to an orthonormal frame $\left\{\boldsymbol{\hat{e}}_1,\boldsymbol{\hat{e}}_2,\boldsymbol{\hat{e}}_3\right\}$ in the $\boldsymbol{\hat{e}}_1\wedge\boldsymbol{\hat{e}}_2$ plane.}\label{Fig: Rotating frame}
\end{figure}

\noindent We consider an orthonormal frame $\left\{\boldsymbol{\hat{f}}_1,\boldsymbol{\hat{f}}_2,\boldsymbol{\hat{f}}_3\right\}$ moving with respect to a fixed orthonormal frame $\left\{\boldsymbol{\hat{e}}_1,\boldsymbol{\hat{e}}_2,\boldsymbol{\hat{e}}_3\right\}$. A translation does not affect a vector. Only a rotation of the moving frame changes the basis vectors of the frame. Thus, we consider a rotation by an angle $\theta$ in a plane along the constant unit bivector $\boldsymbol{\hat{B}}$ (Fig.~\ref{Fig: Rotating frame}).

\begin{figure}[!ht]
\begin{center}
\includegraphics[scale=0.55]{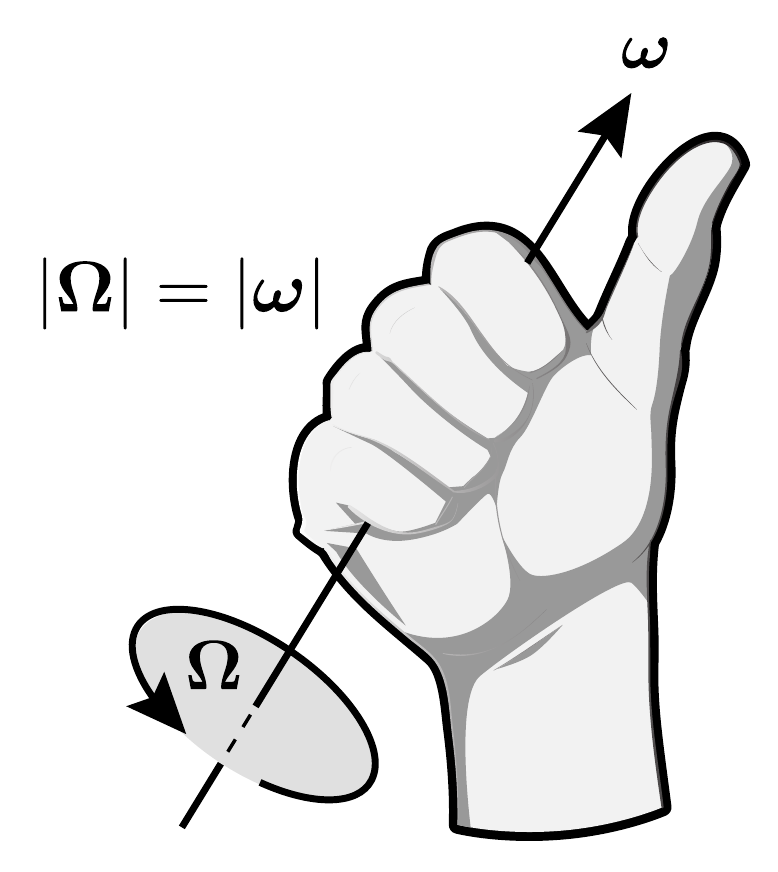}
\end{center}
\caption{Duality between the angular velocity bivector $\boldsymbol{\Omega}$ and the angular velocity pseudovector $\boldsymbol{\omega}$ illustrated by the right hand rule.}\label{Fig: Angular velocity}
\end{figure}

\noindent According to relation~\eqref{rotor angle theta} and~\eqref{rotation automorphism}, the unit vector $\boldsymbol{\hat{f}}_i$ is expressed in terms of the unit vector $\boldsymbol{\hat{e}}_i$ for $i=1,2,3$ as follows,
\begin{equation}\label{rotation frames}
\boldsymbol{\hat{f}}_i = R\,\boldsymbol{\hat{e}}_i\,R^{\dag}
\end{equation}
where the time dependent rotor and its reverse are given by,
\begin{equation}\label{rotor and reverse frames}
R = e^{-\,\boldsymbol{\hat{B}}\theta/2} \qquad\text{and}\qquad R^{\dag} = e^{-\,\boldsymbol{\hat{B}}\theta/2}
\end{equation}
The time derivative of the rotating basis vector~\eqref{rotation frames} is written as,
\begin{equation}\label{rotation frames derivative}
\boldsymbol{\dot{\hat{f}}}_i = \dot{R}\,\boldsymbol{\hat{e}}_i\,R^{\dag} + R\,\boldsymbol{\hat{e}}_i\,\dot{R}^{\dag}
\end{equation}
In view of relation~\eqref{rotation frames}, it can be recast as,
\begin{equation}\label{rotation frames derivative bis}
\boldsymbol{\dot{\hat{f}}}_i = \dot{R}\,R^{\dag}\,\boldsymbol{\hat{f}}_i + \boldsymbol{\hat{f}}_i\,R\,\dot{R}^{\dag}
\end{equation}
In view of relation~\eqref{modulus rotor}, the normalisation condition for a time dependent rotor reads,
\begin{equation}\label{unit rotor time dependent}
R\,R^{\dag} = 1
\end{equation}
The time derivative of the condition~\eqref{unit rotor time dependent} is written as,
\begin{equation}\label{unit rotor time derivative}
\dot{R}\,R^{\dag} + R\,\dot{R}^{\dag} = 0
\end{equation}
Using relation~\eqref{unit rotor time derivative}, the time derivative of the rotating basis vector~\eqref{rotation frames derivative} becomes,
\begin{equation}\label{rotation frames derivative ter}
\boldsymbol{\dot{\hat{f}}}_i = \dot{R}\,R^{\dag}\,\boldsymbol{\hat{f}}_i -\,\boldsymbol{\hat{f}}_i\,\dot{R}\,R^{\dag}
\end{equation}
In view of relation~\eqref{rotor and reverse frames}, we obtain the identity,
\begin{equation}\label{derivative identity}
\dot{R}\,R^{\dag} = -\,\frac{\dot{\theta}}{2}\,\boldsymbol{\hat{B}}\,e^{-\,\boldsymbol{\hat{B}}\theta/2}\,e^{\boldsymbol{\hat{B}}\theta/2} = -\,\frac{\dot{\theta}}{2}\,\boldsymbol{\hat{B}}
\end{equation}
The angular velocity is a bivector is defined as,
\begin{equation}\label{angular velocity bivector}
\boldsymbol{\Omega} = \dot{\theta}\,\boldsymbol{\hat{B}}
\end{equation}
which implies that,
\begin{equation}\label{derivative identity bis}
\dot{R}\,R^{\dag} = -\,\frac{1}{2}\,\boldsymbol{\Omega}
\end{equation}
and can be recast as,
\begin{equation}\label{derivative identity ter}
\dot{R} = -\,\frac{1}{2}\,\boldsymbol{\Omega}\,R
\end{equation}
In view of the reverse~\eqref{reverse B} of the bivector $\boldsymbol{\Omega}$, the reverse of relation~\eqref{derivative identity ter} is written as,
\begin{equation}\label{derivative identity quad}
\dot{R}^{\dag} = \frac{1}{2}\,R^{\dag}\,\boldsymbol{\Omega}
\end{equation}
Using relation~\eqref{derivative identity bis}, the time derivative of the rotating basis vector~\eqref{rotation frames derivative ter} is expressed in terms of the angular velocity $\boldsymbol{\Omega}$ as,
\begin{equation}\label{rotation frames derivative quad}
\boldsymbol{\dot{\hat{f}}}_i = \frac{1}{2}\Big(\boldsymbol{\hat{f}}_i\cdot\boldsymbol{\Omega} -\,\boldsymbol{\Omega}\cdot\boldsymbol{\hat{f}}_i\!\Big)
\end{equation}
In view of the inner product~\eqref{inner product v B antisymmetric} between the vector $\boldsymbol{\hat{f}}_i$ and the bivector $\boldsymbol{\Omega}$, the time derivative of the rotating basis vector~\eqref{rotation frames derivative ter} reduces to,
\begin{equation}\label{Poisson formula}
\boldsymbol{\dot{\hat{f}}}_i = \boldsymbol{\hat{f}}_i\cdot\boldsymbol{\Omega}
\end{equation}
which is the Poisson formula in geometric algebra. To recover the Poisson formula in vector algebra, we define the angular velocity pseudovector $\boldsymbol{\omega}$ as the dual of the angular velocity bivector $\boldsymbol{\Omega}$ using the relations~\eqref{dual v} and~\eqref{dual B},
\begin{equation}\label{angular velocity duality}
\boldsymbol{\omega} = \boldsymbol{\Omega}^{\ast} \equiv \boldsymbol{\Omega}\,I^{-1} \qquad\text{and}\qquad \boldsymbol{\Omega} \equiv -\,\boldsymbol{\omega}^{\ast} = -\,\boldsymbol{\omega}\,I^{-1}
\end{equation}
where the unit pseudoscalar $I^{-1} = -\,I = -\,\boldsymbol{\hat{e}}_1\boldsymbol{\hat{e}}_2\boldsymbol{\hat{e}}_3$ and the dual of the dual is the opposite of the identity, i.e. $\left(\boldsymbol{\Omega}^{\ast}\right)^{\ast} = -\,\boldsymbol{\Omega}$, since $I^2 = -\,1$. Note that the duality preserves the modulus of a multivector. Thus, the plan area covered by the angular velocity bivector $\boldsymbol{\Omega}$ is equal to the length of the angular velocity pseudovector $\boldsymbol{\omega}$,
\begin{equation}\label{duality modulus}
\vert\boldsymbol{\Omega}\vert = \vert\boldsymbol{\omega}\vert
\end{equation}
The geometric interpretation of this duality is the following : if the palm of the right hand is oriented along the angular velocity bivector $\boldsymbol{\Omega}$ in the plane of rotation, then the thumb is oriented along the angular velocity pseudovector $\boldsymbol{\omega}$ (Fig.~\ref{Fig: Angular velocity}). 

\noindent Using identities~\eqref{wedge and cross products duality},~\eqref{vector duality} and~\eqref{angular velocity duality}, the right-hand side of the Poisson formula~\eqref{Poisson formula} is recast as,
\begin{equation}\label{Poisson identity}
\boldsymbol{\hat{f}}_i\cdot\boldsymbol{\Omega} = -\,\boldsymbol{\hat{f}}_i\cdot\boldsymbol{\omega}^{\ast} = -\,\Big(\boldsymbol{\hat{f}}_i\wedge\boldsymbol{\omega}\Big)^{\ast} = -\,\boldsymbol{\hat{f}}_i\times\boldsymbol{\omega}
\end{equation}
In view of the identity~\eqref{Poisson identity}, since the vector cross product is antisymmetric,
\begin{equation}\label{Poisson cross product}
-\,\boldsymbol{\hat{f}}_i\times\boldsymbol{\omega} = \boldsymbol{\omega}\times\boldsymbol{\hat{f}}_i
\end{equation}
the Poisson formula~\eqref{Poisson formula} is recast in vector algebra as,
\begin{equation}\label{Poisson formula vector algebra}
\boldsymbol{\dot{\hat{f}}}_i = \boldsymbol{\omega}\times\boldsymbol{\hat{f}}_i
\end{equation}
as expected. It is worth emphasising that the Poisson formula in geometric algebra~\eqref{Poisson formula} is in a sense much more natural than the Poisson formula in vector algebra~\eqref{Poisson formula vector algebra}. In geometric algebra, the geometric entities are contained entirely in the rotation plane whereas in vector algebra the angular velocity pseudovector is orthogonal to the plane. In geometric algebra, the angular velocity bivector $\boldsymbol{\Omega}$ rotates the basis vector $\boldsymbol{\hat{f}}_i$ by a $90^{\circ}$ angle in the rotation direction defined by the unit bivector $\boldsymbol{\hat{B}}$, this is the geometric interpretation of the inner product $\boldsymbol{\hat{f}}_i\cdot\boldsymbol{\Omega}$.
%


\section{Rotating cylindrical frame}
\label{Rotating cylindrical frame}

\begin{figure}[!ht]
\begin{center}
\includegraphics[scale=0.85]{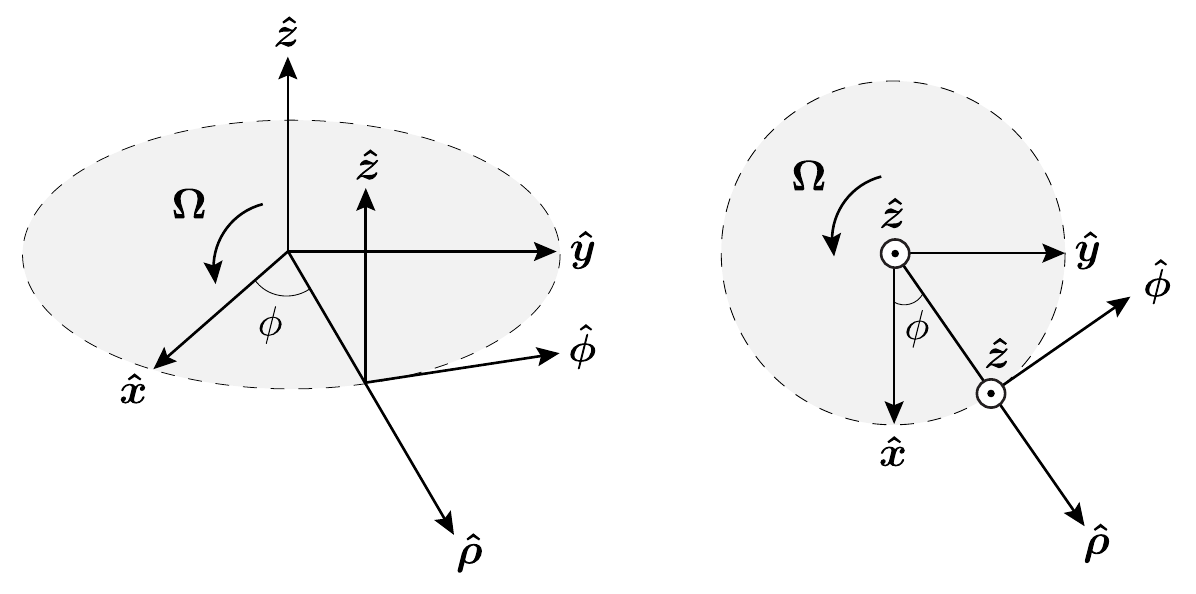}
\end{center}
\caption{Cylindrical frame $\left\{\boldsymbol{\hat{\rho}},\boldsymbol{\hat{\phi}},\boldsymbol{\hat{z}}\right\}$ rotating with an angular velocity $\boldsymbol{\Omega}$ around a fixed Cartesian frame $\left\{\boldsymbol{\hat{x}},\boldsymbol{\hat{y}},\boldsymbol{\hat{z}}\right\}$ in the $\boldsymbol{\hat{x}}\wedge\boldsymbol{\hat{y}}$ plane.}\label{Fig: cylindrical frame}
\end{figure}

\noindent We consider an cylindrical frame $\left\{\boldsymbol{\hat{\rho}},\boldsymbol{\hat{\phi}},\boldsymbol{\hat{z}}\right\}$ rotating around a fixed Cartesian frame $\left\{\boldsymbol{\hat{x}},\boldsymbol{\hat{y}},\boldsymbol{\hat{z}}\right\}$. This rotation is characterised by an azimuthal angle $\phi$ in the horizontal plane (Fig.~\ref{Fig: cylindrical frame}).

\noindent The unit bivector oriented along the rotation in the horizontal plane $\boldsymbol{\hat{B}}_{\phi}$ is written in the Cartesian frame as,
\begin{equation}\label{unit bivector Cartesian}
\boldsymbol{\hat{B}}_{\phi} = \boldsymbol{\hat{x}}\wedge\boldsymbol{\hat{y}} = \boldsymbol{\hat{x}}\,\boldsymbol{\hat{y}}
\end{equation}
where the index $\phi$ refers to the azimuthal angle in the plane of the bivector. In view of the unit bivector~\eqref{unit bivector Cartesian}, the rotation that maps the Cartesian frame $\left\{\boldsymbol{\hat{x}},\boldsymbol{\hat{y}},\boldsymbol{\hat{z}}\right\}$ onto the cylindrical frame $\left\{\boldsymbol{\hat{r}},\boldsymbol{\hat{\theta}},\boldsymbol{\hat{\phi}}\right\}$ is described by the rotor $R_{\phi}$ in the horizontal rotation plane,
\begin{equation}\label{rotor cylindrical}
R_{\phi} = e^{-\,\boldsymbol{\hat{B}}_{\phi}\,\phi/2} = e^{-\,\boldsymbol{\hat{x}}\wedge\boldsymbol{\hat{y}}\,\phi/2} = e^{-\,\boldsymbol{\hat{x}}\,\boldsymbol{\hat{y}}\,\phi/2}
\end{equation}
The time derivative of the rotor~\eqref{rotor cylindrical} is written as,
\begin{equation}\label{time derivative rotor cylindrical}
\dot{R}_{\phi} = -\,\frac{\dot{\phi}}{2}\,\boldsymbol{\hat{x}}\,\boldsymbol{\hat{y}}\,R_{\phi}
\end{equation}
In view of relations~\eqref{unit rotor time dependent},~\eqref{derivative identity bis},~\eqref{unit bivector Cartesian},~\eqref{rotor cylindrical} and~\eqref{time derivative rotor cylindrical}, the angular velocity is given by,
\begin{equation}\label{angular velocity cylindrical}
\boldsymbol{\Omega} = -\,2\,\dot{R}_{\phi}\,R^{\dag}_{\phi} = \dot{\phi}\,\boldsymbol{\hat{x}}\,\boldsymbol{\hat{y}}\,R_{\phi}\,R^{\dag}_{\phi} = \dot{\phi}\,\boldsymbol{\hat{x}}\,\boldsymbol{\hat{y}} = \dot{\phi}\,\boldsymbol{\hat{B}}_{\phi}
\end{equation}
The unit vectors $\boldsymbol{\hat{x}}$ and $\boldsymbol{\hat{y}}$ are written in the cylindrical frame as (Fig.~\ref{Fig: cylindrical frame}),
\begin{equation}\label{rho phi}
\begin{split}
&\boldsymbol{\hat{x}} = \cos\theta\,\boldsymbol{\hat{\rho}} -\,\sin\theta\,\boldsymbol{\hat{\phi}}\\
&\boldsymbol{\hat{y}} = \sin\theta\,\boldsymbol{\hat{\rho}} + \cos\theta\,\boldsymbol{\hat{\phi}}
\end{split}
\end{equation}
In view of the change of basis~\eqref{rho z}, the unit bivector $\boldsymbol{\hat{B}}_{\phi}$ is recast in the cylindrical frame as,
\begin{equation}\label{B rho phi}
\boldsymbol{\hat{B}}_{\phi} = \boldsymbol{\hat{x}}\,\boldsymbol{\hat{y}} = \left(\cos\theta\,\boldsymbol{\hat{\rho}} -\,\sin\theta\,\boldsymbol{\hat{\phi}}\right)\left(\sin\theta\,\boldsymbol{\hat{\rho}} + \cos\theta\,\boldsymbol{\hat{\phi}}\right) = \boldsymbol{\hat{\rho}}\,\boldsymbol{\hat{\phi}}
\end{equation}
The angular velocity bivector~\eqref{angular velocity cylindrical} is recast in the cylindrical frame as,
\begin{equation}\label{angular velocity cylindrical bis}
\boldsymbol{\Omega} = \dot{\phi}\,\boldsymbol{\hat{B}}_{\phi} = \dot{\phi}\,\boldsymbol{\hat{\rho}}\,\boldsymbol{\hat{\phi}}
\end{equation}
The Poisson formula~\eqref{Poisson formula} for the radial unit vector $\boldsymbol{\hat{\rho}}$ and the azimuthal unit vector $\boldsymbol{\hat{\phi}}$ are written as,
\begin{equation}\label{Poisson cylindrical}
\begin{split}
&\boldsymbol{\dot{\hat{\rho}}} = \boldsymbol{\hat{\rho}}\cdot\boldsymbol{\Omega} = \boldsymbol{\hat{\rho}}\ \boldsymbol{\Omega} = \boldsymbol{\hat{\rho}}\left(\dot{\phi}\,\boldsymbol{\hat{\rho}}\,\boldsymbol{\hat{\phi}}\right) = \dot{\phi}\,\boldsymbol{\hat{\rho}}\,\boldsymbol{\hat{\rho}}\,\boldsymbol{\hat{\phi}} = \dot{\phi}\,\boldsymbol{\hat{\phi}}\\
&\boldsymbol{\dot{\hat{\phi}}} = \boldsymbol{\hat{\phi}}\cdot\boldsymbol{\Omega} = \boldsymbol{\hat{\phi}}\ \boldsymbol{\Omega} = \boldsymbol{\hat{\phi}}\left(\dot{\phi}\,\boldsymbol{\hat{\rho}}\,\boldsymbol{\hat{\phi}}\right) = \dot{\phi}\,\boldsymbol{\hat{\phi}}\,\boldsymbol{\hat{\rho}}\,\boldsymbol{\hat{\phi}} = -\,\dot{\phi}\,\boldsymbol{\hat{\rho}}
\end{split}
\end{equation}
where $\,\boldsymbol{\hat{\rho}}\,\boldsymbol{\hat{\rho}}\,\boldsymbol{\hat{\phi}} = \boldsymbol{\hat{\rho}}^2\,\boldsymbol{\hat{\phi}} = \boldsymbol{\hat{\phi}}\,$ and $\,\boldsymbol{\hat{\phi}}\,\boldsymbol{\hat{\rho}}\,\boldsymbol{\hat{\phi}} = -\,\boldsymbol{\hat{\phi}}^2\,\boldsymbol{\hat{\rho}} = -\,\boldsymbol{\hat{\rho}}$. In view of the time derivatives of the unit vectors~\eqref{Poisson cylindrical}, the time derivatives of the unit bivectors in the cylindrical frame are given by,
\begin{equation}\label{Poisson cylindrical bivectors}
\begin{split}
&\left(\boldsymbol{\hat{\rho}}\,\boldsymbol{\hat{\phi}}\right)^{\boldsymbol{\dotp}} = \boldsymbol{\dot{\hat{\rho}}}\,\boldsymbol{\hat{\phi}} + \boldsymbol{\hat{\rho}}\,\boldsymbol{\dot{\hat{\phi}}} = \dot{\phi}\,\boldsymbol{\hat{\phi}}\,\boldsymbol{\hat{\phi}} -\,\dot{\phi}\,\boldsymbol{\hat{\rho}}\,\boldsymbol{\hat{\rho}} = 0\\
&\left(\boldsymbol{\hat{\phi}}\,\boldsymbol{\hat{z}}\right)^{\boldsymbol{\dotp}} = \boldsymbol{\dot{\hat{\phi}}}\,\boldsymbol{\hat{z}} =  -\,\dot{\phi}\,\boldsymbol{\hat{\rho}}\,\boldsymbol{\hat{z}} = \dot{\phi}\,\boldsymbol{\hat{z}}\,\boldsymbol{\hat{\rho}}\\
&\left(\boldsymbol{\hat{z}}\,\boldsymbol{\hat{\rho}}\right)^{\boldsymbol{\dotp}} = \boldsymbol{\hat{z}}\,\boldsymbol{\dot{\hat{\rho}}} =  \dot{\phi}\,\boldsymbol{\hat{z}}\,\boldsymbol{\hat{\phi}} = -\,\dot{\phi}\,\boldsymbol{\hat{\phi}}\,\boldsymbol{\hat{z}}
\end{split}
\end{equation}
In view of the duality~\eqref{dual B} of a bivector and the pseudoscalar $I = \boldsymbol{\hat{\rho}}\,\boldsymbol{\hat{\phi}}\,\boldsymbol{\hat{z}}$, the dual of the unit bivectors in the cylindrical frame are given by,
\begin{equation}\label{dual unit cylindrical bivectors}
\begin{split}
&\left(\boldsymbol{\hat{\rho}}\,\boldsymbol{\hat{\phi}}\right)^{\ast} = -\,\boldsymbol{\hat{\rho}}\,\boldsymbol{\hat{\phi}}\,I = -\,\boldsymbol{\hat{\rho}}\,\boldsymbol{\hat{\phi}}\,\boldsymbol{\hat{\rho}}\,\boldsymbol{\hat{\phi}}\,\boldsymbol{\hat{z}} = \boldsymbol{\hat{\rho}}\,\boldsymbol{\hat{\rho}}\,\boldsymbol{\hat{\phi}}\,\boldsymbol{\hat{\phi}}\,\boldsymbol{\hat{z}} = \boldsymbol{\hat{z}}\\
&\left(\boldsymbol{\hat{\phi}}\,\boldsymbol{\hat{z}}\right)^{\ast} = -\,\boldsymbol{\hat{\phi}}\,\boldsymbol{\hat{z}}\,I = -\,\boldsymbol{\hat{\phi}}\,\boldsymbol{\hat{z}}\,\boldsymbol{\hat{\rho}}\,\boldsymbol{\hat{\phi}}\,\boldsymbol{\hat{z}} = \boldsymbol{\hat{\phi}}\,\boldsymbol{\hat{\phi}}\,\boldsymbol{\hat{z}}\,\boldsymbol{\hat{z}}\,\boldsymbol{\hat{\rho}} = \boldsymbol{\hat{\rho}}\\
&\left(\boldsymbol{\hat{z}}\,\boldsymbol{\hat{\rho}}\right)^{\ast} = -\,\boldsymbol{\hat{z}}\,\boldsymbol{\hat{\rho}}\,I = -\,\boldsymbol{\hat{z}}\,\boldsymbol{\hat{\rho}}\,\boldsymbol{\hat{\rho}}\,\boldsymbol{\hat{\phi}}\,\boldsymbol{\hat{z}} = \boldsymbol{\hat{z}}\,\boldsymbol{\hat{z}}\,\boldsymbol{\hat{\rho}}\,\boldsymbol{\hat{\rho}}\,\boldsymbol{\hat{\phi}} = \boldsymbol{\hat{\phi}}
\end{split}
\end{equation}
Since the pseudoscalar $I$ is a unit frame invariant trivector, 
\begin{equation}\label{pseudovector cylindrical}
I = \boldsymbol{\hat{x}}\,\boldsymbol{\hat{y}}\,\boldsymbol{\hat{z}} = \boldsymbol{\hat{\rho}}\,\boldsymbol{\hat{\phi}}\,\boldsymbol{\hat{z}}
\end{equation}
it is constant, i.e $\dot{I} = 0$. Thus, for any bivector $\boldsymbol{B}$, the time derivation commutes with the duality,
\begin{equation}\label{duality time derivative bivector}
\left(\left(\boldsymbol{B}\right)^{\ast}\right)^{\boldsymbol{\dotp}} = \left(-\,\boldsymbol{B}\,I\right)^{\boldsymbol{\dotp}} = -\left(\boldsymbol{B}\right)^{\boldsymbol{\dotp}}I = \left(\left(\boldsymbol{B}\right)^{\boldsymbol{\dotp}}\right)^{\ast}
\end{equation}
In view of the duality~\eqref{dual unit cylindrical bivectors} and the identity~\eqref{duality time derivative bivector}, the time derivatives of the unit vectors~\eqref{Poisson cylindrical} are the dual of the time derivatives of the unit bivectors~\eqref{Poisson cylindrical bivectors},
\begin{equation}\label{dual unit vector bivectors cylindrical}
\begin{split}
&\boldsymbol{\dot{\hat{\rho}}} = \left(\left(\boldsymbol{\hat{\phi}}\,\boldsymbol{\hat{z}}\right)^{\ast}\right)^{\boldsymbol{\dotp}} = \left(\left(\boldsymbol{\hat{\phi}}\,\boldsymbol{\hat{z}}\right)^{\boldsymbol{\dotp}}\right)^{\ast} = \left(\dot{\phi}\,\boldsymbol{\hat{z}}\,\boldsymbol{\hat{\rho}}\right)^{\ast} = \dot{\phi}\left(\boldsymbol{\hat{z}}\,\boldsymbol{\hat{\rho}}\right)^{\ast} = \dot{\phi}\,\boldsymbol{\hat{\phi}}\\
&\boldsymbol{\dot{\hat{\phi}}} = \left(\left(\boldsymbol{\hat{z}}\,\boldsymbol{\hat{\rho}}\right)^{\ast}\right)^{\boldsymbol{\dotp}} = \left(\left(\boldsymbol{\hat{z}}\,\boldsymbol{\hat{\rho}}\right)^{\boldsymbol{\dotp}}\right)^{\ast} = \left(-\,\dot{\phi}\,\boldsymbol{\hat{\phi}}\,\boldsymbol{\hat{z}}\right)^{\ast} = -\,\dot{\phi}\left(\boldsymbol{\hat{\phi}}\,\boldsymbol{\hat{z}}\right)^{\ast} = -\,\dot{\phi}\,\boldsymbol{\hat{\rho}}
\end{split}
\end{equation}
as expected. 


\section{Rotating spherical frame}
\label{Rotating spherical frame}

\begin{figure}[!ht]
\begin{center}
\includegraphics[scale=0.85]{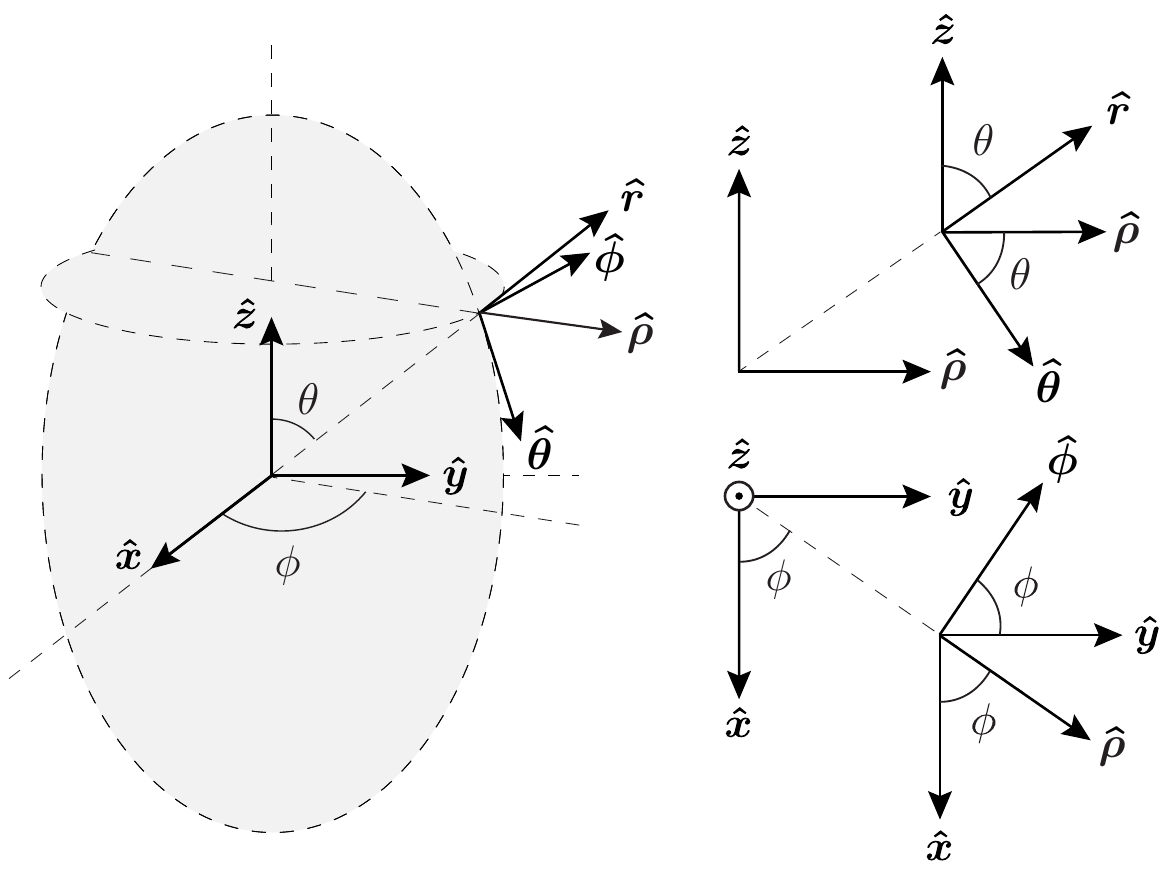}
\end{center}
\caption{Spherical frame $\left\{\boldsymbol{\hat{r}},\boldsymbol{\hat{\theta}},\boldsymbol{\hat{\phi}}\right\}$ rotating around a fixed Cartesian frame $\left\{\boldsymbol{\hat{x}},\boldsymbol{\hat{y}},\boldsymbol{\hat{z}}\right\}$ with an auxiliary horizontal radial vector $\boldsymbol{\hat{\rho}}$.}\label{Fig: spherical frame}
\end{figure}

\noindent We consider an spherical frame $\left\{\boldsymbol{\hat{r}},\boldsymbol{\hat{\theta}},\boldsymbol{\hat{\phi}}\right\}$ rotating around a fixed Cartesian frame $\left\{\boldsymbol{\hat{x}},\boldsymbol{\hat{y}},\boldsymbol{\hat{z}}\right\}$. This rotation is characterised by an azimuthal angle $\phi$ in the horizontal rotation plane and a polar angle $\theta$ in a vertical radial rotation plane (Fig.~\ref{Fig: spherical frame}).
The unit bivector oriented along the rotation in the horizontal plane $\boldsymbol{\hat{B}}_{\phi}$ is written as,
\begin{equation}\label{unit bivector horizontal}
\boldsymbol{\hat{B}}_{\phi} = \boldsymbol{\hat{x}}\wedge\boldsymbol{\hat{y}} = \boldsymbol{\hat{x}}\,\boldsymbol{\hat{y}}
\end{equation}
The unit bivector oriented along the rotation in the vertical radial plane $\boldsymbol{\hat{B}}_{\theta}$ is obtained by rotating the reference vertical unit bivector $\boldsymbol{\hat{z}}\wedge\boldsymbol{\hat{x}} = \boldsymbol{\hat{z}}\,\boldsymbol{\hat{x}}$ by an angle $\phi$ in the horizontal plane,
\begin{equation}\label{unit bivector vertical}
\boldsymbol{\hat{B}}_{\theta} = R_{\phi}\left(\boldsymbol{\hat{z}}\wedge\boldsymbol{\hat{x}}\right)R^{\dag}_{\phi} = R_{\phi}\,\boldsymbol{\hat{z}}\,\boldsymbol{\hat{x}}\,R^{\dag}_{\phi}
\end{equation}
The rotation that maps the Cartesian frame $\left\{\boldsymbol{\hat{x}},\boldsymbol{\hat{y}},\boldsymbol{\hat{z}}\right\}$ onto the spherical frame $\left\{\boldsymbol{\hat{r}},\boldsymbol{\hat{\theta}},\boldsymbol{\hat{\phi}}\right\}$ described by a composition of rotors in the horizontal and radial vertical rotation planes,
\begin{equation}\label{rotor spherical}
R = R_{\theta}\,R_{\phi}
\end{equation}
In view of the unit bivector~\eqref{unit bivector horizontal}, the azimuthal rotor $R_{\phi}$ is written as,
\begin{equation}\label{rotor azimuthal}
R_{\phi} = e^{-\,\boldsymbol{\hat{B}}_{\phi}\,\phi/2} = e^{-\,\boldsymbol{\hat{x}}\,\boldsymbol{\hat{y}}\,\phi/2}
\end{equation}
In view of the unit bivector~\eqref{unit bivector vertical} and the Euler formula~\eqref{Euler formula}, the vertical rotor $R_{\theta}$ is written as,
\begin{equation}\label{rotor vertical}
R_{\theta} = e^{-\,\boldsymbol{\hat{B}}_{\theta}\,\theta/2} = e^{-\,R_{\phi}\,\boldsymbol{\hat{z}}\,\boldsymbol{\hat{x}}\,R^{\dag}_{\phi}\,\theta/2} = R_{\phi}\,e^{-\,\boldsymbol{\hat{z}}\,\boldsymbol{\hat{x}}\,\theta/2}\,R^{\dag}_{\phi}
\end{equation}
Using the rotor~\eqref{rotor azimuthal}, the rotor~\eqref{rotor vertical} is recast as,
\begin{equation}\label{rotor vertical ter}
R_{\theta} = R_{\phi}\,e^{-\,\boldsymbol{\hat{z}}\,\boldsymbol{\hat{x}}\,\theta/2}\,R^{\dag}_{\phi} = e^{-\,\boldsymbol{\hat{x}}\,\boldsymbol{\hat{y}}\,\phi/2}\,e^{-\,\boldsymbol{\hat{z}}\,\boldsymbol{\hat{x}}\,\theta/2}\,R^{\dag}_{\phi}
\end{equation}
In view of the rotors~\eqref{rotor azimuthal}~\eqref{rotor vertical ter}, the rotor~\eqref{rotor spherical} reduces to,
\begin{equation}\label{rotor spherical bis}
R = R_{\phi}\,R_{\theta} = e^{-\,\boldsymbol{\hat{x}}\,\boldsymbol{\hat{y}}\,\phi/2}\,e^{-\,\boldsymbol{\hat{z}}\,\boldsymbol{\hat{x}}\,\theta/2}
\end{equation}
The time derivative of the rotor~\eqref{rotor spherical bis} is written as,
\begin{equation}\label{time derivative rotor spherical}
\dot{R} = \left(-\,\frac{\dot{\phi}}{2}\,\boldsymbol{\hat{x}}\,\boldsymbol{\hat{y}}\right)R + R\left(-\,\frac{\dot{\theta}}{2}\,\boldsymbol{\hat{z}}\,\boldsymbol{\hat{x}}\right)  
\end{equation}
In view of relations~\eqref{unit rotor time dependent},~\eqref{derivative identity bis},~\eqref{unit bivector horizontal},~\eqref{unit bivector vertical} and~\eqref{time derivative rotor spherical}, the angular velocity is given by,
\begin{equation}\label{angular velocity spherical}
\boldsymbol{\Omega} = -\,2\,\dot{R}\,R^{\dag} = \dot{\phi}\,\boldsymbol{\hat{x}}\,\boldsymbol{\hat{y}} + \dot{\theta}\,R\,\boldsymbol{\hat{z}}\,\boldsymbol{\hat{x}}\,R^{\dag}
\end{equation}
In view of the rotors~\eqref{rotor azimuthal},~\eqref{rotor vertical} and~\eqref{rotor spherical bis}, we obtain the following identity,
\begin{equation}\label{rotation bivector vertical radial 0}
R\,\boldsymbol{\hat{z}}\,\boldsymbol{\hat{x}}\,R^{\dag} = R_{\phi}\,R_{\theta}\,\boldsymbol{\hat{z}}\,\boldsymbol{\hat{x}}\,R_{\theta}^{\dag}\,R_{\phi}^{\dag} = R_{\phi}\,e^{-\,\boldsymbol{\hat{z}}\,\boldsymbol{\hat{x}}\,\theta/2}\,\boldsymbol{\hat{z}}\,\boldsymbol{\hat{x}}\,e^{\boldsymbol{\hat{z}}\,\boldsymbol{\hat{x}}\,\theta/2}\,R_{\phi}^{\dag}
\end{equation}
In view of the Euler formula~\eqref{Euler formula},
\begin{equation}\label{rotation bivector vertical radial 1}
e^{-\,\boldsymbol{\hat{z}}\,\boldsymbol{\hat{x}}\,\theta/2}\,\boldsymbol{\hat{z}}\,\boldsymbol{\hat{x}}\,e^{\boldsymbol{\hat{z}}\,\boldsymbol{\hat{x}}\,\theta/2} = \boldsymbol{\hat{z}}\,\boldsymbol{\hat{x}}\,e^{-\,\boldsymbol{\hat{z}}\,\boldsymbol{\hat{x}}\,\theta/2}\,e^{\boldsymbol{\hat{z}}\,\boldsymbol{\hat{x}}\,\theta/2} = \boldsymbol{\hat{z}}\,\boldsymbol{\hat{x}}
\end{equation}
which means that the bivector $\boldsymbol{\hat{z}}\,\boldsymbol{\hat{x}}$ is invariant under a rotation in the $\boldsymbol{\hat{z}}\,\boldsymbol{\hat{x}}$ plane, as expected. In view of the orthonormality condition~\eqref{unit rotor time dependent} and the transformation law~\eqref{rotation bivector vertical radial 1}, the bivector~\eqref{rotation bivector vertical radial 0} becomes,
\begin{equation}\label{rotation bivector vertical radial sec}
R\,\boldsymbol{\hat{z}}\,\boldsymbol{\hat{x}}\,R^{\dag} = R_{\phi}\,\boldsymbol{\hat{z}}\,\boldsymbol{\hat{x}}\,R_{\phi}^{\dag} = \left(R_{\phi}\,\boldsymbol{\hat{z}}\,R_{\phi}^{\dag}\right)\left(R_{\phi}\,\boldsymbol{\hat{x}}\,R_{\phi}^{\dag}\right)
\end{equation}
Since the vector $\boldsymbol{\hat{z}}$ is orthogonal to the plan of rotation, this vector is invariant under a rotation~\eqref{unit rotor time dependent}
\begin{equation}\label{invariant z cylindrical}
R_{\phi}\,\boldsymbol{\hat{z}}\,R_{\phi}^{\dag} = \boldsymbol{\hat{z}}\,R_{\phi}\,R_{\phi}^{\dag} = \boldsymbol{\hat{z}}
\end{equation}
In view of the Euler formula~\eqref{Euler formula}, since the vector $\boldsymbol{\hat{x}}$ is in the plan of rotation, the rotation of this vector is a right-handed transformation~\eqref{rotation rotor phi projection pent},
\begin{equation}\label{single-handed transformation cylindrical}
R_{\phi}\,\boldsymbol{\hat{x}}\,R_{\phi}^{\dag} = \boldsymbol{\hat{x}}\,R_{\phi}^{\dag\,2}
\end{equation}
In view of the Euler formula~\eqref{Euler formula} and the rotor~\eqref{rotor cylindrical}, the rotor reversed squared characterising the rotation by an angle $\phi$ of a vector in the rotation plane is given by,
\begin{equation}\label{rotor in plane cylindrical}
\begin{split}
R_{\phi}^{\dag\,2} = \left(e^{-\,\boldsymbol{\hat{B}}_{\phi}\,\phi/2}\right)^{\dag\,2} = e^{\boldsymbol{\hat{B}}_{\phi}\,\phi} = e^{\phi\,\boldsymbol{\hat{x}}\wedge\boldsymbol{\hat{y}}} = e^{\phi\,\boldsymbol{\hat{x}}\,\boldsymbol{\hat{y}}} = \cos\phi + \sin\phi\,\boldsymbol{\hat{x}}\,\boldsymbol{\hat{y}}
\end{split}
\end{equation}
In view of relations~\eqref{single-handed transformation cylindrical} and~\eqref{rotor in plane cylindrical}, we obtain the following relation,
\begin{equation}\label{rotation unit vectors cylindrical}
R_{\phi}\,\boldsymbol{\hat{x}}\,R_{\phi}^{\dag} = \boldsymbol{\hat{x}}\,R_{\phi}^{\dag\,2} = \cos\phi\,\boldsymbol{\hat{x}} + \sin\phi\,\boldsymbol{\hat{x}}\,\boldsymbol{\hat{x}}\,\boldsymbol{\hat{y}} = \cos\phi\,\boldsymbol{\hat{x}} + \sin\phi\,\boldsymbol{\hat{y}} = \boldsymbol{\hat{\rho}}
\end{equation}
where $\,\boldsymbol{\hat{x}}\,\boldsymbol{\hat{x}}\,\boldsymbol{\hat{y}} = \boldsymbol{\hat{x}}^2\,\boldsymbol{\hat{y}} = \boldsymbol{\hat{y}}\,$. In view of the transformation laws~\eqref{invariant z cylindrical} and~\eqref{single-handed transformation cylindrical}, the bivector~\eqref{rotation bivector vertical radial sec} becomes,
\begin{equation}\label{rotation bivector vertical radial}
R\,\boldsymbol{\hat{z}}\,\boldsymbol{\hat{x}}\,R^{\dag} = R_{\phi}\,\boldsymbol{\hat{z}}\,\boldsymbol{\hat{x}}\,R_{\phi}^{\dag} = \left(R_{\phi}\,\boldsymbol{\hat{z}}\,R_{\phi}^{\dag}\right)\left(R_{\phi}\,\boldsymbol{\hat{x}}\,R_{\phi}^{\dag}\right) = \boldsymbol{\hat{z}}\,\boldsymbol{\hat{\rho}} 
\end{equation}
In view of the bivectors~\eqref{B rho phi},~\eqref{unit bivector vertical} and~\eqref{rotation bivector vertical radial}, the angular velocity~\eqref{angular velocity spherical} is recast,
\begin{equation}\label{angular velocity spherical bis}
\boldsymbol{\Omega} = \dot{\phi}\,\boldsymbol{\hat{B}}_{\phi} + \dot{\theta}\,\boldsymbol{\hat{B}}_{\theta} = \dot{\phi}\,\boldsymbol{\hat{\rho}}\,\boldsymbol{\hat{\phi}} + \dot{\theta}\,\boldsymbol{\hat{z}}\,\boldsymbol{\hat{\rho}}
\end{equation}
The auxiliary vector $\boldsymbol{\hat{\rho}}$ and the vertical vector $\boldsymbol{\hat{z}}$ are written in the spherical frame as (Fig.~\ref{Fig: spherical frame}),
\begin{equation}\label{rho z}
\begin{split}
&\boldsymbol{\hat{\rho}} = \sin\theta\,\boldsymbol{\hat{r}} + \cos\theta\,\boldsymbol{\hat{\theta}}\\
&\boldsymbol{\hat{z}} = \cos\theta\,\boldsymbol{\hat{r}} -\,\sin\theta\,\boldsymbol{\hat{\theta}}
\end{split}
\end{equation}
In view of the change of basis~\eqref{rho z}, the unit bivector $\boldsymbol{\hat{B}}_{\phi}$ is recast in the spherical frame as,
\begin{equation}\label{phi spherical}
\boldsymbol{\hat{B}}_{\phi} = \boldsymbol{\hat{\rho}}\,\boldsymbol{\hat{\phi}} = \left(\sin\theta\,\boldsymbol{\hat{r}} + \cos\theta\,\boldsymbol{\hat{\theta}}\right)\boldsymbol{\hat{\phi}} = \cos\theta\,\boldsymbol{\hat{\theta}}\,\boldsymbol{\hat{\phi}} -\,\sin\theta\,\boldsymbol{\hat{\phi}}\,\boldsymbol{\hat{r}}
\end{equation}
In view of the change of basis~\eqref{rho z}, the unit bivector $\boldsymbol{\hat{B}}_{\theta}$ is recast in the spherical frame as,
\begin{equation}\label{theta spherical}
\boldsymbol{\hat{B}}_{\theta} = \boldsymbol{\hat{z}}\,\boldsymbol{\hat{\rho}} = \left(\cos\theta\,\boldsymbol{\hat{r}} -\,\sin\theta\,\boldsymbol{\hat{\theta}}\right)\left(\sin\theta\,\boldsymbol{\hat{r}} + \cos\theta\,\boldsymbol{\hat{\theta}}\right) = \boldsymbol{\hat{r}}\,\boldsymbol{\hat{\theta}}
\end{equation}
where we used the trigonometric identity $\sin^2\theta + \cos^2\theta = 1$. Using the unit bivectors~\eqref{phi spherical} and~\eqref{theta spherical}, the angular velocity bivector~\eqref{angular velocity spherical bis} is recast in the spherical frame as, 
\begin{equation}\label{angular velocity spherical ter}
\boldsymbol{\Omega} = \dot{\phi}\,\boldsymbol{\hat{B}}_{\phi} + \dot{\theta}\,\boldsymbol{\hat{B}}_{\theta} = \dot{\phi}\,\cos\theta\,\boldsymbol{\hat{\theta}}\,\boldsymbol{\hat{\phi}} -\,\dot{\phi}\,\sin\theta\,\boldsymbol{\hat{\phi}}\,\boldsymbol{\hat{r}} + \dot{\theta}\,\boldsymbol{\hat{r}}\,\boldsymbol{\hat{\theta}}
\end{equation}
The Poisson formula~\eqref{Poisson formula} for the radial unit vector $\boldsymbol{\hat{r}}$ reads,
\begin{equation}\label{Poisson spherical radial}
\begin{split}
&\boldsymbol{\dot{\hat{r}}} = \boldsymbol{\hat{r}}\cdot\boldsymbol{\Omega} = \boldsymbol{\hat{r}}\cdot\left(\dot{\phi}\,\cos\theta\,\boldsymbol{\hat{\theta}}\,\boldsymbol{\hat{\phi}} -\,\dot{\phi}\,\sin\theta\,\boldsymbol{\hat{\phi}}\,\boldsymbol{\hat{r}} + \dot{\theta}\,\boldsymbol{\hat{r}}\,\boldsymbol{\hat{\theta}}\right)\\
&\phantom{\boldsymbol{\dot{\hat{r}}}} = -\,\dot{\phi}\,\sin\theta\,\boldsymbol{\hat{r}}\,\boldsymbol{\hat{\phi}}\,\boldsymbol{\hat{r}} + \dot{\theta}\,\boldsymbol{\hat{r}}\,\boldsymbol{\hat{r}}\,\boldsymbol{\hat{\theta}} = \dot{\theta}\,\boldsymbol{\hat{\theta}} + \dot{\phi}\,\sin\theta\,\boldsymbol{\hat{\phi}} 
\end{split}
\end{equation}
where the vector $\boldsymbol{\hat{r}}$ is orthogonal to the bivector $\boldsymbol{\hat{\theta}}\,\boldsymbol{\hat{\phi}}$ and $\,\boldsymbol{\hat{r}}\,\boldsymbol{\hat{\phi}}\,\boldsymbol{\hat{r}} = -\,\boldsymbol{\hat{r}}^2\,\boldsymbol{\hat{\phi}} = -\,\boldsymbol{\hat{\phi}}\,$ and also $\,\boldsymbol{\hat{r}}\,\boldsymbol{\hat{r}}\,\boldsymbol{\hat{\theta}} = \boldsymbol{\hat{r}}^2\,\boldsymbol{\hat{\theta}} = \boldsymbol{\hat{\theta}}$. The Poisson formula~\eqref{Poisson formula} for the polar unit vector $\boldsymbol{\hat{\theta}}$ reads,
\begin{equation}\label{Poisson spherical polar}
\begin{split}
&\boldsymbol{\dot{\hat{\theta}}} = \boldsymbol{\hat{\theta}}\cdot\boldsymbol{\Omega} = \boldsymbol{\hat{\theta}}\cdot\left(\dot{\phi}\,\cos\theta\,\boldsymbol{\hat{\theta}}\,\boldsymbol{\hat{\phi}} -\,\dot{\phi}\,\sin\theta\,\boldsymbol{\hat{\phi}}\,\boldsymbol{\hat{r}} + \dot{\theta}\,\boldsymbol{\hat{r}}\,\boldsymbol{\hat{\theta}}\right)\\
&\phantom{\boldsymbol{\dot{\hat{\theta}}}} = \dot{\phi}\,\cos\theta\,\boldsymbol{\hat{\theta}}\,\boldsymbol{\hat{\theta}}\,\boldsymbol{\hat{\phi}} + \dot{\theta}\,\boldsymbol{\hat{\theta}}\,\boldsymbol{\hat{r}}\,\boldsymbol{\hat{\theta}} = -\,\dot{\theta}\,\boldsymbol{\hat{r}} + \dot{\phi}\,\cos\theta\,\boldsymbol{\hat{\phi}} 
\end{split}
\end{equation}
where the vector $\boldsymbol{\hat{\theta}}$ is orthogonal to the bivector $\boldsymbol{\hat{\phi}}\,\boldsymbol{\hat{r}}$ and $\,\boldsymbol{\hat{\theta}}\,\boldsymbol{\hat{\theta}}\,\boldsymbol{\hat{\phi}} = \boldsymbol{\hat{\theta}}^2\,\boldsymbol{\hat{\phi}} = \boldsymbol{\hat{\phi}}\,$ and also $\,\boldsymbol{\hat{\theta}}\,\boldsymbol{\hat{r}}\,\boldsymbol{\hat{\theta}} = -\,\boldsymbol{\hat{\theta}}^2\,\boldsymbol{\hat{r}} = -\,\boldsymbol{\hat{r}}$. The Poisson formula~\eqref{Poisson formula} for the azimuthal unit vector $\boldsymbol{\hat{\phi}}$ reads,
\begin{equation}\label{Poisson spherical azimuthal}
\begin{split}
&\boldsymbol{\dot{\hat{\phi}}} = \boldsymbol{\hat{\phi}}\cdot\boldsymbol{\Omega} = \boldsymbol{\hat{\phi}}\cdot\left(\dot{\phi}\,\cos\theta\,\boldsymbol{\hat{\theta}}\,\boldsymbol{\hat{\phi}} -\,\dot{\phi}\,\sin\theta\,\boldsymbol{\hat{\phi}}\,\boldsymbol{\hat{r}} + \dot{\theta}\,\boldsymbol{\hat{r}}\,\boldsymbol{\hat{\theta}}\right)\\
&\phantom{\boldsymbol{\dot{\hat{\phi}}}} = \dot{\phi}\,\cos\theta\,\boldsymbol{\hat{\phi}}\,\boldsymbol{\hat{\theta}}\,\boldsymbol{\hat{\phi}} -\,\dot{\phi}\,\sin\theta\,\boldsymbol{\hat{\phi}}\,\boldsymbol{\hat{\phi}}\,\boldsymbol{\hat{r}} = -\,\dot{\phi}\left(\sin\theta\,\boldsymbol{\hat{r}} + \cos\theta\,\boldsymbol{\hat{\theta}}\right)
\end{split}
\end{equation}
where the vector $\boldsymbol{\hat{\phi}}$ is orthogonal to the bivector $\boldsymbol{\hat{r}}\,\boldsymbol{\hat{\theta}}$ and $\,\boldsymbol{\hat{\phi}}\,\boldsymbol{\hat{\theta}}\,\boldsymbol{\hat{\phi}} = -\,\boldsymbol{\hat{\phi}}^2\,\boldsymbol{\hat{\theta}} = -\,\boldsymbol{\hat{\theta}}\,$ and also $\,\boldsymbol{\hat{\phi}}\,\boldsymbol{\hat{\phi}}\,\boldsymbol{\hat{r}} = \boldsymbol{\hat{\phi}}^2\,\boldsymbol{\hat{r}} = \boldsymbol{\hat{r}}$. In view of the time derivatives of the unit vectors~\eqref{Poisson spherical radial},~\eqref{Poisson spherical polar} and~\eqref{Poisson spherical azimuthal}, the time derivatives of the unit bivectors in the spherical frame are given by,
\begin{equation}\label{Poisson spherical bivectors}
\begin{split}
&\left(\boldsymbol{\hat{r}}\,\boldsymbol{\hat{\theta}}\right)^{\boldsymbol{\dotp}} = \boldsymbol{\dot{\hat{r}}}\,\boldsymbol{\hat{\theta}} + \boldsymbol{\hat{r}}\,\boldsymbol{\dot{\hat{\theta}}} = -\,\dot{\phi}\left(\sin\theta\,\boldsymbol{\hat{\theta}}\,\boldsymbol{\hat{\phi}} + \cos\theta\,\boldsymbol{\hat{\phi}}\,\boldsymbol{\hat{r}}\right)\\
&\left(\boldsymbol{\hat{\theta}}\,\boldsymbol{\hat{\phi}}\right)^{\boldsymbol{\dotp}} = \boldsymbol{\dot{\hat{\theta}}}\,\boldsymbol{\hat{\phi}} + \boldsymbol{\hat{\theta}}\,\boldsymbol{\dot{\hat{\phi}}} = \dot{\theta}\,\boldsymbol{\hat{\phi}}\,\boldsymbol{\hat{r}} + \dot{\phi}\,\sin\theta\,\boldsymbol{\hat{r}}\,\boldsymbol{\hat{\theta}}\\
&\left(\boldsymbol{\hat{\phi}}\,\boldsymbol{\hat{r}}\right)^{\boldsymbol{\dotp}} = \boldsymbol{\dot{\hat{\phi}}}\,\boldsymbol{\hat{r}} + \boldsymbol{\hat{\phi}}\,\boldsymbol{\dot{\hat{r}}} = -\,\dot{\theta}\,\boldsymbol{\hat{\theta}}\,\boldsymbol{\hat{\phi}} + \dot{\phi}\cos\theta\,\boldsymbol{\hat{r}}\,\boldsymbol{\hat{\theta}}
\end{split}
\end{equation}
In view of the duality~\eqref{dual B} of a bivector and the pseudoscalar $I = \boldsymbol{\hat{r}}\,\boldsymbol{\hat{\theta}}\,\boldsymbol{\hat{\phi}}$, the dual of the unit bivectors in the spherical frame are given by,
\begin{equation}\label{dual unit spherical bivectors}
\begin{split}
&\left(\boldsymbol{\hat{r}}\,\boldsymbol{\hat{\theta}}\right)^{\ast} = -\,\boldsymbol{\hat{r}}\,\boldsymbol{\hat{\theta}}\,I = -\,\boldsymbol{\hat{r}}\,\boldsymbol{\hat{\theta}}\,\boldsymbol{\hat{r}}\,\boldsymbol{\hat{\theta}}\,\boldsymbol{\hat{\phi}} = \boldsymbol{\hat{r}}\,\boldsymbol{\hat{r}}\,\boldsymbol{\hat{\theta}}\,\boldsymbol{\hat{\theta}}\,\boldsymbol{\hat{\phi}} = \boldsymbol{\hat{\phi}}\\
&\left(\boldsymbol{\hat{\theta}}\,\boldsymbol{\hat{\phi}}\right)^{\ast} = -\,\boldsymbol{\hat{\theta}}\,\boldsymbol{\hat{\phi}}\,I = -\,\boldsymbol{\hat{\theta}}\,\boldsymbol{\hat{\phi}}\,\boldsymbol{\hat{r}}\,\boldsymbol{\hat{\theta}}\,\boldsymbol{\hat{\phi}} = \boldsymbol{\hat{\theta}}\,\boldsymbol{\hat{\theta}}\,\boldsymbol{\hat{\phi}}\,\boldsymbol{\hat{\phi}}\,\boldsymbol{\hat{r}} = \boldsymbol{\hat{r}}\\
&\left(\boldsymbol{\hat{\phi}}\,\boldsymbol{\hat{r}}\right)^{\ast} = -\,\boldsymbol{\hat{\phi}}\,\boldsymbol{\hat{r}}\,I = -\,\boldsymbol{\hat{\phi}}\,\boldsymbol{\hat{r}}\,\boldsymbol{\hat{r}}\,\boldsymbol{\hat{\theta}}\,\boldsymbol{\hat{\phi}} = \boldsymbol{\hat{\phi}}\,\boldsymbol{\hat{\phi}}\,\boldsymbol{\hat{r}}\,\boldsymbol{\hat{r}}\,\boldsymbol{\hat{\theta}} = \boldsymbol{\hat{\theta}}
\end{split}
\end{equation}
Since the pseudoscalar $I$ is a unit frame invariant trivector, 
\begin{equation}\label{pseudovector spherical}
I = \boldsymbol{\hat{x}}\,\boldsymbol{\hat{y}}\,\boldsymbol{\hat{z}} = \boldsymbol{\hat{r}}\,\boldsymbol{\hat{\theta}}\,\boldsymbol{\hat{\phi}}
\end{equation}
it is constant, i.e $\dot{I} = 0$. In view of the duality~\eqref{dual unit spherical bivectors} and the identity~\eqref{duality time derivative bivector}, the time derivative of the unit radial vector~\eqref{Poisson spherical radial},~\eqref{Poisson spherical polar} and~\eqref{Poisson spherical azimuthal} is the dual of the time derivative of the corresponding unit bivector~\eqref{Poisson spherical bivectors},
\begin{equation}\label{dual unit vector bivectors spherical radial}
\begin{split}
&\boldsymbol{\dot{\hat{r}}} = \left(\left(\boldsymbol{\hat{\theta}}\,\boldsymbol{\hat{\phi}}\right)^{\ast}\right)^{\boldsymbol{\dotp}} = \left(\left(\boldsymbol{\hat{\theta}}\,\boldsymbol{\hat{\phi}}\right)^{\boldsymbol{\dotp}}\right)^{\ast} = \left(\dot{\theta}\,\boldsymbol{\hat{\phi}}\,\boldsymbol{\hat{r}} + \dot{\phi}\,\sin\theta\,\boldsymbol{\hat{r}}\,\boldsymbol{\hat{\theta}}\right)^{\ast} \\
&\phantom{\boldsymbol{\dot{\hat{r}}}} = \dot{\theta}\left(\boldsymbol{\hat{\phi}}\,\boldsymbol{\hat{r}}\right)^{\ast} + \dot{\phi}\,\sin\theta\left(\boldsymbol{\hat{r}}\,\boldsymbol{\hat{\theta}}\right)^{\ast} = \dot{\theta}\,\boldsymbol{\hat{\theta}} + \dot{\phi}\,\sin\theta\,\boldsymbol{\hat{\phi}}
\end{split}
\end{equation}
The time derivative of the unit polar vector~\eqref{Poisson spherical radial},~\eqref{Poisson spherical polar} and~\eqref{Poisson spherical azimuthal} is the dual of the time derivative of the corresponding unit bivector~\eqref{Poisson spherical bivectors},
\begin{equation}\label{dual unit vector bivectors spherical polar}
\begin{split}
&\boldsymbol{\dot{\hat{\theta}}} = \left(\left(\boldsymbol{\hat{\phi}}\,\boldsymbol{\hat{r}}\right)^{\ast}\right)^{\boldsymbol{\dotp}} = \left(\left(\boldsymbol{\hat{\phi}}\,\boldsymbol{\hat{r}}\right)^{\boldsymbol{\dotp}}\right)^{\ast} = \left(-\,\dot{\theta}\,\boldsymbol{\hat{\theta}}\,\boldsymbol{\hat{\phi}} + \dot{\phi}\cos\theta\,\boldsymbol{\hat{r}}\,\boldsymbol{\hat{\theta}}\right)^{\ast} \\
&\phantom{\boldsymbol{\dot{\hat{\theta}}}} = -\,\dot{\theta}\left(\boldsymbol{\hat{\theta}}\,\boldsymbol{\hat{\phi}}\right)^{\ast} + \dot{\phi}\,\cos\theta\left(\boldsymbol{\hat{r}}\,\boldsymbol{\hat{\theta}}\right)^{\ast} = -\,\dot{\theta}\,\boldsymbol{\hat{r}} + \dot{\phi}\,\cos\theta\,\boldsymbol{\hat{\phi}}
\end{split}
\end{equation}
The time derivative of the unit polar vector~\eqref{Poisson spherical radial},~\eqref{Poisson spherical polar} and~\eqref{Poisson spherical azimuthal} is the dual of the time derivative of the corresponding unit bivector~\eqref{Poisson spherical bivectors},
\begin{equation}\label{dual unit vector bivectors spherical azimuthal}
\begin{split}
&\boldsymbol{\dot{\hat{\phi}}} = \left(\left(\boldsymbol{\hat{r}}\,\boldsymbol{\hat{\theta}}\right)^{\ast}\right)^{\boldsymbol{\dotp}} = \left(\left(\boldsymbol{\hat{r}}\,\boldsymbol{\hat{\theta}}\right)^{\boldsymbol{\dotp}}\right)^{\ast} = \left(-\,\dot{\phi}\left(\sin\theta\,\boldsymbol{\hat{\theta}}\,\boldsymbol{\hat{\phi}} + \cos\theta\,\boldsymbol{\hat{\phi}}\,\boldsymbol{\hat{r}}\right)\right)^{\ast} \\
&\phantom{\boldsymbol{\dot{\hat{\phi}}}} = -\,\dot{\phi}\left(\sin\theta\left(\boldsymbol{\hat{\theta}}\,\boldsymbol{\hat{\phi}}\right)^{\ast} + \cos\theta\left(\boldsymbol{\hat{\phi}}\,\boldsymbol{\hat{r}}\right)^{\ast}\right) = -\,\dot{\phi}\left(\sin\theta\,\boldsymbol{\hat{r}} + \cos\theta\,\boldsymbol{\hat{\theta}}\right)
\end{split}
\end{equation}
as expected.


\section{Point particle motion}
\label{Point particle motion}

\begin{figure}[!ht]
\begin{center}
\includegraphics[scale=0.55]{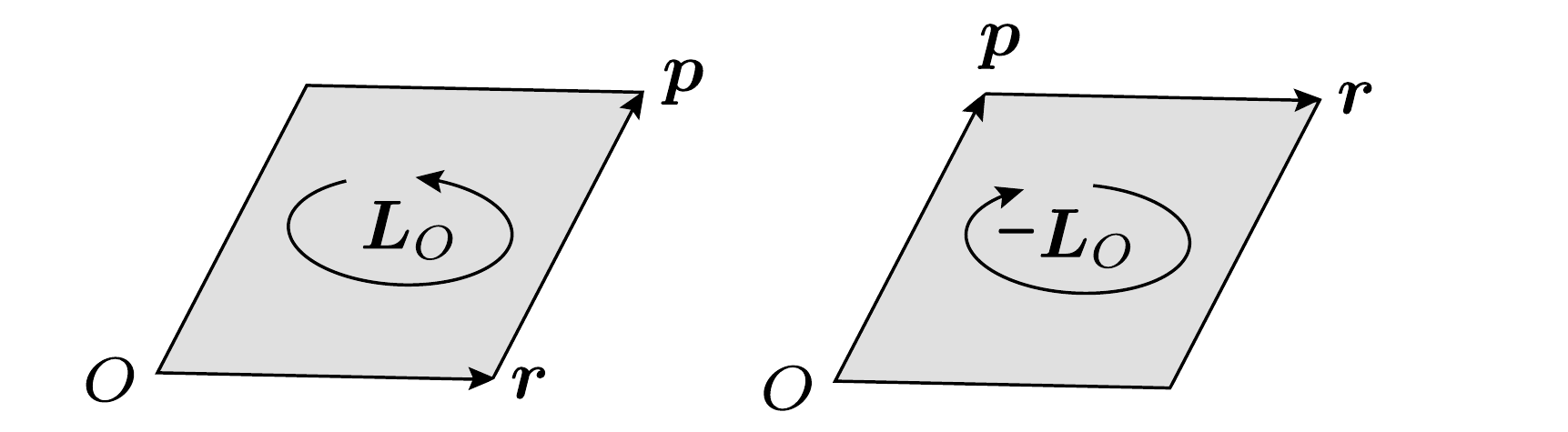}
\end{center}
\caption{Antisymmetry of the bivectors $\boldsymbol{L}_O = \boldsymbol{r}\wedge\boldsymbol{p}$ and $-\,\boldsymbol{L}_O = \boldsymbol{p}\wedge\boldsymbol{r}$.}\label{Fig: Angular momentum antisymmetric}
\end{figure}

\begin{figure}[!ht]
\begin{center}
\includegraphics[scale=0.55]{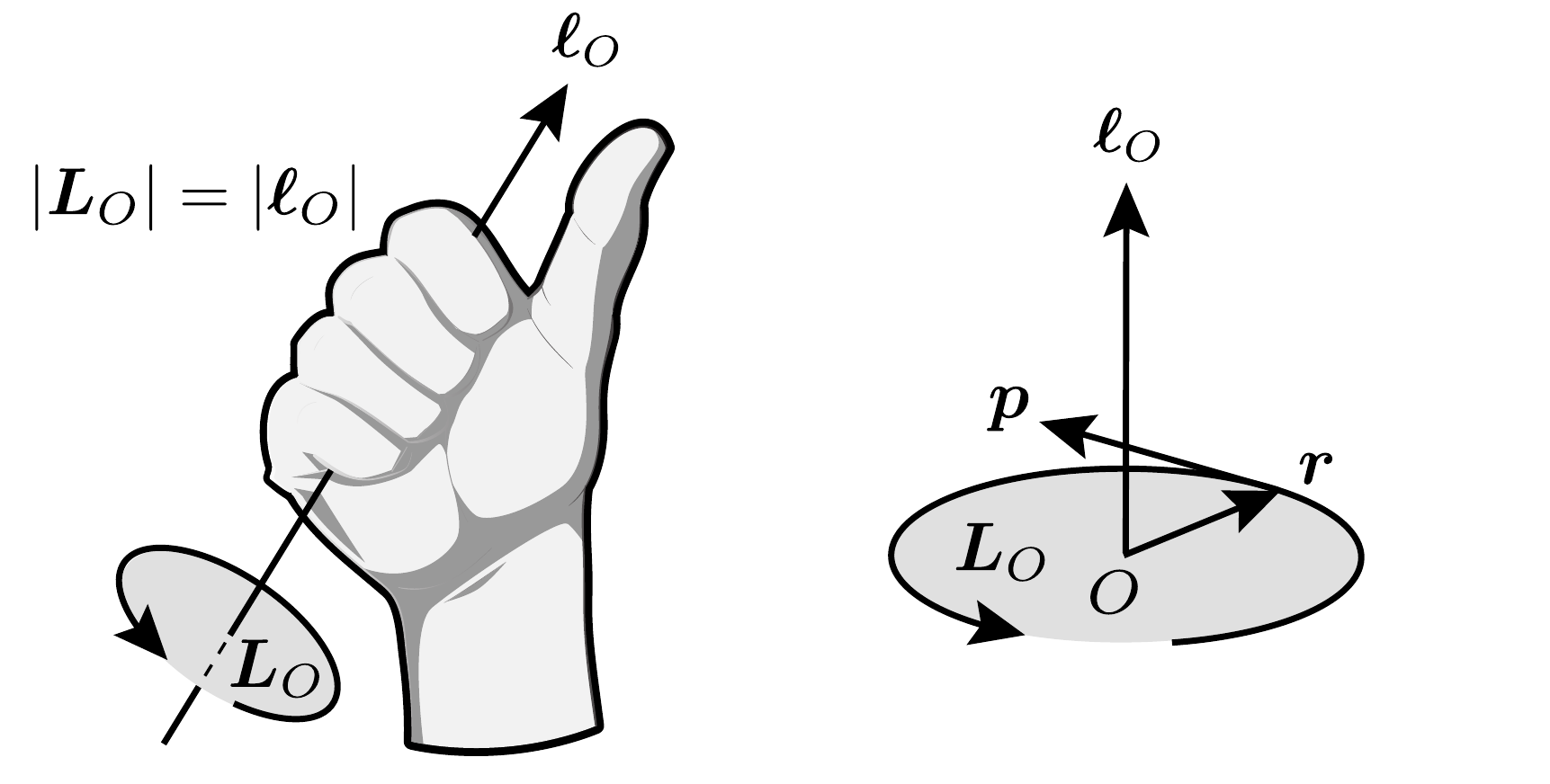}
\end{center}
\caption{Duality between the angular momentum bivector $\boldsymbol{L}_O$ and the angular momentum pseudovector $\boldsymbol{\ell}_O$ illustrated by the right hand rule.}\label{Fig: Angular momentum}
\end{figure}

\noindent We consider the motion of a point particle $P$ of mass $m$ with respect to an inertial frame of reference. We describe this motion with respect to a fixed orthonormal frame $\left\{\boldsymbol{\hat{e}}_1,\boldsymbol{\hat{e}}_2,\boldsymbol{\hat{e}}_3\right\}$. The motion of the point particle is given by Newton's second law where the external forces $\boldsymbol{f}^{\,\text{ext}}$ are the cause of the time variation of the momentum $\boldsymbol{p}$ according to,
\begin{equation}\label{Newton second law}
\sum\,\boldsymbol{f}^{\,\text{ext}} = \boldsymbol{\dot{p}}
\end{equation}
which has the same structure in geometric algebra as in vector algebra since the geometric entities involved are only vectors. The rotational motion of the point particle $P$ is described in the plane spanned by the position $\boldsymbol{r}$ and the momentum $\boldsymbol{p}$. In order to describe such a motion, we define the angular momentum bivector in the plane of motion as,
\begin{equation}\label{angular momentum bivector}
\boldsymbol{L}_O = \boldsymbol{r}\wedge\boldsymbol{p}
\end{equation}
It is antisymmetric under the permutation of the position $\boldsymbol{r}$ and the momentum $\boldsymbol{p}$ 
\begin{equation}\label{angular momentum bivector antisymmetric}
\boldsymbol{L}_O = \boldsymbol{r}\wedge\boldsymbol{p} = -\,\boldsymbol{p}\wedge\boldsymbol{r} = -\,\boldsymbol{L}_O
\end{equation}
The geometric meaning is that if the order of two vectors in an outer product is changed the resulting bivectors turns the other way around (Fig.~\ref{Fig: Angular momentum antisymmetric}) 
In the fixed orthonormal frame, the position $\boldsymbol{r}$ and the momentum $\boldsymbol{p} = m\,\boldsymbol{v}$ are written in components as,
\begin{equation}\label{position and momentum}
\begin{split}
&\boldsymbol{r} = x_1\,\boldsymbol{\hat{e}}_1 + x_2\,\boldsymbol{\hat{e}}_2 + x_3\,\boldsymbol{\hat{e}}_3\\
&\boldsymbol{p} = m\,\dot{x}_1\,\boldsymbol{\hat{e}}_1 + m\,\dot{x}_2\,\boldsymbol{\hat{e}}_2 + m\,\dot{x}_3\,\boldsymbol{\hat{e}}_3
\end{split}
\end{equation}
and the angular momentum~\eqref{angular momentum bivector} reads,
\begin{equation}\label{angular momentum bivector bis}
\boldsymbol{L}_O = m\left(x_1\,\dot{x}_2 -\,x_2\,\dot{x}_1\right)\boldsymbol{\hat{e}}_1\,\boldsymbol{\hat{e}}_2 + m\left(x_2\,\dot{x}_3 -\,x_3\,\dot{x}_2\right)\boldsymbol{\hat{e}}_2\,\boldsymbol{\hat{e}}_3 + m\left(x_1\,\dot{x}_3 -\,x_3\,\dot{x}_1\right)\boldsymbol{\hat{e}}_3\,\boldsymbol{\hat{e}}_1
\end{equation}
where the unit vectors $\boldsymbol{\hat{e}}_1\,\boldsymbol{\hat{e}}_2 = \boldsymbol{\hat{e}}_1\wedge\boldsymbol{\hat{e}}_2$, $\boldsymbol{\hat{e}}_2\,\boldsymbol{\hat{e}}_3 = \boldsymbol{\hat{e}}_2\wedge\boldsymbol{\hat{e}}_3$ and $\boldsymbol{\hat{e}}_3\,\boldsymbol{\hat{e}}_1 = \boldsymbol{\hat{e}}_3\wedge\boldsymbol{\hat{e}}_1$. According to identity~\eqref{wedge and cross products duality}, the dual of the angular momentum bivector is the angular momentum pseudovector $\boldsymbol{\ell}_O$,
\begin{equation}\label{angular momentum duality}
\boldsymbol{L}_O^{\ast} = \left(\boldsymbol{r}\wedge\boldsymbol{p}\right)^{\ast} = \boldsymbol{r}\times\boldsymbol{p} = \boldsymbol{\ell}_O
\end{equation}
which is the definition of the angular momentum pseudovector in vector algebra. In view of the duality~\eqref{dual B} of a bivector,
\begin{equation}\label{duality bivector unit vector}
\left(\boldsymbol{\hat{e}}_1\,\boldsymbol{\hat{e}}_2\right)^{\ast} = \boldsymbol{\hat{e}}_3 \quad\text{and}\quad
\left(\boldsymbol{\hat{e}}_2\,\boldsymbol{\hat{e}}_3\right)^{\ast} = \boldsymbol{\hat{e}}_1 \quad\text{and}\quad
\left(\boldsymbol{\hat{e}}_3\,\boldsymbol{\hat{e}}_1\right)^{\ast} = \boldsymbol{\hat{e}}_2
\end{equation}
which implies that the angular momentum pseudovector~\eqref{angular momentum duality} is written in vector algebra as,
\begin{equation}\label{angular momentum pseudovector}
\boldsymbol{\ell}_O =  m\left(x_2\,\dot{x}_3 -\,x_3\,\dot{x}_2\right)\boldsymbol{\hat{e}}_1 + m\left(x_1\,\dot{x}_3 -\,x_3\,\dot{x}_1\right)\boldsymbol{\hat{e}}_2 + m\left(x_1\,\dot{x}_2 -\,x_2\,\dot{x}_1\right)\boldsymbol{\hat{e}}_3
\end{equation}
as expected. The plan area covered by the angular momentum bivector~\eqref{angular momentum bivector bis} is equal to the length of the angular momentum pseudovector~\eqref{angular momentum pseudovector},
\begin{equation}\label{duality modulus angular momentum}
\vert\boldsymbol{L}_O\vert = \vert\boldsymbol{\ell}_O\vert
\end{equation}
The geometric interpretation of this duality is the following : if the palm of the right hand is oriented along the angular momentum bivector $\boldsymbol{L}_O$ in the plane of rotation, then the thumb is oriented along the angular momentum pseudovector $\boldsymbol{\ell}_O$ (Fig.~\ref{Fig: Angular momentum}). 

\begin{figure}[!ht]
\begin{center}
\includegraphics[scale=0.55]{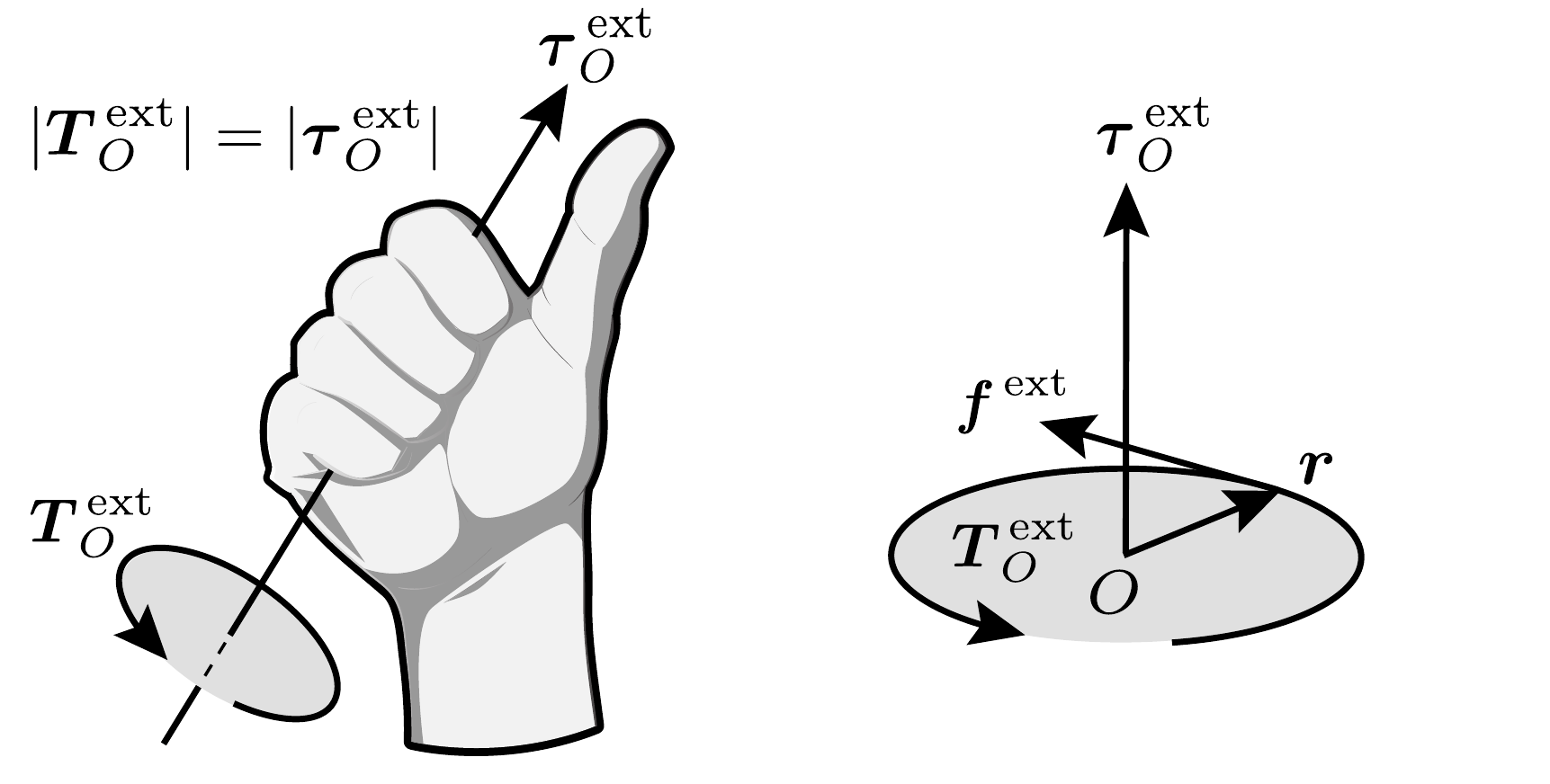}
\end{center}
\caption{Duality between the external torque bivector $\boldsymbol{T}^{\,\text{ext}}_O$ and the external torque pseudovector $\boldsymbol{\tau}^{\,\text{ext}}_O$ illustrated by the right hand rule.}\label{Fig: External torque}
\end{figure}

\noindent The time derivative of the angular momentum bivector~\eqref{angular momentum bivector} is given by,
\begin{equation}\label{angular momentum bivector time derivative}
\boldsymbol{\dot{L}}_O = \boldsymbol{\dot{r}}\wedge\boldsymbol{p} + \boldsymbol{r}\wedge\boldsymbol{\dot{p}}
\end{equation}
Since the momentum $\boldsymbol{p}$ is collinear to the velocity $\boldsymbol{\dot{r}}$, the bivector spanned by these vectors vanishes,
\begin{equation}\label{null bivector}
\boldsymbol{\dot{r}}\wedge\boldsymbol{p} = \boldsymbol{v}\wedge\left(m\,\boldsymbol{v}\right) = m\,\boldsymbol{v}\wedge\boldsymbol{v} = 0 
\end{equation}
In view of the identity~\eqref{null bivector} and Newton's second law~\eqref{Newton second law}, the time derivative of the angular momentum bivector becomes,
\begin{equation}\label{angular momentum bivector time derivative bis}
\boldsymbol{\dot{L}}_O = \boldsymbol{r}\wedge\boldsymbol{\dot{p}} = \boldsymbol{r}\wedge\sum\,\boldsymbol{f}^{\,\text{ext}} = \sum\,\boldsymbol{r}\wedge\boldsymbol{f}^{\,\text{ext}}
\end{equation}
The external torque bivector $\boldsymbol{T}_O^{\,\text{ext}}$ due to the external force $\boldsymbol{f}^{\,\text{ext}}$ acting of the point particle $P$ is defined as the oriented area spanned by the position $\boldsymbol{r} = \boldsymbol{OP}$ and the external force $\boldsymbol{f}^{\,\text{ext}}$,
\begin{equation}\label{external torque}
\boldsymbol{T}_O^{\,\text{ext}} = \boldsymbol{r}\wedge\boldsymbol{f}^{\,\text{ext}}
\end{equation}
In the fixed orthonormal frame, the external force $\boldsymbol{f}^{\,\text{ext}}$ is written in components as,
\begin{equation}\label{external force}
\boldsymbol{f}^{\,\text{ext}} = f^{\,\text{ext}}_1\,\boldsymbol{\hat{e}}_1 + f^{\,\text{ext}}_2\,\boldsymbol{\hat{e}}_2 + f^{\,\text{ext}}_3\,\boldsymbol{\hat{e}}_3
\end{equation}
and in view of the position~\eqref{position and momentum}, the external torque~\eqref{external torque} reads,
\begin{equation}\label{external torque bis}
\begin{split}
&\boldsymbol{T}_O^{\,\text{ext}} = \left(x_1\,f^{\,\text{ext}}_2 -\,x_2\,f^{\,\text{ext}}_1\right)\boldsymbol{\hat{e}}_1\,\boldsymbol{\hat{e}}_2 + \left(x_2\,f^{\,\text{ext}}_3 -\,x_3\,f^{\,\text{ext}}_2\right)\boldsymbol{\hat{e}}_2\,\boldsymbol{\hat{e}}_3\\
&\phantom{\boldsymbol{T}_O^{\,\text{ext}} =} + \left(x_1\,f^{\,\text{ext}}_3 -\,x_3\,f^{\,\text{ext}}_1\right)\boldsymbol{\hat{e}}_3\,\boldsymbol{\hat{e}}_1
\end{split}
\end{equation}
According to identity~\eqref{wedge and cross products duality}, the dual of the external torque bivector is the external torque pseudovector $\boldsymbol{\tau}_O$,
\begin{equation}\label{external torque duality}
\boldsymbol{T}_O^{\,\text{ext}\,\ast} = \left(\boldsymbol{r}\wedge\boldsymbol{f}^{\,\text{ext}}\right)^{\ast} = \boldsymbol{r}\times\boldsymbol{f}^{\,\text{ext}} = \boldsymbol{\tau}_O^{\,\text{ext}}
\end{equation}
which is the definition of the external torque pseudovector in vector algebra. In view of the duals of the unit bivectors~\eqref{duality bivector unit vector}, the external torque pseudovector~\eqref{external torque duality} is written in vector algebra as,
\begin{equation}\label{external torque pseudovector}
\begin{split}
&\boldsymbol{\tau}_O^{\,\text{ext}} = \left(x_2\,f^{\,\text{ext}}_3 -\,x_3\,f^{\,\text{ext}}_2\right)\boldsymbol{\hat{e}}_1 + \left(x_1\,f^{\,\text{ext}}_3 -\,x_3\,f^{\,\text{ext}}_1\right)\boldsymbol{\hat{e}}_2\\
&\phantom{\boldsymbol{\tau}_O^{\,\text{ext}} =} + \left(x_1\,f^{\,\text{ext}}_2 -\,x_2\,f^{\,\text{ext}}_1\right)\boldsymbol{\hat{e}}_3
\end{split}
\end{equation}
as expected. The plan area covered by the external torque bivector~\eqref{external torque bis} is equal to the length of the external torque pseudovector~\eqref{external torque pseudovector},
\begin{equation}\label{duality modulus external torque}
\vert\boldsymbol{T}^{\,\text{ext}}_O\vert = \vert\boldsymbol{\tau}^{\,\text{ext}}_O\vert
\end{equation}
The geometric interpretation of this duality is the following : if the palm of the right hand is oriented along the external torque bivector $\boldsymbol{T}^{\,\text{ext}}_O$, then the thumb is oriented along the external torque pseudovector $\boldsymbol{\tau}^{\,\text{ext}}_O$ (Fig.~\ref{Fig: External torque}). 

\noindent In view of the time derivative of the angular momentum bivector~\eqref{angular momentum bivector time derivative bis} and the external torque bivector~\eqref{external torque}, we obtain the angular momentum theorem,
\begin{equation}\label{angular momentum theorem}
\sum\,\boldsymbol{T}_O^{\,\text{ext}} = \boldsymbol{\dot{L}}_O
\end{equation}
Using the identity~\eqref{duality time derivative bivector} and the duality~\eqref{angular momentum duality}, the dual of the time derivative of the angular momentum bivector is the time derivative of the angular momentum pseudovector, 
\begin{equation}\label{duality time derivative angular momentum}
\left(\boldsymbol{\dot{L}}_O\right)^{\ast} = \left(\boldsymbol{L}_O^{\ast}\right)^{\boldsymbol{\dotp}} = \boldsymbol{\dot{\ell}}_O
\end{equation}
In view of the dualities~\eqref{external torque duality} and~\eqref{duality time derivative angular momentum}, the dual of the angular momentum theorem~\eqref{angular momentum theorem} in geometric algebra is the angular momentum theorem in vector algebra,
\begin{equation}\label{angular momentum theorem vspace}
\sum\,\boldsymbol{\tau}_O^{\,\text{ext}} = \boldsymbol{\dot{\ell}}_O
\end{equation}
as expected.


\section{Rigid body motion}
\label{Rigid body motion}

\begin{figure}[!ht]
\begin{center}
\includegraphics[scale=0.50]{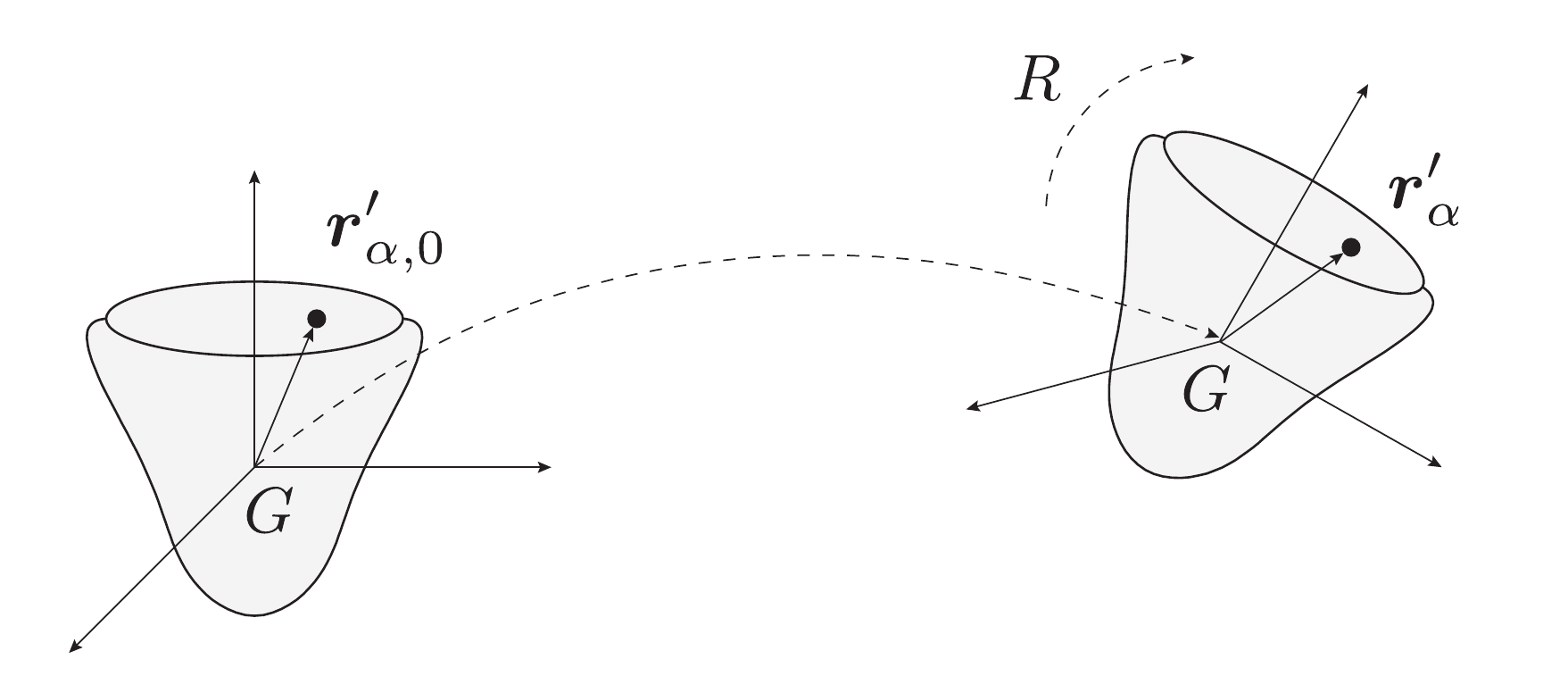}
\end{center}
\caption{Rotation of the rigid body described by the rotor $R$. The relative position of the point particle $P_{\alpha}$ rotates from $\boldsymbol{r}^{\prime}_{\alpha,0}$ to $\boldsymbol{r}^{\prime}_{\alpha}$ around the centre of mass $G$.}\label{Fig: Rotation rigid body}
\end{figure}

\begin{figure}[!ht]
\begin{center}
\includegraphics[scale=1.00]{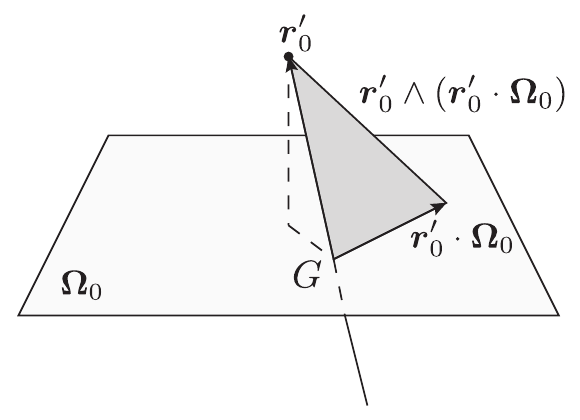}
\end{center}
\caption{The inertia map $\boldsymbol{I}_G\left(\boldsymbol{\Omega}_{0}\right)$ is obtained by integrating the local bivectors $\boldsymbol{r}_{0}^{\prime}\wedge\left(\boldsymbol{r}_{0}^{\prime}\cdot\boldsymbol{\Omega}_{0}\right)$ over the volume of the rigid body. These bivectors are spanned by the vectors $\boldsymbol{r}_{0}^{\prime}$ and $\boldsymbol{r}_{0}^{\prime}\cdot\boldsymbol{\Omega}_{0}$.}\label{Fig: Inertia}
\end{figure}

\noindent A rigid body is a set of point particles $P_{\alpha}$ such that the relative distance between every couple of points remains constant over time. The position $\boldsymbol{r}_{\alpha}$ of every point particle $P_{\alpha}$ is the sum of the position of the centre of mass $\boldsymbol{r}_{G}$ and the relative position of $\boldsymbol{r}_{\alpha}^{\prime}$,
\begin{equation}\label{position alpha}
\boldsymbol{r}_{\alpha} = \boldsymbol{r}_{G} + \boldsymbol{r}^{\prime}_{\alpha}
\end{equation}
The intrinsic rotation of the rigid body is characterised by a rotor $R$. Thus, the relative position $\boldsymbol{r}_{\alpha}^{\prime}$ of point $P_{\alpha}$ is related to the initial relative position $\boldsymbol{r}^{\prime}_{\alpha,0}$ through the rotation (Fig.~\ref{Fig: Rotation rigid body}),
\begin{equation}\label{relative position alpha}
\boldsymbol{r}^{\prime}_{\alpha} = R\ \boldsymbol{r}^{\prime}_{\alpha,0}\,R^{\dag}
\end{equation}
The relative velocity $\boldsymbol{v}_{\alpha}$ of point $P_{\alpha}$ is obtained by taking the time derivative of the relative position~\eqref{relative position alpha},
\begin{equation}\label{relative velocity alpha}
\boldsymbol{v}^{\prime}_{\alpha} = \dot{R}\ \boldsymbol{r}^{\prime}_{\alpha,0}\,R^{\dag} + R\ \boldsymbol{r}^{\prime}_{\alpha,0}\,\dot{R}^{\dag}
\end{equation}
In view of the rotor equations~\eqref{derivative identity ter} and~\eqref{derivative identity quad}, the relative velocity~\eqref{relative velocity alpha} is recast as,
\begin{equation}\label{relative velocity alpha bis}
\boldsymbol{v}^{\prime}_{\alpha} = -\,\frac{1}{2}\,\boldsymbol{\Omega}\,R\ \boldsymbol{r}^{\prime}_{\alpha,0}\,R^{\dag} + \frac{1}{2}\,R\ \boldsymbol{r}^{\prime}_{\alpha,0}\,R^{\dag}\,\boldsymbol{\Omega}
\end{equation}
Using the relative position~\eqref{relative position alpha}, the relative velocity~\eqref{relative velocity alpha bis} becomes,
\begin{equation}\label{relative velocity alpha ter}
\boldsymbol{v}^{\prime}_{\alpha} = \frac{1}{2}\left(\boldsymbol{r}_{\alpha}^{\prime}\,\boldsymbol{\Omega} -\,\boldsymbol{\Omega}\,\boldsymbol{r}_{\alpha}^{\prime}\right)
\end{equation}
Using the inner product~\eqref{inner product antisymmetry v B} between the relative position vector $\boldsymbol{r}_{\alpha}^{\prime}$ and the angular velocity bivector $\boldsymbol{\Omega}$, the relative velocity~\eqref{relative velocity alpha ter} reduces to,
\begin{equation}\label{relative velocity alpha quad}
\boldsymbol{v}^{\prime}_{\alpha} = \boldsymbol{r}_{\alpha}^{\prime}\cdot\boldsymbol{\Omega}
\end{equation}
The angular momentum of the rigid body evaluated at the centre of mass $G$ is the sum of the relative angular momenta of all the point particles of mass $m_{\alpha}$, relative position $\boldsymbol{r}_{\alpha}^{\prime}$ and relative momentum $\boldsymbol{p}^{\prime}_{\alpha} = m_{\alpha}\,\boldsymbol{v}^{\prime}_{\alpha}$ that belong to the rigid body. In view of the relative velocity~\eqref{relative velocity alpha quad} and the angular momentum~\eqref{angular momentum bivector}, the angular momentum bivector of the rigid body evaluated at the centre of mass $G$ reads,
\begin{equation}\label{relative angular momentum}
\boldsymbol{L}_{G} = \sum_{\alpha}\,\boldsymbol{r}_{\alpha}^{\prime}\wedge\boldsymbol{p}_{\alpha}^{\prime} = \sum_{\alpha}\,m_{\alpha}\,\boldsymbol{r}_{\alpha}^{\prime}\wedge\boldsymbol{v}_{\alpha}^{\prime} = \sum_{\alpha}\,m_{\alpha}\,\boldsymbol{r}_{\alpha}^{\prime}\wedge\left(\boldsymbol{r}_{\alpha}^{\prime}\cdot\boldsymbol{\Omega}\right)
\end{equation}
The linear mapping of the angular velocity bivector $\boldsymbol{\Omega}$ to the angular momentum bivector $\boldsymbol{L}_{G}$~\eqref{relative angular momentum} is a rotation of the inertia tensor of the rigid body. This mapping is time dependent since the relative position vector $\boldsymbol{r}_{\alpha}^{\prime}$ and the angular velocity bivector $\boldsymbol{\Omega}$ are time dependent. The angular velocity bivector $\boldsymbol{\Omega}$ is related to the initial angular velocity bivector $\boldsymbol{\Omega}_{0}$ through the rotation,
\begin{equation}\label{intial angular velocity}
\boldsymbol{\Omega} = R\ \boldsymbol{\Omega}_{0}\,R^{\dag}
\end{equation}
In view of the orthonormality condition~\eqref{unit rotor time dependent}, the rotations~\eqref{relative position alpha} and~\eqref{intial angular velocity}, the relative velocity~\eqref{relative velocity alpha ter} is recast as,
\begin{equation}\label{relative velocity alpha pent}
\boldsymbol{v}^{\prime}_{\alpha} = \frac{1}{2}\left(R\ \boldsymbol{r}_{\alpha,0}^{\prime}\,\boldsymbol{\Omega}_{0}\,R^{\dag} -\,R\ \boldsymbol{\Omega}_{0}\,\boldsymbol{r}_{\alpha,0}^{\prime}\,R^{\dag}\right) = R\left(\boldsymbol{r}_{\alpha,0}^{\prime}\cdot\boldsymbol{\Omega}_{0}\right)R^{\dag}
\end{equation}
as expected in view of relation~\eqref{relative velocity alpha quad}. Using the rotations~\eqref{relative position alpha} and~\eqref{intial angular velocity}, the angular momentum~\eqref{relative angular momentum} becomes,
\begin{equation}\label{relative angular momentum bis}
\boldsymbol{L}_{G} = \sum_{\alpha}\,m_{\alpha}\,R\ \boldsymbol{r}^{\prime}_{\alpha,0}\,R^{\dag}\wedge R\left(\boldsymbol{r}_{\alpha,0}^{\prime}\cdot\boldsymbol{\Omega}_{0}\right)R^{\dag}
\end{equation}
Using the orthonormality condition~\eqref{unit rotor time dependent}, the angular momentum~\eqref{relative angular momentum bis} is recast as,
\begin{equation}\label{relative angular momentum ter}
\boldsymbol{L}_{G} = \frac{1}{2}\,\sum_{\alpha}\,m_{\alpha}\,R\,\Big(\boldsymbol{r}^{\prime}_{\alpha,0}\left(\boldsymbol{r}_{\alpha,0}^{\prime}\cdot\boldsymbol{\Omega}_{0}\right) -\,\left(\boldsymbol{r}_{\alpha,0}^{\prime}\cdot\boldsymbol{\Omega}_{0}\right)\boldsymbol{r}^{\prime}_{\alpha,0}\Big)\,R^{\dag}
\end{equation}
and reduces to,
\begin{equation}\label{relative angular momentum quad}
\boldsymbol{L}_{G} = R\,\left(\sum_{\alpha}\,m_{\alpha}\,\boldsymbol{r}_{\alpha,0}^{\prime}\wedge\left(\boldsymbol{r}_{\alpha,0}^{\prime}\cdot\boldsymbol{\Omega}_{0}\right)\right)\,R^{\dag}
\end{equation}
The linear mapping of the initial angular velocity bivector $\boldsymbol{\Omega}_{0}$ to the initial angular momentum bivector is the inertia map,
\begin{equation}\label{inertia tensor}
\boldsymbol{I}_G\left(\boldsymbol{\Omega}_{0}\right) = \sum_{\alpha}\,m_{\alpha}\,\boldsymbol{r}_{\alpha,0}^{\prime}\wedge\left(\boldsymbol{r}_{\alpha,0}^{\prime}\cdot\boldsymbol{\Omega}_{0}\right)
\end{equation}
Indeed, in view of the inertia map~\eqref{inertia tensor}, the angular momentum~\eqref{relative angular momentum quad} is recast as,
\begin{equation}\label{relative angular momentum pent}
\boldsymbol{L}_{G} = R\ \boldsymbol{L}_{G,0}\,R^{\dag} = R\ \boldsymbol{I}_G\left(\boldsymbol{\Omega}_{0}\right)R^{\dag}
\end{equation}
where the initial angular momentum $\boldsymbol{L}_{G,0}$ coincides with the inertia map $\boldsymbol{I}_G\left(\boldsymbol{\Omega}_{0}\right)$ of the initial angular velocity $\boldsymbol{\Omega}_{0}$. Afterwards, the angular momentum $\boldsymbol{L}_{G}$ is obtained by performing a rotation of the inertia map $\boldsymbol{I}_G\left(\boldsymbol{\Omega}_{0}\right)$ using the time dependent rotor $R$. For a rigid body, the initial inertia map~\eqref{inertia tensor} is a constant that depends on the mass distribution. In the continuum limit, the point particles are replaced by an infinitesimal volume $dV$ of mass density $\rho\left(\boldsymbol{r}_{0}^{\prime}\right)$. In this limit, the initial inertia map~\eqref{inertia tensor} is recast as an integral over the volume $V$ of the rigid body,
\begin{equation}\label{inertia tensor bis}
\boldsymbol{I}_G\left(\boldsymbol{\Omega}_{0}\right) = \int_V\,dV\,\rho\left(\boldsymbol{r}_{0}^{\prime}\right)\,\boldsymbol{r}_{0}^{\prime}\wedge\left(\boldsymbol{r}_{0}^{\prime}\cdot\boldsymbol{\Omega}_{0}\right)
\end{equation}
The orientation of the inertia map $\boldsymbol{I}_G\left(\boldsymbol{\Omega}_{0}\right)$ is locally determined the bivector $\boldsymbol{r}_{0}^{\prime}\wedge\left(\boldsymbol{r}_{0}^{\prime}\cdot\boldsymbol{\Omega}_{0}\right)$ around each point $\boldsymbol{r}_{0}^{\prime}$ (Fig.~\ref{Fig: Inertia}).
The linearity of the inertia map~\eqref{inertia tensor bis} as a function of the initial angular velocity bivector $\boldsymbol{\Omega}_{0}$ is straightforward to show,
\begin{equation}\label{inertia tensor linear}
\begin{split}
&\boldsymbol{I}_G\left(\lambda_1\,\boldsymbol{\Omega}_{1} + \lambda_2\,\boldsymbol{\Omega}_{2}\right) =  \int_V\,dV\,\rho\left(\boldsymbol{r}_{0}^{\prime}\right)\,\boldsymbol{r}_{0}^{\prime}\wedge\Big(\boldsymbol{r}_{0}^{\prime}\cdot\left(\lambda_1\,\boldsymbol{\Omega}_{1} + \lambda_2\,\boldsymbol{\Omega}_{2}\right)\Big)\\
& = \lambda_1\int_V\,dV\,\rho\left(\boldsymbol{r}_{0}^{\prime}\right)\,\boldsymbol{r}_{0}^{\prime}\wedge\left(\boldsymbol{r}_{0}^{\prime}\cdot\boldsymbol{\Omega}_{1}\right) + \lambda_2\int_V\,dV\,\rho\left(\boldsymbol{r}_{0}^{\prime}\right)\,\boldsymbol{r}_{0}^{\prime}\wedge\left(\boldsymbol{r}_{0}^{\prime}\cdot\boldsymbol{\Omega}_{2}\right)\\
& = \lambda_1\,\boldsymbol{I}_G\left(\boldsymbol{\Omega}_{1}\right) + \lambda_2\,\boldsymbol{I}_G\left(\boldsymbol{\Omega}_{2}\right)
\end{split}
\end{equation}
Using the identity, we write,
\begin{equation}\label{identity inertia}
\begin{split}
&\boldsymbol{r}_{0}^{\prime}\wedge\left(\boldsymbol{r}_{0}^{\prime}\cdot\boldsymbol{\Omega}_{0}\right) = \frac{1}{2}\Big(\boldsymbol{r}_{0}^{\prime}\left(\boldsymbol{r}_{0}^{\prime}\cdot\boldsymbol{\Omega}_{0}\right) -\,\left(\boldsymbol{r}_{0}^{\prime}\cdot\boldsymbol{\Omega}_{0}\right)\boldsymbol{r}_{0}^{\prime}\Big)\\
& = \frac{1}{4}\Big(\boldsymbol{r}_{0}^{\prime}\,\boldsymbol{r}_{0}^{\prime}\,\boldsymbol{\Omega}_{0} -\,2\,\boldsymbol{r}_{0}^{\prime}\,\boldsymbol{\Omega}_{0}\,\boldsymbol{r}_{0}^{\prime} + \boldsymbol{\Omega}_{0}\,\boldsymbol{r}_{0}^{\prime}\,\boldsymbol{r}_{0}^{\prime}\Big)
\end{split}
\end{equation}
which is recast as,
\begin{equation}\label{identity inertia bis}
\boldsymbol{r}_{0}^{\prime}\wedge\left(\boldsymbol{r}_{0}^{\prime}\cdot\boldsymbol{\Omega}_{0}\right) = \frac{1}{2}\,\Big(\boldsymbol{r}_{0}^{\prime\,2}\,\boldsymbol{\Omega}_{0} -\,\boldsymbol{r}_{0}^{\prime}\,\boldsymbol{\Omega}_{0}\,\boldsymbol{r}_{0}^{\prime}\Big)
\end{equation}
In view of the identity~\eqref{identity inertia bis}, the initial inertia map is recast as,
\begin{equation}\label{inertia tensor ter}
\boldsymbol{I}_G\left(\boldsymbol{\Omega}_{0}\right) = \frac{1}{2}\,\int_V\,dV\,\rho\left(\boldsymbol{r}_{0}^{\prime}\right)\,\Big(\boldsymbol{r}_{0}^{\prime\,2}\,\boldsymbol{\Omega}_{0} -\,\boldsymbol{r}_{0}^{\prime}\,\boldsymbol{\Omega}_{0}\,\boldsymbol{r}_{0}^{\prime}\Big)
\end{equation}
%


\section{Inertia and principal body frame}
\label{Inertia and principal body frame}

\begin{figure}[!ht]
\begin{center}
\includegraphics[scale=0.50]{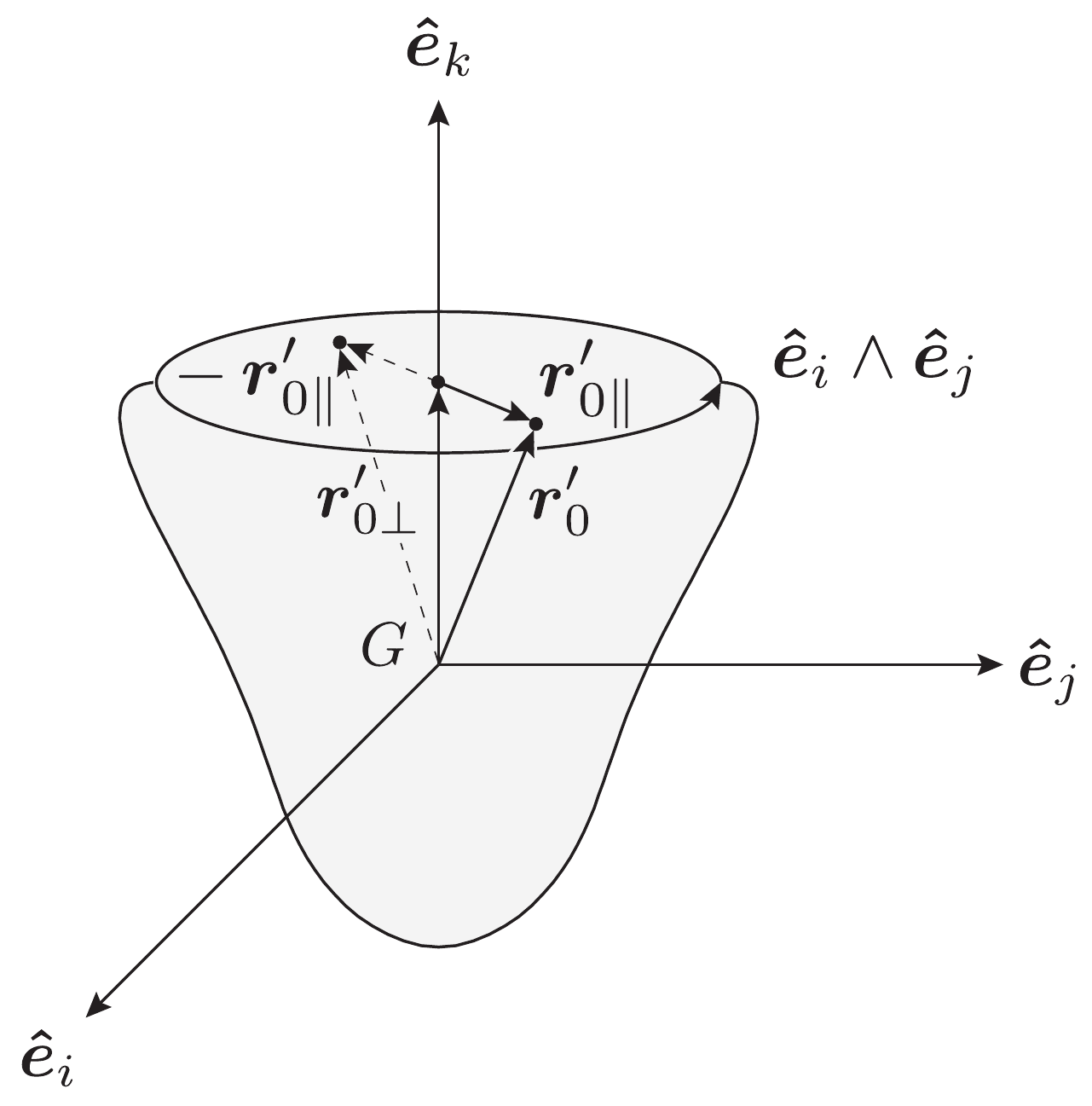}
\end{center}
\caption{For a rigid body with an axis of symmetry $G\boldsymbol{\hat{e}}_k$, for every point described by a relative parallel position vector $\boldsymbol{r}_{0\parallel}^{\prime}$ there is a symmetric point described by a relative parallel position vector $-\,\boldsymbol{r}_{0\parallel}^{\prime}$.}\label{Fig: Body_frame_condition}
\end{figure}

\noindent In an orthonormal frame $\left\{\boldsymbol{\hat{e}}_1,\boldsymbol{\hat{e}}_2,\boldsymbol{\hat{e}}_3\right\}$ initially in the frame of reference of the rigid body and attached to its centre of mass $G$, the angular velocity bivector $\boldsymbol{\Omega}_{0}$ is a linear combination of the inertia bivectors,
\begin{equation}\label{angular velocity basis}
\boldsymbol{\Omega}_{0} = \frac{1}{2}\sum_{i,j = 1}^{3}\,\Omega_{ij}\,\boldsymbol{\hat{e}}_{i}\wedge\boldsymbol{\hat{e}}_{j}
\end{equation}
where $\Omega_{ij} = -\,\Omega_{ji}$ is the initial scalar angular velocity in the oriented plane defined by the unit bivector $\boldsymbol{\hat{e}}_{i}\wedge\boldsymbol{\hat{e}}_{j}$ defined as,
\begin{equation}\label{scalar angular velocity}
\Omega_{ij} = \boldsymbol{\hat{e}}_{i}\cdot\boldsymbol{\Omega}_{0}\cdot\boldsymbol{\hat{e}}_{j} = \left(\boldsymbol{\hat{e}}_{j}\wedge\boldsymbol{\hat{e}}_{i}\right)\cdot\boldsymbol{\Omega}_{0} = \boldsymbol{\Omega}_{0}\cdot\left(\boldsymbol{\hat{e}}_{j}\wedge\boldsymbol{\hat{e}}_{i}\right)
\end{equation}
Using the initial scalar angular velocity~\eqref{angular velocity basis}, the initial inertia map can be written as a linear combination of inertia bivectors $\boldsymbol{I}_G\left(\boldsymbol{\hat{e}}_i\wedge\boldsymbol{\hat{e}}_j\right)$,
\begin{equation}\label{inertia map}
\boldsymbol{I}_G\left(\boldsymbol{\Omega}_{0}\right) = \frac{1}{2}\sum_{i,j = 1}^{3}\,\Omega_{ij}\,\boldsymbol{I}_G\left(\boldsymbol{\hat{e}}_{i}\wedge\boldsymbol{\hat{e}}_{j}\right)
\end{equation}
In view the inertia maps~\eqref{inertia tensor ter} and~\eqref{inertia map}, the inertia bivectors are written as,
\begin{equation}\label{inertia bivector}
\boldsymbol{I}_G\left(\boldsymbol{\hat{e}}_{i}\wedge\boldsymbol{\hat{e}}_{j}\right) = \frac{1}{2}\,\int_V\,dV\,\rho\left(\boldsymbol{r}_{0}^{\prime}\right)\,\Big(\boldsymbol{r}_{0}^{\prime\,2}\left(\boldsymbol{\hat{e}}_{i}\wedge\boldsymbol{\hat{e}}_{j}\right) -\,\boldsymbol{r}_{0}^{\prime}\left(\boldsymbol{\hat{e}}_{i}\wedge\boldsymbol{\hat{e}}_{j}\right)\boldsymbol{r}_{0}^{\prime}\Big)
\end{equation}
which shows that the inertia bivector is antisymmetric,
\begin{equation}\label{inertia bivector antisymmetric}
\boldsymbol{I}_G\left(\boldsymbol{\hat{e}}_{i}\wedge\boldsymbol{\hat{e}}_{j}\right) = -\,\boldsymbol{I}_G\left(\boldsymbol{\hat{e}}_{j}\wedge\boldsymbol{\hat{e}}_{i}\right)
\end{equation}
as expected. In order to account for the respective orientation of the vector $\boldsymbol{r}_{0}^{\prime}$ with respect to the oriented plane spanned by the unit bivector $\boldsymbol{\hat{e}}_{i}\wedge\boldsymbol{\hat{e}}_{j}$, we decompose the relative position vector $\boldsymbol{r}_{0}^{\prime}$ into a parallel part $\boldsymbol{r}_{0\parallel}^{\prime}$ and an orthogonal part $\boldsymbol{r}_{0\perp}^{\prime}$ using the projection~\eqref{projection automorphism plane} and the rejection~\eqref{rejection automorphism plane},
\begin{equation}\label{decomposition of unit vector inertia}
\boldsymbol{r}_{0}^{\prime} = \boldsymbol{r}_{0\parallel}^{\prime} + \boldsymbol{r}_{0\perp}^{\prime}
\end{equation}
where,
\begin{equation}\label{projection and rejection vector inertia}
\begin{split}
&\boldsymbol{r}_{0\parallel}^{\prime} = \mathsf{P}_{\boldsymbol{\hat{e}}_{i}\wedge\boldsymbol{\hat{e}}_{j}}\left(\boldsymbol{r}_{0}^{\prime}\right) = \left(\boldsymbol{\hat{e}}_{j}\wedge\boldsymbol{\hat{e}}_{i}\right)\cdot\Big(\left(\boldsymbol{\hat{e}}_{i}\wedge\boldsymbol{\hat{e}}_{j}\right)\cdot\boldsymbol{r}_{0}\Big)\\
&\boldsymbol{r}_{0\perp}^{\prime} = \mathsf{\bar{P}}_{\boldsymbol{\hat{e}}_{i}\wedge\boldsymbol{\hat{e}}_{j}}\left(\boldsymbol{r}_{0}^{\prime}\right) = \left(\boldsymbol{\hat{e}}_{j}\wedge\boldsymbol{\hat{e}}_{i}\right)\cdot\Big(\left(\boldsymbol{\hat{e}}_{i}\wedge\boldsymbol{\hat{e}}_{j}\right)\wedge\boldsymbol{r}_{0}\Big)
\end{split}
\end{equation}
and thus, $\boldsymbol{r}_{0\parallel}^{\prime}\cdot\boldsymbol{r}_{0\perp}^{\prime} = 0$. In view of the decomposition~\eqref{decomposition of unit vector inertia}, we have,
\begin{equation}\label{decomposition of unit vector inertia squared}
\boldsymbol{r}_{0}^{\prime\,2} = \boldsymbol{r}_{0\parallel}^{\prime\,2} + \boldsymbol{r}_{0\perp}^{\prime\,2}
\end{equation}
and,
\begin{equation}\label{decomposition of reflection inertia}
\begin{split}
&\boldsymbol{r}_{0}^{\prime}\left(\boldsymbol{\hat{e}}_{i}\wedge\boldsymbol{\hat{e}}_{j}\right)\boldsymbol{r}_{0}^{\prime} = \boldsymbol{r}_{0\parallel}^{\prime}\left(\boldsymbol{\hat{e}}_{i}\wedge\boldsymbol{\hat{e}}_{j}\right)\boldsymbol{r}_{0\parallel}^{\prime} + \boldsymbol{r}_{0\perp}^{\prime}\left(\boldsymbol{\hat{e}}_{i}\wedge\boldsymbol{\hat{e}}_{j}\right)\boldsymbol{r}_{0\perp}^{\prime}\\
&\phantom{\boldsymbol{r}_{0}^{\prime}\left(\boldsymbol{\hat{e}}_{i}\wedge\boldsymbol{\hat{e}}_{j}\right)\boldsymbol{r}_{0}^{\prime} = } + \boldsymbol{r}_{0\parallel}^{\prime}\left(\boldsymbol{\hat{e}}_{i}\wedge\boldsymbol{\hat{e}}_{j}\right)\boldsymbol{r}_{0\perp}^{\prime} + \boldsymbol{r}_{0\perp}^{\prime}\left(\boldsymbol{\hat{e}}_{i}\wedge\boldsymbol{\hat{e}}_{j}\right)\boldsymbol{r}_{0\parallel}^{\prime}
\end{split}
\end{equation}
The vectors $\boldsymbol{r}_{0\parallel}^{\prime}$ and $\boldsymbol{r}_{0\perp}^{\prime}$ are written in components in the orthonormal frame as,
\begin{equation}\label{vectors inertia components}
\boldsymbol{r}_{0\parallel}^{\prime} = r_{\parallel,i}^{\prime}\,\boldsymbol{\hat{e}}_{i} + r_{\parallel,j}^{\prime}\,\boldsymbol{\hat{e}}_{j} \qquad\text{and}\qquad \boldsymbol{r}_{0\perp}^{\prime} = r_{\perp,k}^{\prime}\,\boldsymbol{\hat{e}}_{k}
\end{equation}
where the unit vector $\boldsymbol{\hat{e}}_{k}$ is the dual of the unit bivector $\boldsymbol{\hat{e}}_{i}\cdot\boldsymbol{\hat{e}}_{j}$ according to, 
\begin{equation}\label{dual inertia frame}
\boldsymbol{\hat{e}}_{k} = \left(\boldsymbol{\hat{e}}_{i}\wedge\boldsymbol{\hat{e}}_{j}\right)^{\ast} \quad\text{and}\quad \boldsymbol{\hat{e}}_{i}\wedge\boldsymbol{\hat{e}}_{j} = -\,\boldsymbol{\hat{e}}_{k}^{\ast} \quad\text{thus}\quad \boldsymbol{\hat{e}}_{i}\cdot\boldsymbol{\hat{e}}_{k} = \boldsymbol{\hat{e}}_{j}\cdot\boldsymbol{\hat{e}}_{k} = 0
\end{equation}
Since the vector $\boldsymbol{r}_{0\perp}^{\prime} = r_{\perp,k}^{\prime}\,\boldsymbol{\hat{e}}_{k}$ is orthogonal to the oriented plane described by the unit bivector $\boldsymbol{\hat{e}}_{i}\wedge\boldsymbol{\hat{e}}_{j}$, the vector $\boldsymbol{r}_{0\perp}^{\prime}$ commutes with this bivector,
\begin{equation}\label{commutation r perp bivector}
\boldsymbol{r}_{0\perp}^{\prime}\left(\boldsymbol{\hat{e}}_{i}\wedge\boldsymbol{\hat{e}}_{j}\right) = r_{\perp,k}^{\prime}\,\boldsymbol{\hat{e}}_{k}\,\boldsymbol{\hat{e}}_{i}\,\boldsymbol{\hat{e}}_{j} = r_{\perp,k}^{\prime}\,\boldsymbol{\hat{e}}_{i}\,\boldsymbol{\hat{e}}_{j}\,\boldsymbol{\hat{e}}_{k} = \left(\boldsymbol{\hat{e}}_{i}\wedge\boldsymbol{\hat{e}}_{j}\right)\boldsymbol{r}_{0\perp}^{\prime}
\end{equation}
Since the vector $\boldsymbol{r}_{0\parallel}^{\prime} = r_{\parallel,i}^{\prime}\,\boldsymbol{\hat{e}}_{i} + r_{\parallel,j}^{\prime}\,\boldsymbol{\hat{e}}_{j}$ is coplanar to the oriented plane described by the unit bivector $\boldsymbol{\hat{e}}_{i}\wedge\boldsymbol{\hat{e}}_{j}$, the vector $\boldsymbol{r}_{0\parallel}^{\prime}$ anticommutes with this bivector,
\begin{equation}\label{anticommutation r parallel bivector}
\begin{split}
&\boldsymbol{r}_{0\parallel}^{\prime}\left(\boldsymbol{\hat{e}}_{i}\wedge\boldsymbol{\hat{e}}_{j}\right) = r_{\parallel,i}^{\prime}\,\boldsymbol{\hat{e}}_{i}\,\boldsymbol{\hat{e}}_{i}\,\boldsymbol{\hat{e}}_{j} + r_{\parallel,j}^{\prime}\,\boldsymbol{\hat{e}}_{j}\,\boldsymbol{\hat{e}}_{i}\,\boldsymbol{\hat{e}}_{j}\\
&\phantom{\boldsymbol{r}_{0\parallel}^{\prime}\left(\boldsymbol{\hat{e}}_{i}\wedge\boldsymbol{\hat{e}}_{j}\right)} = -\,r_{\parallel,i}^{\prime}\,\boldsymbol{\hat{e}}_{i}\,\boldsymbol{\hat{e}}_{j}\,\boldsymbol{\hat{e}}_{i} -\,r_{\parallel,j}^{\prime}\,\boldsymbol{\hat{e}}_{i}\,\boldsymbol{\hat{e}}_{j}\,\boldsymbol{\hat{e}}_{j} = -\,\left(\boldsymbol{\hat{e}}_{i}\wedge\boldsymbol{\hat{e}}_{j}\right)\boldsymbol{r}_{0\parallel}^{\prime}
\end{split}
\end{equation}
In view of the identities~\eqref{commutation r perp bivector} and~\eqref{anticommutation r parallel bivector}, the identity~\eqref{decomposition of reflection inertia} becomes,
\begin{equation}\label{decomposition of reflection inertia bis}
\begin{split}
&\boldsymbol{r}_{0}^{\prime}\left(\boldsymbol{\hat{e}}_{i}\wedge\boldsymbol{\hat{e}}_{j}\right)\boldsymbol{r}_{0}^{\prime} = \left(-\,\boldsymbol{r}_{0\parallel}^{\prime}\,\boldsymbol{r}_{0\parallel}^{\prime} + \boldsymbol{r}_{0\perp}^{\prime}\,\boldsymbol{r}_{0\perp}^{\prime} -\,\boldsymbol{r}_{0\perp}^{\prime}\,\boldsymbol{r}_{0\parallel}^{\prime} + \boldsymbol{r}_{0\parallel}^{\prime}\,\boldsymbol{r}_{0\perp}^{\prime} \right)\left(\boldsymbol{\hat{e}}_{i}\wedge\boldsymbol{\hat{e}}_{j}\right)\\
&= -\,\left(\boldsymbol{r}_{0\parallel}^{\prime\,2}-\,\boldsymbol{r}_{0\perp}^{\prime\,2}\right)\left(\boldsymbol{\hat{e}}_{i}\wedge\boldsymbol{\hat{e}}_{j}\right) + 2\,\boldsymbol{r}_{0\parallel}^{\prime}\cdot\Big(\boldsymbol{r}_{0\perp}^{\prime}\wedge\left(\boldsymbol{\hat{e}}_{i}\wedge\boldsymbol{\hat{e}}_{j}\right)\Big)
\end{split}
\end{equation}
since the vector $\boldsymbol{r}_{0\perp}^{\prime}$ is orthogonal to the bivector $\boldsymbol{\hat{e}}_{i}\wedge\boldsymbol{\hat{e}}_{j}$ and the vector $\boldsymbol{r}_{0\parallel}^{\prime}$ belongs to the oriented volume element defined by the trivector $\boldsymbol{r}_{0\perp}^{\prime}\wedge\left(\boldsymbol{\hat{e}}_{i}\wedge\boldsymbol{\hat{e}}_{j}\right)$. Using identities~\eqref{decomposition of unit vector inertia squared} and~\eqref{decomposition of reflection inertia bis}, the inertia bivector~\eqref{inertia bivector} becomes,
\begin{equation}\label{inertia bivector bis}
\begin{split}
&\boldsymbol{I}_G\left(\boldsymbol{\hat{e}}_{i}\wedge\boldsymbol{\hat{e}}_{j}\right) = \int_V\,dV\,\rho\left(\boldsymbol{r}_{0}^{\prime}\right)\,\boldsymbol{r}_{0\parallel}^{\prime\,2}\ \boldsymbol{\hat{e}}_{i}\wedge\boldsymbol{\hat{e}}_{j}\\
&\phantom{\boldsymbol{I}_G\left(\boldsymbol{\hat{e}}_{i}\wedge\boldsymbol{\hat{e}}_{j}\right) =} -\,\int_V\,dV\,\rho\left(\boldsymbol{r}_{0}^{\prime}\right)\boldsymbol{r}_{0\parallel}^{\prime}\cdot\Big(\boldsymbol{r}_{0\perp}^{\prime}\wedge\left(\boldsymbol{\hat{e}}_{i}\wedge\boldsymbol{\hat{e}}_{j}\right)\Big)
\end{split}
\end{equation}
where the second integral is a bivector orthogonal to the vector $\boldsymbol{r}_{0\parallel}^{\prime}$. It is useful to introduce the inertia dual vector $\boldsymbol{i}_G\left(\boldsymbol{\hat{e}}_k\right)$ defined as the dual of the inertia bivector $\boldsymbol{I}_G\left(\boldsymbol{\hat{e}}_{i}\wedge\boldsymbol{\hat{e}}_{j}\right)$,
\begin{equation}\label{inertia duality}
\boldsymbol{i}_G\left(\boldsymbol{\hat{e}}_k\right) = \boldsymbol{I}_G^{\,\ast}\left(\boldsymbol{\hat{e}}_{i}\wedge\boldsymbol{\hat{e}}_{j}\right) \qquad\text{and}\qquad \boldsymbol{I}_G\left(\boldsymbol{\hat{e}}_{i}\wedge\boldsymbol{\hat{e}}_{j}\right) = -\,\boldsymbol{i}_G^{\,\ast}\left(\boldsymbol{\hat{e}}_k\right)
\end{equation}
In view of relations~\eqref{inertia bivector bis} and~\eqref{inertia duality}, the inertia dual vector is written as,
\begin{equation}\label{inertia vector}
\begin{split}
&\boldsymbol{i}_G\left(\boldsymbol{\hat{e}}_{k}\right) = \int_V\,dV\,\rho\left(\boldsymbol{r}_{0}^{\prime}\right)\,\boldsymbol{r}_{0\parallel}^{\prime\,2}\left(\boldsymbol{\hat{e}}_{i}\wedge\boldsymbol{\hat{e}}_{j}\right)^{\ast}\\
&\phantom{\boldsymbol{i}_G\left(\boldsymbol{\hat{e}}_{k}\right) =} -\,\int_V\,dV\,\rho\left(\boldsymbol{r}_{0}^{\prime}\right)\bigg(\boldsymbol{r}_{0\parallel}^{\prime}\cdot\Big(\boldsymbol{r}_{0\perp}^{\prime}\wedge\left(\boldsymbol{\hat{e}}_{i}\wedge\boldsymbol{\hat{e}}_{j}\right)\Big)\bigg)^{\ast}
\end{split}
\end{equation}
In view of the dualities~\eqref{dual vector trivector inner} and~\eqref{scalar duality}, we obtain the identity,
\begin{equation}\label{duality inertia orthogonal}
\begin{split}
&\bigg(\boldsymbol{r}_{0\parallel}^{\prime}\cdot\Big(\boldsymbol{r}_{0\perp}^{\prime}\wedge\left(\boldsymbol{\hat{e}}_{i}\wedge\boldsymbol{\hat{e}}_{j}\right)\Big)\bigg)^{\ast} = \boldsymbol{r}_{0\parallel}^{\prime}\Big(\boldsymbol{r}_{0\perp}^{\prime}\wedge\left(\boldsymbol{\hat{e}}_{i}\wedge\boldsymbol{\hat{e}}_{j}\right)\Big)^{\ast}\\
&= \boldsymbol{r}_{0\parallel}^{\prime}\Big(\boldsymbol{r}_{0\perp}^{\prime}\cdot\left(\boldsymbol{\hat{e}}_{i}\wedge\boldsymbol{\hat{e}}_{j}\right)^{\ast}\Big) = \boldsymbol{r}_{0\parallel}^{\prime}\left(\boldsymbol{r}_{0\perp}^{\prime}\cdot\boldsymbol{\hat{e}}_{k}\right)
\end{split}
\end{equation}
In view of the dualities~\eqref{dual inertia frame} and~\eqref{duality inertia orthogonal}, the inertia dual vector~\eqref{inertia vector} reduces to,
\begin{equation}\label{inertia vector bis}
\boldsymbol{i}_G\left(\boldsymbol{\hat{e}}_{k}\right) = \int_V\,dV\,\rho\left(\boldsymbol{r}_{0}^{\prime}\right)\,\boldsymbol{r}_{0\parallel}^{\prime\,2}\,\boldsymbol{\hat{e}}_{k} -\,\int_V\,dV\,\rho\left(\boldsymbol{r}_{0}^{\prime}\right)\boldsymbol{r}_{0\parallel}^{\prime}\left(\boldsymbol{r}_{0}^{\prime}\cdot\boldsymbol{\hat{e}}_k\right)
\end{equation}
For a unit vector $\boldsymbol{\hat{e}}_{\ell} = \cos\phi\,\boldsymbol{\hat{e}}_i + \sin\phi\,\boldsymbol{\hat{e}}_j$, where the angle $\phi\in[0,2\pi)$, in the oriented plane spanned by the bivector $\boldsymbol{\hat{e}}_{i}\wedge\boldsymbol{\hat{e}}_{j}$ orthogonal to the unit vector $\boldsymbol{\hat{e}}_{k}$, the projection of the inertia dual vector~\eqref{inertia vector bis} along the axis $G\,\boldsymbol{\hat{e}}_{\ell}$ yields the matrix element,
\begin{equation}\label{inertia vector ter}
i_{G,k\ell} = \boldsymbol{i}_G\left(\boldsymbol{\hat{e}}_{k}\right)\cdot\boldsymbol{\hat{e}}_{\ell} = \int_V\,dV\,\rho\left(\boldsymbol{r}_{0}^{\prime}\right)\Big(\,\boldsymbol{r}_{0\parallel}^{\prime\,2}\left(\boldsymbol{\hat{e}}_{k}\cdot\boldsymbol{\hat{e}}_{\ell}\right) -\,\left(\boldsymbol{r}_{0}^{\prime}\cdot\boldsymbol{\hat{e}}_k\right)\left(\boldsymbol{r}_{0}^{\prime}\cdot\boldsymbol{\hat{e}}_{\ell}\right)\Big)
\end{equation}
since $\boldsymbol{\hat{e}}_{\ell}\cdot\boldsymbol{\hat{e}}_{k} = 0$ and thus $\boldsymbol{r}_{0}^{\prime}\cdot\boldsymbol{\hat{e}}_{\ell} = \boldsymbol{r}_{0\parallel}^{\prime}\cdot\boldsymbol{\hat{e}}_{\ell}$. In view of relation~\eqref{decomposition of unit vector inertia squared}, for any unit vector $\boldsymbol{\hat{e}}_{\ell}$ belonging to the orthonormal basis $\{\boldsymbol{\hat{e}}_{i},\boldsymbol{\hat{e}}_{j},\boldsymbol{\hat{e}}_{k}\}$, the matrix element~\eqref{inertia vector ter} is the $k\ell$ component of the representation of the inertia tensor of vector algebra in the orthonormal frame,
\begin{equation}\label{inertia vector quad}
i_{G,k\ell} = i_{G,\ell k} = \int_V\,dV\,\rho\left(\boldsymbol{r}_{0}^{\prime}\right)\Big(\,\boldsymbol{r}_{0}^{\prime\,2}\left(\boldsymbol{\hat{e}}_{k}\cdot\boldsymbol{\hat{e}}_{\ell}\right) -\,\left(\boldsymbol{r}_{0}^{\prime}\cdot\boldsymbol{\hat{e}}_k\right)\left(\boldsymbol{r}_{0}^{\prime}\cdot\boldsymbol{\hat{e}}_{\ell}\right)\Big)
\end{equation}
According to the spectral theorem, there is always an orthonormal frame, called a principal body frame, where the inertia tensor is diagonal. Such a frame is not unique but its existence is guaranteed mathematically. From now on, we consider that the unit vectors $\boldsymbol{\hat{e}}_{k}$ with $k=1,2,3$ belong initially to a principal body frame of the rigid body and thus are the unit vectors of the initial principal axis $G\,\boldsymbol{\hat{e}}_{k}$ of the rigid body. This means that the unit vector $\boldsymbol{\hat{e}}_{k}$ is an eigenvector of the inertia dual vector $\boldsymbol{i}_G\left(\boldsymbol{\hat{e}}_{k}\right)$. Thus, in a principal body frame, the inertia dual vector~\eqref{inertia vector bis} reduces to,
\begin{equation}\label{inertia vector pent}
\boldsymbol{i}_G\left(\boldsymbol{\hat{e}}_{k}\right) = i_{G,k}\,\boldsymbol{\hat{e}}_{k} = \int_V\,dV\,\rho\left(\boldsymbol{r}_{0}^{\prime}\right)\,\boldsymbol{r}_{0\parallel}^{\prime\,2}\,\boldsymbol{\hat{e}}_{k}
\end{equation}
where 
\begin{equation}\label{principal axis}
i_{G,k} = \int_V\,dV\,\rho\left(\boldsymbol{r}_{0}^{\prime}\right)\,\boldsymbol{r}_{0\parallel}^{\prime\,2}
\end{equation}
is the initial principal moment of inertia of the rigid body around the principal axis $G\,\boldsymbol{\hat{e}}_{k}$ and the vector $\boldsymbol{r}_{0\parallel}^{\prime}$ is orthogonal of the initial principal axis and parallel to the oriented plane defined by the bivector $\boldsymbol{\hat{e}}_{i}\wedge\boldsymbol{\hat{e}}_{j}$. In view of relations~\eqref{inertia vector bis} and~\eqref{inertia vector pent}, the condition for a frame to be a principal body frame is that every unit vector $\boldsymbol{\hat{e}}_k$ of the frame satisfies the condition,
\begin{equation}\label{condition principal body frame}
\int_V\,dV\,\rho\left(\boldsymbol{r}_{0}^{\prime}\right)\boldsymbol{r}_{0\parallel}^{\prime}\left(\boldsymbol{r}_{0}^{\prime}\cdot\boldsymbol{\hat{e}}_k\right) = 0
\end{equation}
If the axis $G\boldsymbol{\hat{e}}_k$ is an axis of symmetry of the rigid body, the condition~\eqref{condition principal body frame} is clearly satisfied since for every point of the rigid body described by a relative parallel position vector $\boldsymbol{r}_{0\parallel}^{\prime}$ there is a symmetric point described by a relative parallel position vector $-\,\boldsymbol{r}_{0\parallel}^{\prime}$ such that the integral~\eqref{condition principal body frame} vanishes (Fig.~\ref{Fig: Body_frame_condition}). Even if the rigid body has no axis of symmetry, there is always a principal body frame or a class of principal body frames that satisfy the condition~\eqref{condition principal body frame}.

\noindent The rigid body can be sliced into a continuous set of slices of infinitesimal thickness orthogonal to the unit vector $\boldsymbol{\hat{e}}_k$. The material points belonging to each slice are characterised by a constant relative position vector $\boldsymbol{r}_{0\perp}^{\prime}$ orthogonal to the slice. Geometrically, each slice has to satisfy the condition~\eqref{condition principal body frame}, which therefore reduces to,
\begin{equation}\label{condition principal body frame bis}
\int_V\,dV\,\rho\left(\boldsymbol{r}_{0}^{\prime}\right)\boldsymbol{r}_{0\parallel}^{\prime} = 0
\end{equation}
In the principal body frame, in view of the inertia duality~\eqref{inertia duality} and the inertia pseudovector~\eqref{inertia vector pent}, the inertia bivector reduces~\eqref{inertia bivector bis} to,
\begin{equation}\label{inertia bivector ter}
\boldsymbol{I}_G\left(\boldsymbol{\hat{e}}_{i}\wedge\boldsymbol{\hat{e}}_{j}\right) = \int_V\,dV\,\rho\left(\boldsymbol{r}_{0}^{\prime}\right)\,\boldsymbol{r}_{0\parallel}^{\prime\,2}\ \boldsymbol{\hat{e}}_{i}\wedge\boldsymbol{\hat{e}}_{j}
\end{equation}
which means that the unit bivector $\boldsymbol{\hat{e}}_{i}\wedge\boldsymbol{\hat{e}}_{j}$ is an eigenbivector of the inertia bivector $\boldsymbol{I}_G\left(\boldsymbol{\hat{e}}_{i}\wedge\boldsymbol{\hat{e}}_{j}\right)$,
\begin{equation}\label{inertia bivector quad}
\boldsymbol{I}_G\left(\boldsymbol{\hat{e}}_{i}\wedge\boldsymbol{\hat{e}}_{j}\right) = I_{G,ij}\ \boldsymbol{\hat{e}}_{i}\wedge\boldsymbol{\hat{e}}_{j}
\end{equation}
The antisymmetry of the inertia bivector~\eqref{inertia bivector antisymmetric} is written in components as,
\begin{equation}\label{inertia bivector quad}
\boldsymbol{I}_G\left(\boldsymbol{\hat{e}}_{i}\wedge\boldsymbol{\hat{e}}_{j}\right) = I_{G,ij}\ \boldsymbol{\hat{e}}_{i}\wedge\boldsymbol{\hat{e}}_{j} = -\,I_{G,ji}\ \boldsymbol{\hat{e}}_{j}\wedge\boldsymbol{\hat{e}}_{i} = -\,\boldsymbol{I}_G\left(\boldsymbol{\hat{e}}_{j}\wedge\boldsymbol{\hat{e}}_{i}\right)
\end{equation}
In view of the antisymmetry of the unit bivector, i.e. $\boldsymbol{\hat{e}}_{i}\wedge\boldsymbol{\hat{e}}_{j} = -\,\boldsymbol{\hat{e}}_{j}\wedge\boldsymbol{\hat{e}}_{i}$, and of relations~\eqref{Levi Civita},~\eqref{principal axis},~\eqref{inertia bivector ter} and~\eqref{inertia bivector quad}, the principal moment of inertia of the rigid body in the oriented plane defined by the bivector $\boldsymbol{\hat{e}}_{i}\wedge\boldsymbol{\hat{e}}_{j}$ is written as,
\begin{equation}\label{moment of inertia}
I_{G,ij} = I_{G,ji} = \left(\varepsilon_{ijk}\right)^2i_{G,k} = \int_V\,dV\,\rho\left(\boldsymbol{r}_{0}^{\prime}\right)\,\boldsymbol{r}_{0\parallel}^{\prime\,2}
\end{equation}
where the indices $i$, $k$ and $k$ are different and $\boldsymbol{\hat{e}}_{k} = \left(\boldsymbol{\hat{e}}_{i}\wedge\boldsymbol{\hat{e}}_{j}\right)^{\ast}$.


\section{Huygens-Steiner theorem}
\label{Huygens-Steiner theorem}

\noindent In this section, we determine the moments of inertia $I_{A,ij}$ of the rigid body with respect to a point $A$, that may or may not belong to the rigid body, in the rotation plane spanned by the unit bivector $\boldsymbol{\hat{e}}_{i}\wedge\boldsymbol{\hat{e}}_{j}$. In order to do so, we adopt the notation $\boldsymbol{r}_{GP\parallel} = \boldsymbol{r}_{0\parallel}^{\prime}$ for the projection in the plane of rotation of the relative position of a point $P$ belonging to the rigid body. Thus, the moment of inertia~\eqref{moment of inertia} in the oriented plane spanned by the unit bivector $\boldsymbol{\hat{e}}_{i}\wedge\boldsymbol{\hat{e}}_{j}$ is written as, 
\begin{equation}\label{moment of inertia G}
I_{G,ij} = \int_V\,dV\,\rho\left(\boldsymbol{r}_{0}^{\prime}\right)\boldsymbol{r}_{GP\parallel}^{2}
\end{equation}
and the condition~\eqref{condition principal body frame bis} characterising a principal body frame is given by,
\begin{equation}\label{condition principal body frame ter}
\int_V\,dV\,\rho\left(\boldsymbol{r}_{0}^{\prime}\right)\boldsymbol{r}_{GP\parallel} = 0
\end{equation}
In view of the geometric identity,
\begin{equation}\label{geometric identity}
\begin{split}
&\boldsymbol{r}_{GP\parallel}^{2} = \left(\boldsymbol{r}_{AP\parallel}-\,\boldsymbol{r}_{AG}\right)^2 = \boldsymbol{r}_{AP\parallel}^2 -\,2\,\boldsymbol{r}_{AG}\cdot\boldsymbol{r}_{AP\parallel} + \boldsymbol{r}_{AG}^2\\
&\phantom{\boldsymbol{r}_{GP\parallel}^{2}} = \boldsymbol{r}_{AP\parallel}^2 -\,2\,\boldsymbol{r}_{AG}\cdot\boldsymbol{r}_{GP\parallel} -\,\boldsymbol{r}_{AG}^2
\end{split}
\end{equation}
where $\boldsymbol{r}_{AG} = \boldsymbol{r}_{AG\parallel}$ since points $A$ and $G$ belong to the rotation plane. The moment of inertia~\eqref{moment of inertia G} is recast as,
\begin{equation}\label{moment of inertia G bis}
\begin{split}
&I_{G,ij} = \int_V\,dV\,\rho\left(\boldsymbol{r}_{0}^{\prime}\right)\boldsymbol{r}_{AP\parallel}^{2} -\,\int_V\,dV\,\rho\left(\boldsymbol{r}_{0}^{\prime}\right)\boldsymbol{r}_{AG}^2\\
&\phantom{I_{G,ij} =} -\,2\,\boldsymbol{r}_{AG}\cdot\left(\int_V\,dV\,\rho\left(\boldsymbol{r}_{0}^{\prime}\right)\,\boldsymbol{r}_{GP\parallel}\right)
\end{split}
\end{equation}
In view of the moment of inertia with respect at point $G$~\eqref{moment of inertia G}, the moment of inertia with respect to point $A$ in the oriented plane spanned by the unit bivector $\boldsymbol{\hat{e}}_{i}\wedge\boldsymbol{\hat{e}}_{j}$ is given by,
\begin{equation}\label{moment of inertia A}
I_{A,ij} = \int_V\,dV\,\rho\left(\boldsymbol{r}_{0}^{\prime}\right)\boldsymbol{r}_{AP\parallel}^{2}
\end{equation}
The mass $m$ of the rigid body is obtained by integrating the mass density over its volume $V$,
\begin{equation}\label{mass}
m = \int_V\,dV\,\rho\left(\boldsymbol{r}_{0}^{\prime}\right)
\end{equation}
In view of the condition~\eqref{condition principal body frame ter}, the moment of inertia~\eqref{moment of inertia A} and the mass~\eqref{mass}, the identity~\eqref{moment of inertia G bis} yields the Huygens-Steiner theorem,
\begin{equation}\label{Huygens-Steiner theorem bis}
I_{A,ij} = I_{G,ij} + m\,\boldsymbol{r}_{AG}^2
\end{equation}
%


\section{Angular momentum of a rigid body}
\label{Angular momentum of a rigid body}

\noindent In this section, we derive the angular momentum bivector $\boldsymbol{L}_{G}$ and its dual the angular momentum pseudovector $\boldsymbol{\ell}_{G}$ evaluated at the centre of mass of a rigid body. Using the eigenbivector decomposition~\eqref{inertia bivector quad}, the initial inertia map~\eqref{inertia map}, which is the initial angular momentum bivector evaluated at the centre of mass $G$, is written in the principal body frame as,
\begin{equation}\label{inertia map bis}
\boldsymbol{L}_{G,0} = \boldsymbol{I}_G\left(\boldsymbol{\Omega}_{0}\right) = \frac{1}{2}\sum_{i,j = 1}^{3}\,I_{G,ij}\,\Omega_{ij}\ \boldsymbol{\hat{e}}_{i}\wedge\boldsymbol{\hat{e}}_{j}
\end{equation}
In view of the initial inertia map~\eqref{inertia map bis}, the angular momentum bivector of the rigid body~\eqref{relative angular momentum pent} evaluated at the centre of mass $G$ is recast in the principal body frame as,
\begin{equation}\label{relative angular momentum hex}
\boldsymbol{L}_{G} = \frac{1}{2}\sum_{i,j = 1}^{3}\,I_{G,ij}\,\Omega_{ij}\,R\left(\boldsymbol{\hat{e}}_{i}\wedge\boldsymbol{\hat{e}}_{j}\right)R^{\dag}
\end{equation}
Now, we determine the dual of the angular momentum bivector $\boldsymbol{L}_{G}$. In view of the identity~\eqref{commutation B I}, the rotor $R$ and its reverse $R^{\dag}$ which are even multivectors commute with the pseudoscalar $I$, 
\begin{equation}\label{rotor pseudoscalar}
R\,I = I\,R \qquad\text{and}\quad R^{\dag}\,I = I\,R^{\dag}
\end{equation}
Using the duality~\eqref{angular momentum duality} and taking into account the commutation~\eqref{rotor pseudoscalar}, the dual of the angular momentum bivector~\eqref{relative angular momentum pent} is the angular momentum pseudovector,
\begin{equation}\label{relative angular momentum pseudovector}
\begin{split}
&\boldsymbol{\ell}_{G} = \boldsymbol{L}_{G}^{\ast} = \left(R\ \boldsymbol{I}_G\left(\boldsymbol{\Omega}_{0}\right)R^{\dag}\right)^{\ast} = -\,R\ \boldsymbol{I}_G\left(\boldsymbol{\Omega}_{0}\right)R^{\dag}\,I\\
&\phantom{\boldsymbol{\ell}_{G} = \boldsymbol{L}_{G}^{\ast}} = R\left(-\,\boldsymbol{I}_G\left(\boldsymbol{\Omega}_{0}\right)I\right)R^{\dag} = R\ \boldsymbol{I}_G^{\,\ast}\left(\boldsymbol{\Omega}_{0}\right)R^{\dag}
\end{split}
\end{equation}
Using the duality~\eqref{angular velocity duality}, the dual of the initial angular velocity bivector~\eqref{angular velocity basis} is the initial angular velocity pseudovector,
\begin{equation}\label{angular velocity basis dual}
\boldsymbol{\omega}_{0} = \boldsymbol{\Omega}^{\ast}_{0} = \frac{1}{2}\sum_{i,j = 1}^{3}\,\Omega_{ij}\left(\boldsymbol{\hat{e}}_{i}\wedge\boldsymbol{\hat{e}}_{j}\right)^{\ast}
\end{equation}
where,
\begin{equation}\label{angular velocity components dual}
\Omega_{ij} = -\,\Omega_{ji} = \varepsilon_{ijk}\,\omega_{k}
\end{equation}
In view of the dual of the unit bivector~\eqref{dual inertia frame} and the components~\eqref{angular velocity components dual}, the initial angular velocity pseudovector~\eqref{angular velocity basis dual} becomes,
\begin{equation}\label{angular velocity basis dual bis}
\boldsymbol{\omega}_{0} = \frac{1}{2}\sum_{i,j = 1}^{3}\,\varepsilon_{ijk}\,\omega_{k}\,\boldsymbol{\hat{e}}_{k} = \sum_{k = 1}^{3}\,\omega_{k}\,\boldsymbol{\hat{e}}_{k}
\end{equation}
In view of the duality of the bivector~\eqref{inertia duality} and the components~\eqref{angular velocity components dual} and the eigenvector decomposition~\eqref{inertia vector pent}, the dual of the unit bivector~\eqref{dual inertia frame} and the components~\eqref{angular velocity components dual}, the dual of the initial inertia map bivector~\eqref{inertia map} is the initial inertia map pseudovector, which is the initial angular momentum pseudovector evaluated at the centre of mass $G$,
\begin{equation}\label{inertia map dual}
\begin{split}
&\boldsymbol{\ell}_{G,0} = \boldsymbol{i}_G\left(\boldsymbol{\omega}_0\right) = \boldsymbol{I}_G^{\,\ast}\left(\boldsymbol{\Omega}_{0}\right) \\
&\phantom{\boldsymbol{\ell}_{G,0} }= \frac{1}{2}\sum_{i,j = 1}^{3}\,\varepsilon_{ijk}\,\omega_{k}\ \boldsymbol{I}_G^{\,\ast}\left(\boldsymbol{\hat{e}}_{i}\wedge\boldsymbol{\hat{e}}_{j}\right) = \sum_{k = 1}^{3}\,\omega_{k}\,\boldsymbol{i}_G\left(\boldsymbol{\hat{e}}_k\right) = \sum_{k = 1}^{3}\,i_{G,k}\,\omega_{k}\,\boldsymbol{\hat{e}}_k
\end{split}
\end{equation}
In view of the duality between the initial inertia maps~\eqref{inertia map bis}, the angular momentum pseudovector of the rigid body~\eqref{relative angular momentum pseudovector} evaluated at the centre of mass $G$ is recast in the principal body frame as,
\begin{equation}\label{relative angular momentum pseudovector bis}
\boldsymbol{\ell}_{G} = \sum_{k = 1}^{3}\,i_{G,k}\,\omega_{k}\,R\ \boldsymbol{\hat{e}}_{k}\,R^{\dag}
\end{equation}
To get a better picture of the angular momentum, we now introduce the orthonormal frame $\{\boldsymbol{\hat{f}}_1,\boldsymbol{\hat{f}}_2,\boldsymbol{\hat{f}}_3\}$, which is defined as the principal body frame and related to the initial principal body frame $\{\boldsymbol{\hat{e}}_1,\boldsymbol{\hat{e}}_2,\boldsymbol{\hat{e}}_3\}$ through the rotation of the rigid body described by the rotor $R$. The unit vectors of the orthonormal frames are related by,
\begin{equation}\label{orthonormal new frame vector}
\boldsymbol{\hat{f}}_i = R\ \boldsymbol{\hat{e}}_{i}\ R^{\dag}
\end{equation}
In view of the orthonormality condition~\eqref{unit rotor time dependent} and the rotation~\eqref{orthonormal new frame vector}, the unit bivectors of the orthonormal frames are related by,
\begin{equation}\label{orthonormal new frame bivector}
\boldsymbol{\hat{f}}_i\wedge\boldsymbol{\hat{f}}_j = \boldsymbol{\hat{f}}_i\,\boldsymbol{\hat{f}}_j = \left(R\ \boldsymbol{\hat{e}}_{i}\ R^{\dag}\right)\left(R\ \boldsymbol{\hat{e}}_{j}\ R^{\dag}\right) = R\left(\boldsymbol{\hat{e}}_{i}\,\boldsymbol{\hat{e}}_{j}\right)R^{\dag} = R\left(\boldsymbol{\hat{e}}_{i}\wedge\boldsymbol{\hat{e}}_{j}\right)R^{\dag}
\end{equation}
Using the rotation~\eqref{orthonormal new frame vector}, the angular momentum bivector~\eqref{relative angular momentum hex} becomes,
\begin{equation}\label{relative angular momentum hep}
\boldsymbol{L}_{G} = \frac{1}{2}\sum_{i,j = 1}^{3}\,I_{G,ij}\,\Omega_{ij}\,\boldsymbol{\hat{f}}_{i}\wedge\boldsymbol{\hat{f}}_{j}
\end{equation}
Similarly, using the rotation~\eqref{orthonormal new frame bivector}, the angular momentum pseudovector~\eqref{relative angular momentum pseudovector bis} becomes,
\begin{equation}\label{relative angular momentum pseudovector ter}
\boldsymbol{\ell}_{G} = \sum_{k = 1}^{3}\,i_{G,k}\,\omega_{k}\,\boldsymbol{\hat{f}}_{k}
\end{equation}
%


\section{Kinetic energy of a rigid body}
\label{Kinetic energy of a rigid body}

\noindent The kinetic energy of a rigid body reads,
\begin{equation}\label{kinetic energy}
\mathcal{T} = \frac{1}{2}\int_V\,dV\,\rho\left(\boldsymbol{r}_0^{\prime}\right)\,\boldsymbol{v}^2
\end{equation}
where the mass density $\rho\left(\boldsymbol{r}_0^{\prime}\right)$ is a function of the initial relative position $\boldsymbol{r}_0^{\prime}$ and
the velocity $\boldsymbol{v}$ is the sum of the velocity of the centre of mass $\boldsymbol{v}_G$ and the relative velocity $\boldsymbol{v}^{\prime}$
\begin{equation}\label{rigid body velocity}
\boldsymbol{v} = \boldsymbol{v}_G + \boldsymbol{v}^{\prime}
\end{equation}
According to relation~\eqref{relative velocity alpha quad}, the relative velocity is the continuum limit is given by,
\begin{equation}\label{relative velocity}
\boldsymbol{v}^{\prime} = \boldsymbol{r}^{\prime}\cdot\boldsymbol{\Omega}
\end{equation}
In view of the velocity decomposition~\eqref{rigid body velocity} and the relative velocity~\eqref{relative velocity}, the kinetic energy~\eqref{kinetic energy} is recast as,
\begin{equation}\label{kinetic energy bis}
\mathcal{T} = \frac{1}{2}\int_V\,dV\,\rho\left(\boldsymbol{r}_0^{\prime}\right)\Big(\boldsymbol{v}_G^2 + 2\,\boldsymbol{v}_G\cdot\boldsymbol{v}^{\prime} + \left(\boldsymbol{r}^{\prime}\cdot\boldsymbol{\Omega}\right)^2\Big)
\end{equation}
Using the mass~\eqref{mass}, the kinetic energy~\eqref{kinetic energy bis} becomes,
\begin{equation}\label{kinetic energy ter}
\mathcal{T} = \frac{1}{2}\,m\,\boldsymbol{v}_G^2 + \boldsymbol{v}_G\cdot\int_V\,dV\,\rho\left(\boldsymbol{r}_0^{\prime}\right)\boldsymbol{v}^{\prime}
+ \frac{1}{2}\int_V\,dV\,\rho\left(\boldsymbol{r}_0^{\prime}\right)\left(\boldsymbol{r}^{\prime}\cdot\boldsymbol{\Omega}\right)^2
\end{equation}
By definition of the centre of mass, the relative momentum of the rigid body vanishes,
\begin{equation}\label{relative momentum}
\boldsymbol{p}^{\prime} = \int_V\,dV\,\rho\left(\boldsymbol{r}_0^{\prime}\right)\boldsymbol{v}^{\prime} = 0
\end{equation}
which implies that the kinetic energy~\eqref{kinetic energy ter} reduces to,
\begin{equation}\label{kinetic energy quad}
\mathcal{T} = \frac{1}{2}\,m\,\boldsymbol{v}_G^2
+ \frac{1}{2}\int_V\,dV\,\rho\left(\boldsymbol{r}_0^{\prime}\right)\left(\boldsymbol{r}^{\prime}\cdot\boldsymbol{\Omega}\right)^2
\end{equation}
In view of the dualities~\eqref{vector duality bis},~\eqref{bivectorial duality} and~\eqref{bivectorial duality bi}, we obtain the identity,
\begin{equation}\label{energy geometric identity}
\begin{split}
&\left(\left(\boldsymbol{r}^{\prime}\cdot\boldsymbol{\Omega}\right)^{2}\right)^{\ast} =  \Big(\left(\boldsymbol{r}^{\prime}\cdot\boldsymbol{\Omega}\right)\cdot\left(\boldsymbol{r}^{\prime}\cdot\boldsymbol{\Omega}\right)\Big)^{\ast} = \left(\boldsymbol{r}^{\prime}\cdot\boldsymbol{\Omega}\right)\wedge\left(\boldsymbol{r}^{\prime}\cdot\boldsymbol{\Omega}\right)^{\ast}\\
&\phantom{\left(\left(\boldsymbol{r}^{\prime}\cdot\boldsymbol{\Omega}\right)^{2}\right)^{\ast}\,} = \left(\boldsymbol{r}^{\prime}\cdot\boldsymbol{\Omega}\right)\wedge\boldsymbol{r}^{\prime}\wedge\boldsymbol{\Omega}^{\ast} = -\,\boldsymbol{r}^{\prime}\wedge\left(\boldsymbol{r}^{\prime}\cdot\boldsymbol{\Omega}\right)\wedge\boldsymbol{\Omega}^{\ast}\\
&\phantom{\left(\left(\boldsymbol{r}^{\prime}\cdot\boldsymbol{\Omega}\right)^{2}\right)^{\ast}\,} = -\,\Big(\!\left(\boldsymbol{r}^{\prime}\wedge\left(\boldsymbol{r}^{\prime}\cdot\boldsymbol{\Omega}\right)\right)\cdot\boldsymbol{\Omega}\,\Big)^{\ast}
\end{split}
\end{equation}
The dual of the identity~\eqref{energy geometric identity} reads,
\begin{equation}\label{energy geometric identity dual}
\left(\boldsymbol{r}^{\prime}\cdot\boldsymbol{\Omega}\right)^{2} = -\,\Big(\boldsymbol{r}^{\prime}\wedge\left(\boldsymbol{r}^{\prime}\cdot\boldsymbol{\Omega}\right)\!\Big)\cdot\boldsymbol{\Omega}
\end{equation}
Using the identity~\eqref{energy geometric identity dual}, the kinetic energy~\eqref{kinetic energy quad} becomes,
\begin{equation}\label{kinetic energy pent}
\mathcal{T} = \frac{1}{2}\,m\,\boldsymbol{v}_G^2
-\,\frac{1}{2}\int_V\,dV\,\rho\left(\boldsymbol{r}_0^{\prime}\right)\Big(\boldsymbol{r}^{\prime}\wedge\left(\boldsymbol{r}^{\prime}\cdot\boldsymbol{\Omega}\right)\!\Big)\cdot\boldsymbol{\Omega}
\end{equation}
In the continuum limit, the angular momentum bivector~\eqref{relative angular momentum} evaluated at the centre of mass $G$ is recast as,
\begin{equation}\label{relative angular momentum bis}
\boldsymbol{L}_{G} = \int_V\,dV\,\rho\left(\boldsymbol{r}_0^{\prime}\right)\Big(\boldsymbol{r}^{\prime}\wedge\left(\boldsymbol{r}^{\prime}\cdot\boldsymbol{\Omega}\right)\!\Big)
\end{equation}
In view of the total momentum of the rigid body,
\begin{equation}\label{momentum centre of mass}
\boldsymbol{p} = m\,\boldsymbol{v}_{G}
\end{equation}
and the angular momentum~\eqref{relative angular momentum bis}, the kinetic energy~\eqref{kinetic energy pent} reduces to,
\begin{equation}\label{kinetic energy hex}
\mathcal{T} = \frac{1}{2}\ \boldsymbol{p}\cdot\boldsymbol{v}_G
-\,\frac{1}{2}\ \boldsymbol{L}_G\cdot\boldsymbol{\Omega}
\end{equation}
In view of the rotation~\eqref{intial angular velocity} of the initial angular velocity~\eqref{angular velocity basis} and the basis vectors~\eqref{orthonormal new frame vector} of the principal body frame, the angular velocity is written in this frame as,
\begin{equation}\label{angular velocity principal body frame}
\boldsymbol{\Omega} = R\ \boldsymbol{\Omega}_0\,R^{\dag} = \frac{1}{2}\sum_{i,j = 1}^{3}\,\Omega_{ij}\,R\left(\boldsymbol{\hat{e}}_{i}\wedge\boldsymbol{\hat{e}}_{j}\right)R^{\dag} = \frac{1}{2}\sum_{i,j = 1}^{3}\,\Omega_{ij}\,\boldsymbol{\hat{f}}_{i}\wedge\boldsymbol{\hat{f}}_{j}
\end{equation}
The scalar product of the angular momentum and the angular velocity~\eqref{angular velocity principal body frame} is written in the principal body frame as,
\begin{equation}\label{angular momentum velocity dot product}
\boldsymbol{L}_G\cdot\boldsymbol{\Omega} = \frac{1}{4}\sum_{i,j,k,\ell = 1}^{3}\,I_{G,ij}\,\Omega_{ij}\,\Omega_{k\ell}\left(\boldsymbol{\hat{f}}_{i}\wedge\boldsymbol{\hat{f}}_{j}\right)\cdot\left(\boldsymbol{\hat{f}}_{k}\wedge\boldsymbol{\hat{f}}_{\ell}\right)
\end{equation}
where the unit vectors satisfy the scalar identity,
\begin{equation}\label{scalar unit vectors}
\begin{split}
&\left(\boldsymbol{\hat{f}}_{i}\wedge\boldsymbol{\hat{f}}_{j}\right)\cdot\left(\boldsymbol{\hat{f}}_{k}\wedge\boldsymbol{\hat{f}}_{\ell}\right) = \frac{1}{4}\left(\boldsymbol{\hat{f}}_{i}\,\boldsymbol{\hat{f}}_{j}-\,\boldsymbol{\hat{f}}_{j}\,\boldsymbol{\hat{f}}_{i}\right)\cdot\left(\boldsymbol{\hat{f}}_{k}\,\boldsymbol{\hat{f}}_{\ell}-\,\boldsymbol{\hat{f}}_{\ell}\,\boldsymbol{\hat{f}}_{k}\right)\\
&= \delta_{jk}\,\delta_{i\ell} -\,\delta_{ik}\,\delta_{j\ell}
\end{split}
\end{equation}
Using the scalar identity~\eqref{scalar unit vectors}, the scalar product~\eqref{angular momentum velocity dot product bis} reduces to,
\begin{equation}\label{angular momentum velocity dot product bis}
\boldsymbol{L}_G\cdot\boldsymbol{\Omega} = -\,\frac{1}{2}\sum_{i,j = 1}^{3}\,I_{G,ij}\,\Omega_{ij}^2
\end{equation}
In view of the identities~\eqref{momentum centre of mass} and~\eqref{angular momentum velocity dot product bis}, the kinetic energy is recast as,
\begin{equation}\label{kinetic energy hep}
\mathcal{T} = \frac{1}{2}\ m\,\boldsymbol{v}_G^2
+ \frac{1}{4}\sum_{i,j = 1}^{3}\,I_{G,ij}\,\Omega_{ij}^2
\end{equation}
In view of the dualities~\eqref{moment of inertia} and~\eqref{angular velocity components dual} and the property $\varepsilon_{ijk}^2 = 1$ of the Levi-Civita symbol~\eqref{Levi Civita}, the kinetic energy~\eqref{kinetic energy hep} is recast as,
\begin{equation}\label{kinetic energy oct}
\mathcal{T} = \frac{1}{2}\ m\,\boldsymbol{v}_G^2
+ \frac{1}{4}\sum_{i,j,k = 1}^{3}\varepsilon_{ijk}\,i_{G,k}\,\omega_{k}^2
\end{equation}
The antisymmetry of the Levi-Civita symbol~\eqref{Levi Civita} implies that the sum over the different indices $i$ and $j$ yields a factor $2$. Thus, the kinetic energy~\eqref{kinetic energy oct} reduces to,
\begin{equation}\label{kinetic energy nov}
\mathcal{T} = \frac{1}{2}\ m\,\boldsymbol{v}_G^2
+ \frac{1}{2}\sum_{k = 1}^{3}\,i_{G,k}\,\omega_{k}^2
\end{equation}
as expected in vector algebra.


\section{Euler equations}
\label{Euler equations section}

\noindent The rotational dynamics of a rigid body is described by the Euler equations. In order to establish this equation in geometric algebra, we begin by considering the angular momentum theorem~\eqref{angular momentum theorem} evaluated at the centre of mass $G$,
\begin{equation}\label{angular momentum theorem G}
\sum\,\boldsymbol{T}_G^{\,\text{ext}} = \boldsymbol{\dot{L}}_G
\end{equation}
Since the moments of inertia $I_{G,ij}$ are constant for a rigid body, the time derivative of the angular momentum~\eqref{relative angular momentum hex} is given by,
\begin{equation}\label{relative angular momentum time derivative}
\begin{split}
&\boldsymbol{\dot{L}}_{G} = \frac{1}{2}\sum_{i,j = 1}^{3}\,I_{G,ij}\,\dot{\Omega}_{ij}\,R\left(\boldsymbol{\hat{e}}_{i}\wedge\boldsymbol{\hat{e}}_{j}\right)R^{\dag}\\
&\phantom{\boldsymbol{\dot{L}}_{G} =} + \frac{1}{2}\sum_{i,j = 1}^{3}\,I_{G,ij}\,\Omega_{ij}\left(\dot{R}\left(\boldsymbol{\hat{e}}_{i}\wedge\boldsymbol{\hat{e}}_{j}\right)R^{\dag} + R\left(\boldsymbol{\hat{e}}_{i}\wedge\boldsymbol{\hat{e}}_{j}\right)\dot{R}^{\dag}\right)
\end{split}
\end{equation}
In view of the dynamical equations for the rotor~\eqref{derivative identity ter} and its reverse~\eqref{derivative identity quad}, the time derivative of the angular momentum~\eqref{relative angular momentum time derivative} becomes,
\begin{equation}\label{relative angular momentum time derivative bis}
\begin{split}
&\boldsymbol{\dot{L}}_{G} = \frac{1}{2}\sum_{i,j = 1}^{3}\,I_{G,ij}\,\dot{\Omega}_{ij}\,R\left(\boldsymbol{\hat{e}}_{i}\wedge\boldsymbol{\hat{e}}_{j}\right)R^{\dag}\\
&\phantom{\boldsymbol{\dot{L}}_{G} =} + \frac{1}{4}\sum_{i,j = 1}^{3}\,I_{G,ij}\,\Omega_{ij}\Big(R\left(\boldsymbol{\hat{e}}_{i}\wedge\boldsymbol{\hat{e}}_{j}\right)R^{\dag}\,\boldsymbol{\Omega} -\, \boldsymbol{\Omega}\,R\left(\boldsymbol{\hat{e}}_{i}\wedge\boldsymbol{\hat{e}}_{j}\right)R^{\dag}\Big)
\end{split}
\end{equation}
In view of the rotation~\eqref{orthonormal new frame bivector} and the angular momentum~\eqref{relative angular momentum hex}, the time derivative of the angular momentum~\eqref{relative angular momentum time derivative bis} is recast as,
\begin{equation}\label{relative angular momentum time derivative ter}
\boldsymbol{\dot{L}}_{G} = \frac{1}{2}\sum_{i,j = 1}^{3}\,I_{G,ij}\,\dot{\Omega}_{ij}\,\boldsymbol{\hat{f}}_{i}\wedge\boldsymbol{\hat{f}}_{j} + \frac{1}{2}\left(\boldsymbol{L}_{G}\,\boldsymbol{\Omega} -\,\boldsymbol{\Omega}\,\boldsymbol{L}_{G}\right)
\end{equation}
In view of the commutator~\eqref{geometric product bivectors outer} of the bivectors $\boldsymbol{L}_{G}$ and $\boldsymbol{\Omega}$, the time derivative of the angular momentum~\eqref{relative angular momentum time derivative ter} is recast as,
\begin{equation}\label{relative angular momentum time derivative quad}
\boldsymbol{\dot{L}}_{G} = \frac{1}{2}\sum_{i,j = 1}^{3}\,I_{G,ij}\,\dot{\Omega}_{ij}\,\boldsymbol{\hat{f}}_{i}\wedge\boldsymbol{\hat{f}}_{j} + \boldsymbol{L}_{G}\times\boldsymbol{\Omega}
\end{equation}
In view of the initial angular velocity~\eqref{angular velocity basis} and the rotation~\eqref{orthonormal new frame bivector}, the angular velocity pseudovector is written as,
\begin{equation}\label{angular velocity pseudovector}
\boldsymbol{\Omega} = R\ \boldsymbol{\Omega}_0\,R^{\dag} = \frac{1}{2}\,\sum_{k,\ell = 1}^{3}\,\Omega_{k,\ell}\,R\left(\boldsymbol{\hat{e}}_i\wedge\boldsymbol{\hat{e}}_j\right)R^{\dag} = \frac{1}{2}\,\sum_{k,\ell = 1}^{3}\,\Omega_{k,\ell}\,\boldsymbol{\hat{f}}_k\wedge\boldsymbol{\hat{f}}_\ell
\end{equation}
The commutator of the angular momentum bivector~\eqref{relative angular momentum hep} and the angular velocity bivector~\eqref{angular velocity pseudovector} is written in components as, 
\begin{equation}\label{angular momentum velocity commutator}
\begin{split}
&\boldsymbol{L}_{G}\times\boldsymbol{\Omega} = \frac{1}{2}\left(\boldsymbol{L}_{G}\,\boldsymbol{\Omega} -\,\boldsymbol{\Omega}\,\boldsymbol{L}_{G}\right)\\
&\phantom{\boldsymbol{\Omega}\times\boldsymbol{L}_{G}} = \frac{1}{8}\sum_{i,j,k,\ell = 1}^{3}\left(I_{G,ij}-\,I_{G,k\ell}\right)\Omega_{ij}\,\Omega_{k\ell}\left(\boldsymbol{\hat{f}}_{i}\wedge\boldsymbol{\hat{f}}_{j}\right)\left(\boldsymbol{\hat{f}}_{k}\wedge\boldsymbol{\hat{f}}_{\ell}\right)
\end{split}
\end{equation}
where the unit vectors satisfy the bivectorial identity,
\begin{equation}\label{bivector commutator of unit vectors}
\begin{split}
&\left(\boldsymbol{\hat{f}}_{i}\wedge\boldsymbol{\hat{f}}_{j}\right)\left(\boldsymbol{\hat{f}}_{k}\wedge\boldsymbol{\hat{f}}_{\ell}\right) = \frac{1}{4}\left(\boldsymbol{\hat{f}}_{i}\,\boldsymbol{\hat{f}}_{j}-\,\boldsymbol{\hat{f}}_{j}\,\boldsymbol{\hat{f}}_{i}\right)\left(\boldsymbol{\hat{f}}_{k}\,\boldsymbol{\hat{f}}_{\ell}-\,\boldsymbol{\hat{f}}_{\ell}\,\boldsymbol{\hat{f}}_{k}\right)\\
&= \delta_{jk}\,\boldsymbol{\hat{f}}_{i}\wedge\boldsymbol{\hat{f}}_{\ell} -\,\delta_{ik}\,\boldsymbol{\hat{f}}_{j}\wedge\boldsymbol{\hat{f}}_{\ell} + \delta_{i\ell}\,\boldsymbol{\hat{f}}_{j}\wedge\boldsymbol{\hat{f}}_{k} -\,\delta_{j\ell}\,\boldsymbol{\hat{f}}_{i}\wedge\boldsymbol{\hat{f}}_{k}\\
&= \delta_{jk}\,\boldsymbol{\hat{f}}_{i}\wedge\boldsymbol{\hat{f}}_{\ell} + \delta_{ik}\,\boldsymbol{\hat{f}}_{\ell}\wedge\boldsymbol{\hat{f}}_{j} -\,\delta_{j\ell}\,\boldsymbol{\hat{f}}_{i}\wedge\boldsymbol{\hat{f}}_{k} -\,\delta_{i\ell}\,\boldsymbol{\hat{f}}_{k}\wedge\boldsymbol{\hat{f}}_{j}
\end{split}
\end{equation}
Thus, the commutator~\eqref{angular momentum velocity commutator} becomes,
\begin{equation}\label{angular momentum velocity commutator bis}
\begin{split}
&\boldsymbol{L}_{G}\times\boldsymbol{\Omega} = \frac{1}{8}\sum_{i,j,k = 1}^{3}\left(I_{G,ij}-\,I_{G,jk}\right)\Omega_{ij}\,\Omega_{jk}\ \boldsymbol{\hat{f}}_{i}\wedge\boldsymbol{\hat{f}}_{k}\\
&\phantom{\boldsymbol{L}_{G}\times\boldsymbol{\Omega} =} + \frac{1}{8}\sum_{i,j,k = 1}^{3}\left(I_{G,ij}-\,I_{G,ik}\right)\Omega_{ij}\,\Omega_{ik}\ \boldsymbol{\hat{f}}_{k}\wedge\boldsymbol{\hat{f}}_{j}\\
&\phantom{\boldsymbol{L}_{G}\times\boldsymbol{\Omega} =} -\,\frac{1}{8}\sum_{i,j,k = 1}^{3}\left(I_{G,ij}-\,I_{G,kj}\right)\Omega_{ij}\,\Omega_{kj}\ \boldsymbol{\hat{f}}_{i}\wedge\boldsymbol{\hat{f}}_{k}\\
&\phantom{\boldsymbol{L}_{G}\times\boldsymbol{\Omega} =} -\,\frac{1}{8}\sum_{i,j,k = 1}^{3}\left(I_{G,ij}-\,I_{G,ki}\right)\Omega_{ij}\,\Omega_{ki}\ \boldsymbol{\hat{f}}_{k}\wedge\boldsymbol{\hat{f}}_{j}
\end{split}
\end{equation}
where the indices $\ell$ and $k$ were exchanged in the first two terms on the right hand side. In view of the antisymmetric components of the angular velocity bivector, i.e. $\Omega_{jk} = -\,\Omega_{kj}$ and $\Omega_{ik} = -\,\Omega_{ki}$, the commutator~\eqref{angular momentum velocity commutator bis} reduces to,
\begin{equation}\label{angular momentum velocity commutator ter}
\begin{split}
&\boldsymbol{L}_{G}\times\boldsymbol{\Omega} = \frac{1}{4}\sum_{i,j,k = 1}^{3}\left(I_{G,ij}-\,I_{G,jk}\right)\Omega_{ij}\,\Omega_{jk}\ \boldsymbol{\hat{f}}_{i}\wedge\boldsymbol{\hat{f}}_{k}\\
&\phantom{\boldsymbol{L}_{G}\times\boldsymbol{\Omega} =} + \frac{1}{4}\sum_{i,j,k = 1}^{3}\left(I_{G,ij}-\,I_{G,ik}\right)\Omega_{ij}\,\Omega_{ik}\ \boldsymbol{\hat{f}}_{k}\wedge\boldsymbol{\hat{f}}_{j}
\end{split}
\end{equation}
and is recast as,
\begin{equation}\label{angular momentum velocity commutator ter}
\begin{split}
&\boldsymbol{L}_{G}\times\boldsymbol{\Omega} = \frac{1}{4}\sum_{i,j,k = 1}^{3}\left(I_{G,ik}-\,I_{G,kj}\right)\Omega_{ik}\,\Omega_{kj}\ \boldsymbol{\hat{f}}_{i}\wedge\boldsymbol{\hat{f}}_{j}\\
&\phantom{\boldsymbol{L}_{G}\times\boldsymbol{\Omega} =} + \frac{1}{4}\sum_{i,j,k = 1}^{3}\left(I_{G,kj}-\,I_{G,ki}\right)\Omega_{kj}\,\Omega_{ki}\ \boldsymbol{\hat{f}}_{i}\wedge\boldsymbol{\hat{f}}_{j}
\end{split}
\end{equation}
where the indices $j$ and $k$ were exchanged in the first term and the indices $i$ and $k$ were exchanged in the first term on the right hand side. In view of the antisymmetric components of the angular velocity bivector~\eqref{angular velocity components dual}, and the antisymmetric moments of inertia~\eqref{moment of inertia}, the commutator~\eqref{angular momentum velocity commutator bis} reduces to,
\begin{equation}\label{angular momentum velocity commutator quad}
\boldsymbol{L}_{G}\times\boldsymbol{\Omega} = \frac{1}{2}\sum_{i,j = 1}^{3}\left(\,\sum_{k = 1}^{3}\left(I_{G,ik}-\,I_{G,kj}\right)\Omega_{ik}\,\Omega_{kj}\right)\boldsymbol{\hat{f}}_{i}\wedge\boldsymbol{\hat{f}}_{j}
\end{equation}
In view of the commutator~\eqref{angular momentum velocity commutator quad}, the time derivative of the angular momentum~\eqref{relative angular momentum time derivative quad} is recast in components as,
\begin{equation}\label{relative angular momentum time derivative pent}
\boldsymbol{\dot{L}}_{G} = \frac{1}{2}\sum_{i,j = 1}^{3}\left(I_{G,ij}\,\dot{\Omega}_{ij} + \sum_{k = 1}^{3}\left(I_{G,ik}-\,I_{G,kj}\right)\Omega_{ik}\,\Omega_{kj}\right)\boldsymbol{\hat{f}}_{i}\wedge\boldsymbol{\hat{f}}_{j}
\end{equation}
The net external torque bivector $\boldsymbol{T}_G^{\,\text{ext}}$ is written in components in the principal body frame as,
\begin{equation}\label{external torque bivector}
\sum\,\boldsymbol{T}_G^{\,\text{ext}} = \sum\,\frac{1}{2}\sum_{i,j = 1}^{3}T_{G,ij}^{\,\text{ext}}\ \boldsymbol{\hat{f}}_{i}\wedge\boldsymbol{\hat{f}}_{j}
\end{equation}
According to the angular momentum theorem~\eqref{angular momentum theorem G}, the components of the bivectors~\eqref{relative angular momentum time derivative pent} and~\eqref{external torque bivector} are equal,
\begin{equation}\label{Euler equation 0}
\sum\,T_{G,ij}^{\,\text{ext}} = I_{G,ij}\,\dot{\Omega}_{ij} + \sum_{k = 1}^{3}\left(I_{G,ik}-\,I_{G,kj}\right)\Omega_{ik}\,\Omega_{kj}
\end{equation}
For a given unit bivector $\boldsymbol{\hat{f}}_{i}\wedge\boldsymbol{\hat{f}}_{j}$, where $i \neq j$, the index $k$ is the remaining index and no summation over $k$ is needed in equation~\eqref{Euler equation 0}. This equation for the bivector $\boldsymbol{\hat{f}}_{i}\wedge\boldsymbol{\hat{f}}_{j}$ describes the rotation of the rigid body in the oriented plane spanned by this bivector,
\begin{equation}\label{Euler equation}
\sum\,T_{G,ij}^{\,\text{ext}} = I_{G,ij}\,\dot{\Omega}_{ij} + \left(I_{G,ik}-\,I_{G,kj}\right)\Omega_{ik}\,\Omega_{kj}
\end{equation}
Using the antisymmetric components of the angular velocity bivector~\eqref{angular velocity components dual}, and the antisymmetric moments of inertia~\eqref{moment of inertia}, equation~\eqref{Euler equation} evaluated for the couples of indices $ij \in \{12,23,31\}$ yields the Euler equations in geometric algebra,
\begin{equation}\label{Euler equations}
\begin{split}
&\sum\,T_{G,12}^{\,\text{ext}} = I_{G,12}\,\dot{\Omega}_{12} + \left(I_{G,23}-\,I_{G,31}\right)\Omega_{23}\,\Omega_{31}\\
&\sum\,T_{G,23}^{\,\text{ext}} = I_{G,23}\,\dot{\Omega}_{23} + \left(I_{G,31}-\,I_{G,12}\right)\Omega_{31}\,\Omega_{12}\\
&\sum\,T_{G,31}^{\,\text{ext}} = I_{G,31}\,\dot{\Omega}_{31} + \left(I_{G,12}-\,I_{G,23}\right)\Omega_{12}\,\Omega_{23}
\end{split}
\end{equation}
In order to deduce from the Euler equations in geometric algebra the Euler equations in vector algebra, we use the duality of the components of the external torques,
\begin{equation}\label{external torque duality}
T_{G,ij} = -\,T_{G,ji} = \varepsilon_{ijk}\,\tau_{G,k}
\end{equation}
the duality of the components of the angular acceleration,
\begin{equation}\label{angular acceleration duality}
\dot{\Omega}_{ij} = -\,\dot{\Omega}_{ji} = \varepsilon_{ijk}\,\dot{\omega}_{k}
\end{equation}
and the duality of the components of the moments of inertia~\eqref{moment of inertia}. Thus, the Euler equations~\eqref{Euler equations} are recast as,
\begin{equation}\label{Euler equations bis}
\begin{split}
&\sum\,\varepsilon_{123}\,\tau_{G,3}^{\,\text{ext}} = \varepsilon^2_{123}\,i_{G,3}\,\dot{\omega}_{3} + \varepsilon^2_{231}\,\varepsilon^2_{312}\left(i_{G,1}-\,i_{G,2}\right)\omega_{1}\,\omega_{2}\\
&\sum\,\varepsilon_{231}\,\tau_{G,1}^{\,\text{ext}} = \varepsilon^2_{231}\,i_{G,1}\,\dot{\omega}_{1} + \varepsilon^2_{312}\,\varepsilon^2_{123}\left(i_{G,2}-\,i_{G,3}\right)\omega_{2}\,\omega_{3}\\
&\sum\,\varepsilon_{312}\,\tau_{G,2}^{\,\text{ext}} = \varepsilon^2_{312}\,i_{G,2}\,\dot{\omega}_{2} + \varepsilon^2_{123}\,\varepsilon^2_{231}\left(i_{G,3}-\,i_{G,1}\right)\omega_{3}\,\omega_{1}
\end{split}
\end{equation}
Taking into account the invariance of the Levi-Civita symbols~\eqref{Levi Civita} under cyclic permutation,
\begin{equation}\label{Levi Civita cyclic}
\varepsilon_{123} = \varepsilon_{231} = \varepsilon_{312}
\end{equation}
the Euler equations in vector algebra~\eqref{Euler equations bis} reduce to,
\begin{equation}\label{Euler equations ter}
\begin{split}
&\sum\,\tau_{G,1}^{\,\text{ext}} = i_{G,1}\,\dot{\omega}_{1} + \left(i_{G,2}-\,i_{G,3}\right)\omega_{2}\,\omega_{3}\\
&\sum\,\tau_{G,2}^{\,\text{ext}} = i_{G,2}\,\dot{\omega}_{2} + \left(i_{G,3}-\,i_{G,1}\right)\omega_{3}\,\omega_{1}\\
&\sum\,\tau_{G,3}^{\,\text{ext}} = i_{G,3}\,\dot{\omega}_{3} + \left(i_{G,1}-\,i_{G,2}\right)\omega_{1}\,\omega_{2}
\end{split}
\end{equation}
%


\section{Symmetric spinning disk}
\label{Symmetric spinning disk}

\begin{figure}[!ht]
\begin{center}
\includegraphics[scale=0.50]{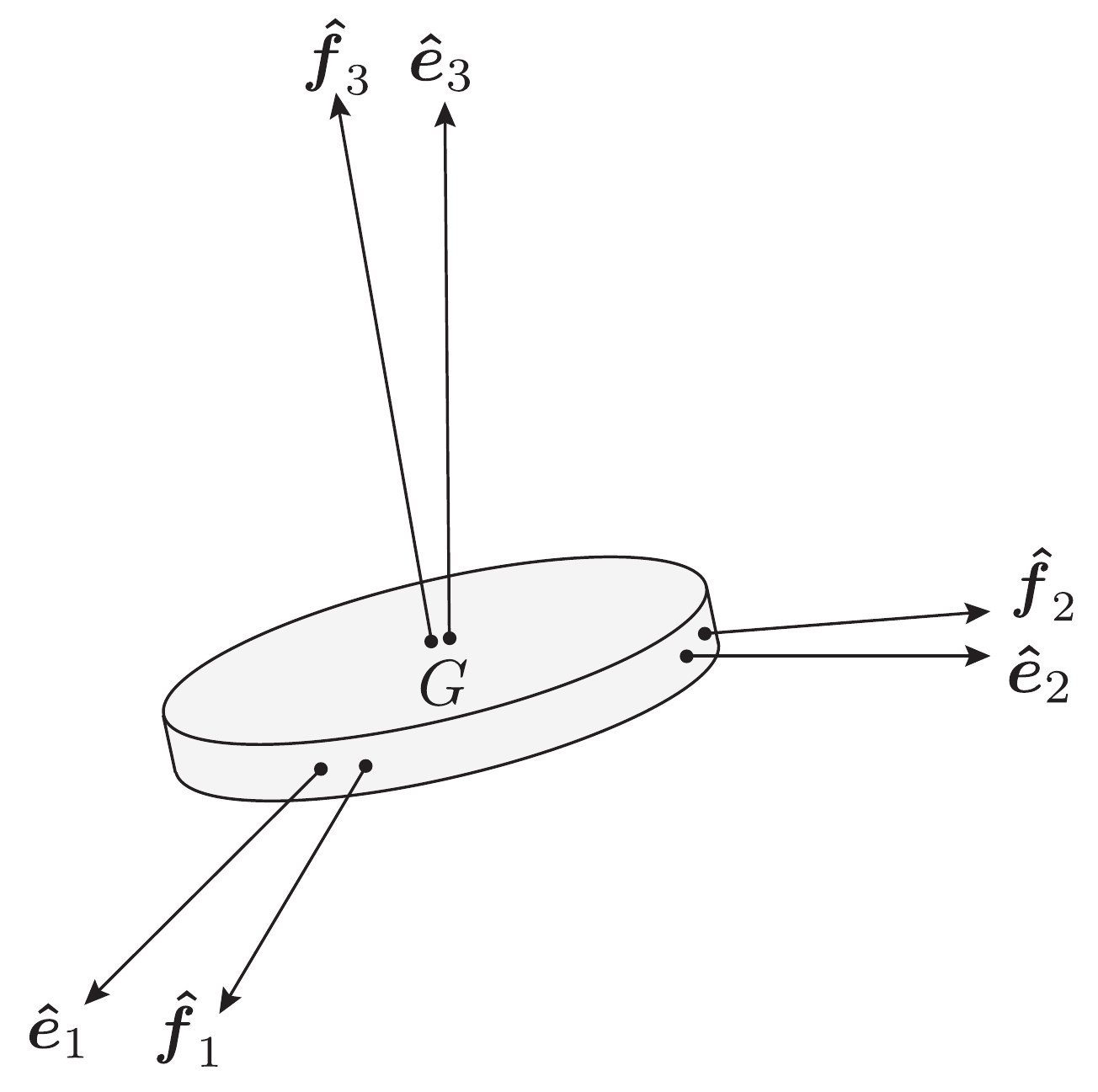}
\end{center}
\caption{Symmetric disk spinning around its centre of mass $G$. The principal body frame $\{\boldsymbol{\hat{f}}_1,\boldsymbol{\hat{f}}_2,\boldsymbol{\hat{f}}_3\}$ is oriented such that the unit vector $\boldsymbol{\hat{f}}_3$ is along the symmetry axis of the disk. The orthonormal frame $\{\boldsymbol{\hat{e}}_1,\boldsymbol{\hat{e}}_2,\boldsymbol{\hat{e}}_3\}$ is fixed.}\label{Fig: Disk}
\end{figure}

\noindent We now consider the torque free motion of a symmetric spinning disk (Fig.~\ref{Fig: Disk}). The disk is a gyroscope that can freely rotate around its center of mass $G$. The principal body frame $\{\boldsymbol{\hat{f}}_1,\boldsymbol{\hat{f}}_2,\boldsymbol{\hat{f}}_3\}$ is oriented such that the unit vector $\boldsymbol{\hat{f}}_3$ is along the symmetry axis of the disk. It coincides initially with the fixed orthonormal frame $\{\boldsymbol{\hat{e}}_1,\boldsymbol{\hat{e}}_2,\boldsymbol{\hat{e}}_3\}$. The plane spanned by the unit bivector $\boldsymbol{\hat{f}}_1\wedge\boldsymbol{\hat{f}}_2$ that is orthogonal to the unit vector $\boldsymbol{\hat{f}}_3$ is the plane of symmetry of the disk. By symmetry, the moments of inertia in the orthogonal planes spanned by the unit bivectors $\boldsymbol{\hat{f}}_2\wedge\boldsymbol{\hat{f}}_3$ and $\boldsymbol{\hat{f}}_3\wedge\boldsymbol{\hat{f}}_1$ containing the unit vector $\boldsymbol{\hat{f}}_3$ are equal,
\begin{equation}\label{symmetry moments of inertia}
I_{G,\parallel} \equiv I_{G,12} \qquad\text{and}\qquad I_{G,\perp} \equiv I_{G,23} = I_{G,31}
\end{equation}
where the indices $\parallel$ and $\perp$ indicate the moment of inertia is defined with respect to a principal axis plane that is either parallel or perpendicular to the plane of symmetry. There is no net external torque bivector evaluated at the center of mass of the freely rotating disk,
\begin{equation}\label{no torque}
\sum\,\boldsymbol{T}_G^{\,\text{ext}} = \boldsymbol{0}
\end{equation}
We now identify the constant bivectors in order to find the rotor that describes entirely the rotator of the symmetric disk. Firstly, in view of the angular momentum theorem~\eqref{angular momentum theorem G}, the angular momentum bivector is constant,
\begin{equation}\label{time deriv angular momentum zero}
\boldsymbol{L}_G = \textbf{const}
\end{equation}
Secondly, in the absence of an external torque~\eqref{no torque}, using the symmetry~\eqref{symmetry moments of inertia}, the projection of the Euler equations~\eqref{Euler equations} in the plane of symmetry reads,
\begin{equation}\label{Euler equations plane of symmetry}
I_{G,\parallel}\,\dot{\Omega}_{12} = 0
\end{equation}
which means that the velocity bivector $\boldsymbol{\Omega}_{0,\parallel}$ that is initially in the plane of symmetry is constant,
\begin{equation}\label{velocity component plane of symmetry}
\boldsymbol{\Omega}_{0,\parallel} \equiv \Omega_{12}\,\boldsymbol{\hat{e}}_1\wedge\boldsymbol{\hat{e}}_2 = \textbf{const}
\end{equation}
In of view of relations~\eqref{relative angular momentum hep} and~\eqref{symmetry moments of inertia}, the angular momentum bivetcor is given by, 
\begin{equation}\label{angular momentum disk}
\boldsymbol{L}_{G} = I_{G,\parallel}\,\Omega_{12}\,\boldsymbol{\hat{f}}_{1}\wedge\boldsymbol{\hat{f}}_{2} + I_{G,\perp}\,\Omega_{23}\,\boldsymbol{\hat{f}}_{2}\wedge\boldsymbol{\hat{f}}_{3} + I_{G,\perp}\,\Omega_{31}\,\boldsymbol{\hat{f}}_{3}\wedge\boldsymbol{\hat{f}}_{1}
\end{equation}
and recast as,
\begin{equation}\label{angular momentum disk bis}
\begin{split}
&\boldsymbol{L}_{G} = I_{G,\perp}\left(\Omega_{12}\,\boldsymbol{\hat{f}}_{1}\wedge\boldsymbol{\hat{f}}_{2} + \Omega_{23}\,\boldsymbol{\hat{f}}_{2}\wedge\boldsymbol{\hat{f}}_{3} + \Omega_{31}\,\boldsymbol{\hat{f}}_{3}\wedge\boldsymbol{\hat{f}}_{1}\right)\\
&\phantom{\boldsymbol{L}_{G} =} + \left(I_{G,\parallel}-\,I_{G,\perp}\right)\Omega_{12}\,\boldsymbol{\hat{f}}_{1}\wedge\boldsymbol{\hat{f}}_{2}
\end{split}
\end{equation}
In view of the angular velocity~\eqref{angular velocity principal body frame}, the angular momentum bivector~\eqref{angular momentum disk bis} reduces to,
\begin{equation}\label{angular momentum disk ter}
\boldsymbol{L}_{G} = I_{G,\perp}\,\boldsymbol{\Omega} + R\,\Big(\!\left(I_{G,\parallel}-\,I_{G,\perp}\right)\Omega_{12}\,\boldsymbol{\hat{e}}_{1}\wedge\boldsymbol{\hat{e}}_{2}\Big)R^{\dag}
\end{equation}
Taking into account the bivector~\eqref{velocity component plane of symmetry}, the angular momentum bivector~\eqref{angular momentum disk ter} becomes,
\begin{equation}\label{angular momentum disk quad}
\boldsymbol{L}_{G} = I_{G,\perp}\,\boldsymbol{\Omega} + R\left(I_{G,\parallel}-\,I_{G,\perp}\right)\boldsymbol{\Omega}_{0,\parallel}\,R^{\dag}
\end{equation}
Thus, the angular velocity bivector $\boldsymbol{\Omega}$ is expressed in terms of the constant bivectors $\boldsymbol{L}_{G}$ and $\boldsymbol{\Omega}_{0,\parallel}$ as,
\begin{equation}\label{angular momentum bivector disk}
\boldsymbol{\Omega} = \frac{1}{I_{G,\perp}}\Big(\boldsymbol{L}_{G} + R\left(I_{G,\parallel}-\,I_{G,\perp}\right)\boldsymbol{\Omega}_{0,\parallel}\,R^{\dag}\Big)
\end{equation}
Using the angular momentum bivector~\eqref{angular momentum bivector disk} and the normalisation condition~\eqref{unit rotor time dependent}, the time evolution equation for the rotor~\eqref{derivative identity ter} is recast as,
\begin{equation}\label{derivative identity disk}
\dot{R} = -\,\frac{1}{2}\,\boldsymbol{\Omega}\,R = -\,\frac{1}{2\,I_{G,\perp}}\Big(\boldsymbol{L}_{G}\,R + R\left(I_{G,\parallel}-\,I_{G,\perp}\right)\boldsymbol{\Omega}_{0,\parallel}\Big)
\end{equation}
Using the two constant bivectors defined as,
\begin{equation}\label{disk const bivectors}
\boldsymbol{\Omega}_{L} \equiv -\,\frac{1}{2\,I_{G,\perp}}\ \boldsymbol{L}_{G} \qquad\text{and}\qquad \boldsymbol{\Omega}_{R} \equiv -\,\frac{I_{G,\parallel}-\,I_{G,\perp}}{2\,I_{G,\perp}}\ \boldsymbol{\Omega}_{0,\parallel}
\end{equation}
the time evolution equation of the rotor~\eqref{derivative identity disk} is recast as,
\begin{equation}\label{derivative identity disk bis}
\dot{R} = -\,\frac{1}{2}\,\boldsymbol{\Omega}_{L}\,R -\,\frac{1}{2}\,R\ \boldsymbol{\Omega}_{R}
\end{equation}
The rotor satisfies the trivial initial condition,
\begin{equation}\label{initial rotor}
R\left(0\right) = 1
\end{equation}
The solution of the time evolution equation~\eqref{derivative identity disk bis} that satisfies the initial condition~\eqref{initial rotor} is the rotor,
\begin{equation}\label{rotor disk}
R\left(t\right) = \exp\left(-\,\frac{1}{2}\,\boldsymbol{\Omega}_{L}\,t\right)\exp\left(-\,\frac{1}{2}\,\boldsymbol{\Omega}_{R}\,t\right)
\end{equation}
as we now show. The time derivative of the rotor~\eqref{rotor disk} is,
\begin{equation}\label{rotor disk deriv}
\begin{split}
&\dot{R}\left(t\right) = \left(-\,\frac{1}{2}\,\boldsymbol{\Omega}_{L}\right)\exp\left(-\,\frac{1}{2}\,\boldsymbol{\Omega}_{L}\,t\right)\exp\left(-\,\frac{1}{2}\,\boldsymbol{\Omega}_{R}\,t\right)\\
&\phantom{\dot{R}\left(t\right) =} + \exp\left(-\,\frac{1}{2}\,\boldsymbol{\Omega}_{L}\,t\right)\exp\left(-\,\frac{1}{2}\,\boldsymbol{\Omega}_{R}\,t\right)\left(-\,\frac{1}{2}\,\boldsymbol{\Omega}_{R}\right)
\end{split}
\end{equation}
which reduces to the dynamical equation~\eqref{derivative identity disk bis} in view of the rotor~\eqref{rotor disk}, as it should.


\section{Conclusion}
\label{Conclusion}

\noindent Rotations can always be expressed as the composition of two reflections of intersecting planes. These planes containing the origin $O$ are determined by their respective orthogonal unit vectors $\boldsymbol{\hat{n}}_1$ and $\boldsymbol{\hat{n}}_2$. The geometric product of these vectors is the rotor $R_{\theta} = \boldsymbol{\hat{n}}_2\,\boldsymbol{\hat{n}}_1 = e^{-\,\boldsymbol{\hat{B}}\,\theta/2} = e^{-\,\boldsymbol{\hat{B}}\,\phi}$ where $\boldsymbol{\hat{B}}$ is the unit bivector and $\phi = \theta/2$ is the angle between the unit vectors $\boldsymbol{\hat{n}}_1$ and $\boldsymbol{\hat{n}}_2$ in the rotation plane. The rotation of a vector $\boldsymbol{v}$ by an angle $\theta = 2\,\phi$ in the rotation plane spanned by the unit bivector $\boldsymbol{\hat{B}}$ is given by the transformation law,
\begin{equation}\label{rotation vector}
\mathsf{R}_{\boldsymbol{\hat{B}}\,\theta}\left(\boldsymbol{v}\right) = R\,\boldsymbol{v}\,R^{\dag} = e^{-\,\boldsymbol{\hat{B}}\theta/2}\,\boldsymbol{v}\,e^{\boldsymbol{\hat{B}}\theta/2}
\end{equation}
where $R$ is the rotor of angle $\theta$ in the rotation plane spanned by the unit bivector $\boldsymbol{\hat{B}}$. Similarly, the rotation of a multivector $M = s + \boldsymbol{v} + \boldsymbol{A} + s^{\prime}\,I$ is given by the transformation law,
\begin{equation}\label{rotation multivector}
\mathsf{R}_{\boldsymbol{\hat{B}}\,\theta}\left(M\right) = R\,M\,R^{\dag} = e^{-\,\boldsymbol{\hat{B}}\theta/2}\,M\,e^{\boldsymbol{\hat{B}}\theta/2}
\end{equation}
The time evolution of the rotor $R$ and its reverse $R^{\dag}$ are expressed in terms of the angular velocity bivector $\boldsymbol{\Omega} = \dot{\theta}\,\boldsymbol{\hat{B}}$ in the rotation plane as,
\begin{equation}\label{time evolution rotor}
\dot{R} = -\,\frac{1}{2}\,\boldsymbol{\Omega}\,R \qquad\text{and}\qquad \dot{R}^{\dag} = \frac{1}{2}\,R^{\dag}\,\boldsymbol{\Omega}
\end{equation}
where the angular bivector $\boldsymbol{\Omega}$ is the dual of the angular pseudovector $\boldsymbol{\omega}$, i.e. $\boldsymbol{\omega} = \boldsymbol{\Omega}^{\ast}$. In geometric algebra, the time evolution of the rotor yields the Poisson formula describing the time evolution of the rotating basis vectors $\boldsymbol{\hat{f}}_i$,
\begin{equation}\label{Poisson formula end}
\boldsymbol{\dot{\hat{f}}}_i = \boldsymbol{\hat{f}}_i\cdot\boldsymbol{\Omega} 
\end{equation}
The geometric meaning of the inner product $\boldsymbol{\hat{f}}_i\cdot\boldsymbol{\Omega}$ is that the angular velocity bivector $\boldsymbol{\Omega}$ rotates the basis vector $\boldsymbol{\hat{f}}_i$ by a $90^{\circ}$ angle in the rotation direction defined by the unit bivector $\boldsymbol{\hat{B}}$.

Rotational dynamics is described by the angular momentum theorem. For a rigid body, this theorem is expressed in the rotation plane in geometric algebra in terms of bivectors evaluated with respect to the centre of mass $G$ as,
\begin{equation}\label{angular momentum theorem GA}
\sum\,\boldsymbol{T}_G^{\,\text{ext}} = \boldsymbol{\dot{L}}_G
\end{equation}
where the angular momentum bivector $\boldsymbol{L}_G$ is the dual of the angular momentum pseudovector $\boldsymbol{\ell}_G$, i.e. $\boldsymbol{\ell}_G = \boldsymbol{L}^{\ast}_G$, and the external torque bivector $\boldsymbol{T}_G^{\,\text{ext}}$ which is the dual of the external torque pseudovector $\boldsymbol{\tau}_G^{\,\text{ext}}$. According to the transformation law~\eqref{rotation multivector} for a multivector under rotation, the time evolution of an angular momentum bivector $\boldsymbol{L}_G$ describing the intrinsic rotation of a rigid body around its centre of mass is written as,
\begin{equation}\label{relative angular momentum rotation}
\boldsymbol{L}_{G} = R\ \boldsymbol{L}_{G,0}\,R^{\dag}
\end{equation}
where the initial angular momentum bivector $\boldsymbol{L}_{G,0}$ is a linear map $\boldsymbol{I}_{G}\left(\boldsymbol{\Omega}_0\right)$ of the initial angular velocity bivector, 
\begin{equation}\label{initial angular velocity bivector end}
\boldsymbol{\Omega}_0 = \frac{1}{2}\sum_{i,j = 1}^{3}\,\Omega_{ij}\ \boldsymbol{\hat{e}}_{i}\wedge\boldsymbol{\hat{e}}_{j} \qquad\text{with}\qquad \Omega_{ji} = -\,\Omega_{ij}
\end{equation}
that can be expressed as,
\begin{equation}\label{inertia map end}
\boldsymbol{L}_{G,0} = \boldsymbol{I}_G\left(\boldsymbol{\Omega}_{0}\right) = \frac{1}{2}\sum_{i,j = 1}^{3}\,I_{G,ij}\,\Omega_{ij}\ \boldsymbol{\hat{e}}_{i}\wedge\boldsymbol{\hat{e}}_{j} \qquad\text{with}\qquad I_{G,ji} = I_{G,ij}
\end{equation}
where $\boldsymbol{\hat{e}}_{i}\wedge\boldsymbol{\hat{e}}_{j}$ are the unit bivectors in the initial principal axis frame of the rigid body, $I_{G,ij} = \left(\varepsilon_{ijk}\right)^2i_{G,k}$ is the dual of the moment of inertia $i_{G,k}$ and $\Omega_{ij}$ is the antisymmetric component of the initial angular moment bivector $\boldsymbol{\Omega}_{0}$ in the rotation plane spanned by the bivectors $\boldsymbol{\hat{e}}_{i}\wedge\boldsymbol{\hat{e}}_{j}$. The vectors $\boldsymbol{\hat{e}}_{i}$ and bivectors $\boldsymbol{\hat{e}}_{i}\wedge\boldsymbol{\hat{e}}_{j}$ of the principal axis frame of the rigid body are related to the vectors $\boldsymbol{\hat{f}}_{i}$ and bivectors $\boldsymbol{\hat{f}}_{i}\wedge\boldsymbol{\hat{f}}_{j}$ of the initial principal axis frame of the rigid body through the rotational transformation laws~\eqref{rotation multivector},
\begin{equation}\label{rotational transformation laws}
\boldsymbol{\hat{f}}_{i} = R\,\boldsymbol{\hat{e}}_{i}\,R^{\dag} \qquad\text{and}\qquad \boldsymbol{\hat{f}}_{i}\wedge\boldsymbol{\hat{f}}_{j} = R\,\boldsymbol{\hat{e}}_{i}\wedge\boldsymbol{\hat{e}}_{j}\,R^{\dag}
\end{equation}
In view the linear inertia map~\eqref{inertia map end} and the rotational transformation laws~\eqref{rotational transformation laws}, the angular momentum bivector is recast in the principal axis frame of the rigid body in terms of the unit bivectors $\boldsymbol{\hat{f}}_{i}\wedge\boldsymbol{\hat{f}}_{j}$ as,
\begin{equation}\label{angular momentum map end}
\boldsymbol{L}_{G} = \frac{1}{2}\sum_{i,j = 1}^{3}\,I_{G,ij}\,\Omega_{ij}\ \boldsymbol{\hat{f}}_{i}\wedge\boldsymbol{\hat{f}}_{j}
\end{equation}
In geometric algebra, the Huygens-Steiner theorem for a rigid body of mass $m$ in the plane spanned by the unit bivector frame $\boldsymbol{\hat{f}}_{i}\wedge\boldsymbol{\hat{f}}_{j}$ reads,
\begin{equation}\label{Huygens-Steiner theorem end}
I_{A,ij} = I_{G,ij} + m\,\boldsymbol{r}_{AG}^2
\end{equation}
and the kinetic energy is given by,
\begin{equation}\label{kinetic energy end}
\mathcal{T} = \frac{1}{2}\ m\,\boldsymbol{v}_G^2
+ \frac{1}{4}\sum_{i,j = 1}^{3}\,I_{G,ij}\,\Omega_{ij}^2
\end{equation}
The Euler equations for the rigid body are expressed in terms of the components of the dynamical bivectors in three orthogonal planes spanned by the unit bivectors $\boldsymbol{\hat{f}}_{1}\wedge\boldsymbol{\hat{f}}_{2}$, $\boldsymbol{\hat{f}}_{2}\wedge\boldsymbol{\hat{f}}_{3}$ and $\boldsymbol{\hat{f}}_{3}\wedge\boldsymbol{\hat{f}}_{1}$ as,
\begin{equation}\label{Euler equations end}
\begin{split}
&\sum\,T_{G,12}^{\,\text{ext}} = I_{G,12}\,\dot{\Omega}_{12} + \left(I_{G,23}-\,I_{G,31}\right)\Omega_{23}\,\Omega_{31}\\
&\sum\,T_{G,23}^{\,\text{ext}} = I_{G,23}\,\dot{\Omega}_{23} + \left(I_{G,31}-\,I_{G,12}\right)\Omega_{31}\,\Omega_{12}\\
&\sum\,T_{G,31}^{\,\text{ext}} = I_{G,31}\,\dot{\Omega}_{31} + \left(I_{G,12}-\,I_{G,23}\right)\Omega_{12}\,\Omega_{23}
\end{split}
\end{equation}
The dynamics of rigid bodies in geometric algebra can be applied to the torque free motion of a gyroscope consisting of a symmetric spinning disk around its centre of mass $G$. It is described by the rotor,
\begin{equation}\label{rotor disk and}
R\left(t\right) = \exp\left(-\,\frac{1}{2}\,\boldsymbol{\Omega}_{L}\,t\right)\exp\left(-\,\frac{1}{2}\,\boldsymbol{\Omega}_{R}\,t\right)
\end{equation}
where the angular velocity bivectors $\boldsymbol{\Omega}_{L}$ and $\boldsymbol{\Omega}_{R}$ are given by,
\begin{equation}\label{disk const bivectors end}
\boldsymbol{\Omega}_{L} \equiv -\,\frac{1}{2\,I_{G,\perp}}\ \boldsymbol{L}_{G} \qquad\text{and}\qquad \boldsymbol{\Omega}_{R} \equiv -\,\frac{I_{G,\parallel}-\,I_{G,\perp}}{2\,I_{G,\perp}}\ \boldsymbol{\Omega}_{0,\parallel}
\end{equation}
$I_{G,\parallel}$ is the moment of inertia and $\boldsymbol{\Omega}_{0,\parallel}$ is the angular velocity in the plane of the disk, and $I_{G,\perp}$ are the moments of inertia in the planes orthogonal to the plane of the disk.


\section*{Acknowledgements}

\noindent The author would like to thank his PhD thesis adviser Anthony Lasenby for introducing him to geometric algebra.


\pagebreak

\appendix

\renewcommand*{\thesection}{\Alph{section}}

\section{Geometric algebra (GA)}
\label{Geometric algebra}

\noindent We consider an orthonormal vector spatial frame $\{\boldsymbol{\hat{e}}_1,\boldsymbol{\hat{e}}_2,\boldsymbol{\hat{e}}_3\}$. The geometric product of two basis vectors reads,~\cite{Lasenby:2003}
\begin{equation}\label{geometric product space 0}
\boldsymbol{\hat{e}}_{i}\,\boldsymbol{\hat{e}}_{j} = \boldsymbol{\hat{e}}_{i}\cdot\boldsymbol{\hat{e}}_{j} + \boldsymbol{\hat{e}}_{i}\wedge\boldsymbol{\hat{e}}_{j} \qquad\text{with}\qquad i,j = 1,2,3
\end{equation}
where,
\begin{equation}\label{square space}
\boldsymbol{\hat{e}}_{1}^2 = \boldsymbol{\hat{e}}_{2}^2 = \boldsymbol{\hat{e}}_{3}^2 = 1 \qquad\text{and}\qquad \boldsymbol{\hat{e}}_{i}\wedge\boldsymbol{\hat{e}}_{i} = \boldsymbol{0}
\end{equation}
The inner product of two basis vectors is symmetric and defined as,
\begin{equation}\label{inner product space}
\boldsymbol{\hat{e}}_{i}\cdot\boldsymbol{\hat{e}}_{j} = \boldsymbol{\hat{e}}_{j}\cdot\boldsymbol{\hat{e}}_{i} = \frac{1}{2}\left(\boldsymbol{\hat{e}}_{i}\,\boldsymbol{\hat{e}}_{j} + \boldsymbol{\hat{e}}_{j}\,\boldsymbol{\hat{e}}_{i}\right)
\end{equation}
and the outer product is antisymmetric,
\begin{equation}\label{outer product space}
\boldsymbol{\hat{e}}_{i}\wedge\boldsymbol{\hat{e}}_{j} = -\,\boldsymbol{\hat{e}}_{j}\wedge\boldsymbol{\hat{e}}_{i} = \frac{1}{2}\left(\boldsymbol{\hat{e}}_{i}\,\boldsymbol{\hat{e}}_{j} -\,\boldsymbol{\hat{e}}_{j}\,\boldsymbol{\hat{e}}_{i}\right)
\end{equation}
The $8$ basis elements of the geometric algebra (GA) $\mathbb{G}^{3}$ called the geometric algebra are :
\begin{itemize}
\item one scalar : $\ 1$
\item three vectors : $\ \boldsymbol{\hat{e}}_1$, $\ \boldsymbol{\hat{e}}_2$, $\ \boldsymbol{\hat{e}}_3$
\item three bivectors : $\ \boldsymbol{\hat{e}}_1\wedge\boldsymbol{\hat{e}}_2$, $\ \boldsymbol{\hat{e}}_2\wedge\boldsymbol{\hat{e}}_3$, $\ \boldsymbol{\hat{e}}_3\wedge\boldsymbol{\hat{e}}_1$
\item one trivector or pseudoscalar : $\ \boldsymbol{\hat{e}}_1\wedge\boldsymbol{\hat{e}}_2\wedge\boldsymbol{\hat{e}}_3$
\end{itemize}
\begin{figure}[!ht]
\begin{center}
\includegraphics[scale=0.55]{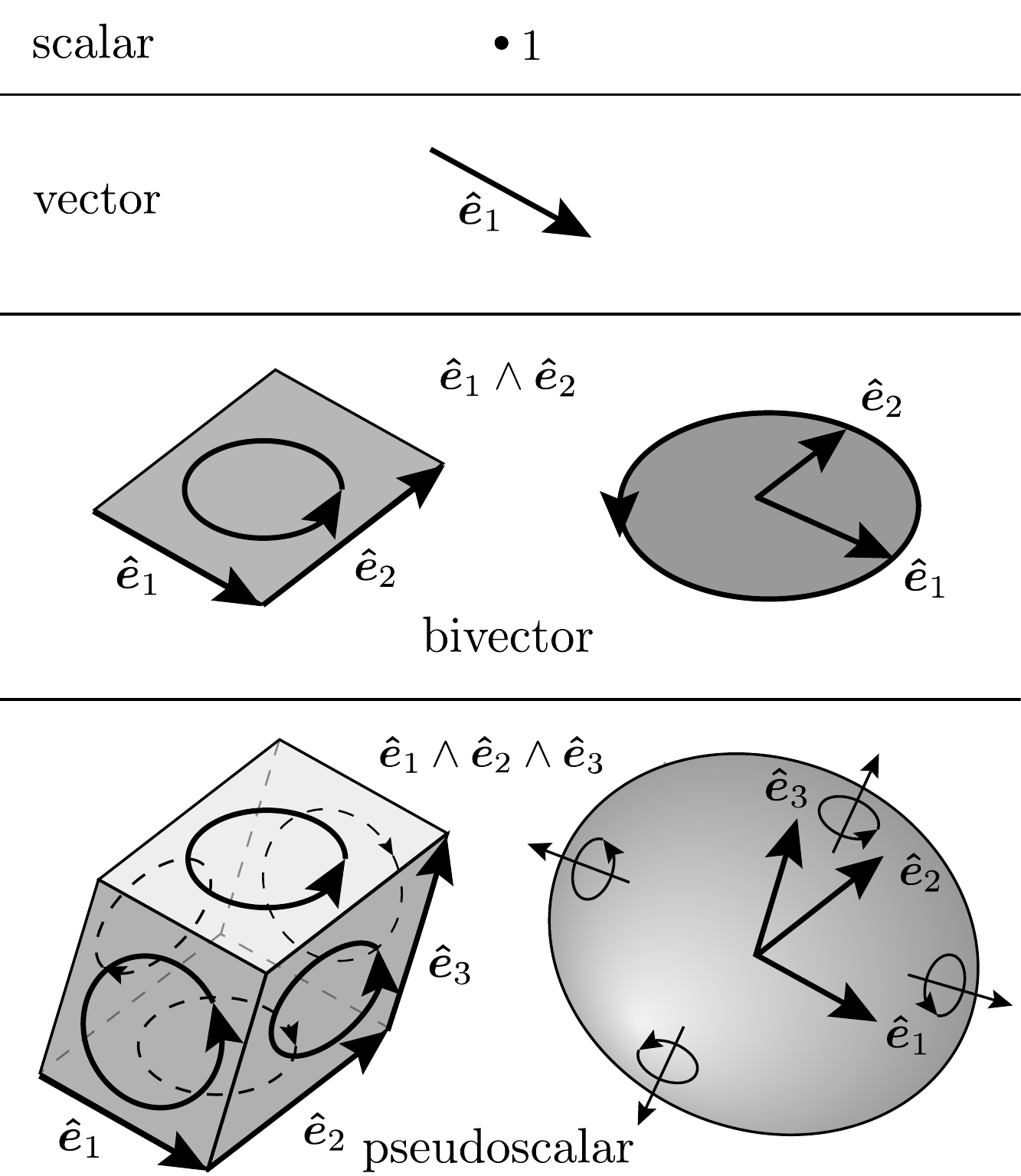}
\end{center}
\caption{A scalar, a vector, a bivector and a pseudoscalar are illustrated graphically. Their orientation and modulus are independent of their shape.}\label{Fig: vectors}
\end{figure}
where the number of elements of each type $1$, $3$, $3$, $1$ are the binomial coefficients in three dimensions. In $\mathbb{G}^{3}$, there are $2^3$ basis elements. The geometric interpretation of the basis elements is clear (Fig.~\ref{Fig: vectors}). A scalar $s$ is an oriented point, a vector is an oriented line element, a bivector is an oriented plane element and a pseudoscalar is an oriented volume element. The notation used is the following : scalars $s$ are written in lower case letters, vectors $\boldsymbol{v}$ are written in lower case bold letters, bivectors $\boldsymbol{B}$ are written in upper case bold letters and pseudoscalars are written as $s^{\prime}I$. A multivector $M$, which is linear combination of elements of $\mathbb{G}^{3}$, is written in upper case letters,
\begin{equation}\label{multivector}
M = s + \boldsymbol{v} + \boldsymbol{B} + s^{\prime}I
\end{equation}
Taking into account the definition~\eqref{geometric product space 0}, the bivectors and the pseudoscalar are also written as geometric products of the basis unit vectors $\boldsymbol{\hat{e}}_1$, $\boldsymbol{\hat{e}}_2$, $\boldsymbol{\hat{e}}_3$ :
\begin{itemize}
\item bivectors : $\ \boldsymbol{\hat{e}}_1\,\boldsymbol{\hat{e}}_2$, $\ \boldsymbol{\hat{e}}_2\,\boldsymbol{\hat{e}}_3$, $\ \boldsymbol{\hat{e}}_3\,\boldsymbol{\hat{e}}_1$
\item pseudoscalar : $\ \boldsymbol{\hat{e}}_1\,\boldsymbol{\hat{e}}_2\,\boldsymbol{\hat{e}}_3$
\end{itemize}
The pseudoscalar $I = \boldsymbol{\hat{e}}_1\,\boldsymbol{\hat{e}}_2\,\boldsymbol{\hat{e}}_3$ behaves as an imaginary number since,
\begin{equation}\label{pseudoscalar squared}
I^2 = \left(\boldsymbol{\hat{e}}_1\,\boldsymbol{\hat{e}}_2\,\boldsymbol{\hat{e}}_3\right)\left(\boldsymbol{\hat{e}}_1\,\boldsymbol{\hat{e}}_2\,\boldsymbol{\hat{e}}_3\right) = \left(\boldsymbol{\hat{e}}_1\,\boldsymbol{\hat{e}}_2\,\boldsymbol{\hat{e}}_1\,\boldsymbol{\hat{e}}_2\right)\left(\boldsymbol{\hat{e}}_3\,\boldsymbol{\hat{e}}_3\right) = -\,\left(\boldsymbol{\hat{e}}_1\,\boldsymbol{\hat{e}}_1\right)\left(\boldsymbol{\hat{e}}_2\,\boldsymbol{\hat{e}}_2\right) = -\,1
\end{equation}

The geometric product of two basis vectors $\boldsymbol{u}$ and $\boldsymbol{v}$ reads,
\begin{equation}\label{geometric product space u v}
\boldsymbol{u}\,\boldsymbol{v} = \boldsymbol{u}\cdot\boldsymbol{v} + \boldsymbol{u}\wedge\boldsymbol{v}
\end{equation}
The inner product between the two basis vectors is symmetric,
\begin{equation}\label{inner product space u v}
\boldsymbol{u}\cdot\boldsymbol{v} = \boldsymbol{v}\cdot\boldsymbol{u} = \frac{1}{2}\left(\boldsymbol{u}\,\boldsymbol{v} + \boldsymbol{v}\,\boldsymbol{u}\right)
\end{equation}
and the outer product is antisymmetric,
\begin{equation}\label{outer product space u v}
\boldsymbol{u}\wedge\boldsymbol{v} = -\,\boldsymbol{v}\wedge\boldsymbol{u} = \frac{1}{2}\left(\boldsymbol{u}\,\boldsymbol{v} -\,\boldsymbol{v}\,\boldsymbol{u}\right)
\end{equation}
The geometric product of a vector $\boldsymbol{v}$ and a bivector $\boldsymbol{B}$ is the sum of the inner product of the outer product,
\begin{equation}\label{geometric product space v B}
\boldsymbol{v}\,\boldsymbol{B} = \boldsymbol{v}\cdot\boldsymbol{B} + \boldsymbol{v}\wedge\boldsymbol{B}
\end{equation}
and the geometric product of a bivector $\boldsymbol{B}$ and a vector $\boldsymbol{v}$ is the sum of the inner product of the outer product,
\begin{equation}\label{geometric product space B v}
\boldsymbol{B}\,\boldsymbol{v} = \boldsymbol{B}\cdot\boldsymbol{v} + \boldsymbol{B}\wedge\boldsymbol{v}
\end{equation}
To determine the symmetry of the these products, the spatial frame is oriented such that the bivector reads $\boldsymbol{B} = B_{12}\,\boldsymbol{\hat{e}}_1\wedge\boldsymbol{\hat{e}}_2 = B_{12}\,\boldsymbol{\hat{e}}_1\,\boldsymbol{\hat{e}}_2$ and the orientation of the vector 
$\boldsymbol{v} = v_1\,\boldsymbol{\hat{e}}_1 + v_2\,\boldsymbol{\hat{e}}_2 + v_3\,\boldsymbol{\hat{e}}_3$ is arbitrary. The geometric product of the vector $\boldsymbol{v}$ and the bivector $\boldsymbol{B}$ reads,
\begin{equation}\label{geometric product v B}
\begin{split}
&\boldsymbol{v}\,\boldsymbol{B} = \left(v_1\,\boldsymbol{\hat{e}}_1 + v_2\,\boldsymbol{\hat{e}}_2 + v_3\,\boldsymbol{\hat{e}}_3\right)\,\left(B_{12}\,\boldsymbol{\hat{e}}_1\,\boldsymbol{\hat{e}}_2\right)\\
&\phantom{\boldsymbol{v}\,\boldsymbol{B}} = v_1\,B_{12}\,\boldsymbol{\hat{e}}_1\,\boldsymbol{\hat{e}}_1\,\boldsymbol{\hat{e}}_2 + v_2\,B_{12}\,\boldsymbol{\hat{e}}_2\,\boldsymbol{\hat{e}}_1\,\boldsymbol{\hat{e}}_2 + v_3\,B_{12}\,\boldsymbol{\hat{e}}_3\,\boldsymbol{\hat{e}}_1\,\boldsymbol{\hat{e}}_2\\
&\phantom{\boldsymbol{v}\,\boldsymbol{B}} = B_{12}\,v_{1}\,\boldsymbol{\hat{e}}_2 -\,B_{12}\,v_{2}\,\boldsymbol{\hat{e}}_1 + B_{12}\,v_{3}\,\boldsymbol{\hat{e}}_1\,\boldsymbol{\hat{e}}_2\,\boldsymbol{\hat{e}}_3
\end{split}
\end{equation}
The geometric product of the bivector $\boldsymbol{B}$ and the vector $\boldsymbol{v}$ reads,
\begin{equation}\label{geometric product B v}
\begin{split}
&\boldsymbol{B}\,\boldsymbol{v} = \left(B_{12}\,\boldsymbol{\hat{e}}_1\,\boldsymbol{\hat{e}}_2\right)\left(v_1\,\boldsymbol{\hat{e}}_1 + v_2\,\boldsymbol{\hat{e}}_2 + v_3\,\boldsymbol{\hat{e}}_3\right)\\
&\phantom{\boldsymbol{B}\,\boldsymbol{v}} = B_{12}\,v_{1}\,\boldsymbol{\hat{e}}_1\,\boldsymbol{\hat{e}}_2\,\boldsymbol{\hat{e}}_1 + B_{12}\,v_{2}\,\boldsymbol{\hat{e}}_1\,\boldsymbol{\hat{e}}_2\,\boldsymbol{\hat{e}}_2 + B_{12}\,v_{3}\,\boldsymbol{\hat{e}}_1\,\boldsymbol{\hat{e}}_2\,\boldsymbol{\hat{e}}_3\\
&\phantom{\boldsymbol{B}\,\boldsymbol{v}} = -\,B_{12}\,v_{1}\,\boldsymbol{\hat{e}}_2 + B_{12}\,v_{2}\,\boldsymbol{\hat{e}}_1 + B_{12}\,v_{3}\,\boldsymbol{\hat{e}}_1\,\boldsymbol{\hat{e}}_2\,\boldsymbol{\hat{e}}_3
\end{split}
\end{equation}
The outer product of the vector $\boldsymbol{v}$ and the bivector $\boldsymbol{B}$ reads,
\begin{align}\label{outer product v B}
&\boldsymbol{v}\wedge\boldsymbol{B} = \left(v_1\,\boldsymbol{\hat{e}}_1 + v_2\,\boldsymbol{\hat{e}}_2 + v_3\,\boldsymbol{\hat{e}}_3\right)\wedge\left(B_{12}\,\boldsymbol{\hat{e}}_1\wedge\boldsymbol{\hat{e}}_2\right)\nonumber\\
&\phantom{\boldsymbol{v}\wedge\boldsymbol{B}} = v_1\,B_{12}\,\boldsymbol{\hat{e}}_1\wedge\boldsymbol{\hat{e}}_1\wedge\boldsymbol{\hat{e}}_2 + v_2\,B_{12}\,\boldsymbol{\hat{e}}_2\wedge\boldsymbol{\hat{e}}_1\wedge\boldsymbol{\hat{e}}_2 + v_3\,B_{12}\,\boldsymbol{\hat{e}}_3\wedge\boldsymbol{\hat{e}}_1\wedge\boldsymbol{\hat{e}}_2\nonumber\\
&\phantom{\boldsymbol{v}\wedge\boldsymbol{B}} = v_3\,B_{12}\,\boldsymbol{\hat{e}}_1\wedge\boldsymbol{\hat{e}}_2\wedge\boldsymbol{\hat{e}}_3
\end{align}
The outer product of the bivector $\boldsymbol{B}$ and the vector $\boldsymbol{v}$ reads,
\begin{align}\label{outer product B v}
&\boldsymbol{B}\wedge\boldsymbol{v} = \left(B_{12}\,\boldsymbol{\hat{e}}_1\wedge\boldsymbol{\hat{e}}_2\right)\wedge\left(v_1\,\boldsymbol{\hat{e}}_1 + v_2\,\boldsymbol{\hat{e}}_2 + v_3\,\boldsymbol{\hat{e}}_3\right)\nonumber\\
&\phantom{\boldsymbol{B}\wedge\boldsymbol{v}} = B_{12}\,v_{1}\,\boldsymbol{\hat{e}}_1\wedge\boldsymbol{\hat{e}}_2\wedge\boldsymbol{\hat{e}}_1 + B_{12}\,v_{2}\,\boldsymbol{\hat{e}}_1\wedge\boldsymbol{\hat{e}}_2\wedge\boldsymbol{\hat{e}}_2 + B_{12}\,v_{3}\,\boldsymbol{\hat{e}}_1\wedge\boldsymbol{\hat{e}}_2\wedge\boldsymbol{\hat{e}}_3\nonumber\\
&\phantom{\boldsymbol{B}\wedge\boldsymbol{v}} = B_{12}\,v_{3}\,\boldsymbol{\hat{e}}_1\wedge\boldsymbol{\hat{e}}_2\wedge\boldsymbol{\hat{e}}_3
\end{align}
From relations~\eqref{outer product v B} and~\eqref{outer product B v} we conclude that the outer product of a vector $\boldsymbol{v}$ and a bivector $\boldsymbol{B}$ is symmetric,
\begin{equation}\label{outer product v B symmetric}
\boldsymbol{v}\wedge\boldsymbol{B} = \boldsymbol{B}\wedge\boldsymbol{v}
\end{equation}
In view of relations\eqref{geometric product space v B},~\eqref{geometric product v B} and~\eqref{outer product v B},
\begin{equation}\label{inner product v B}
\boldsymbol{v}\cdot\boldsymbol{B} = \boldsymbol{v}\,\boldsymbol{B} -\,\boldsymbol{v}\wedge\boldsymbol{B} = -\,B_{12}\,v_{2}\,\boldsymbol{\hat{e}}_1 + B_{12}\,v_{1}\,\boldsymbol{\hat{e}}_2 
\end{equation}
In view of relations\eqref{geometric product space B v},~\eqref{geometric product B v} and~\eqref{outer product B v},
\begin{equation}\label{inner product B v}
\boldsymbol{B}\cdot\boldsymbol{v} = \boldsymbol{B}\,\boldsymbol{v} -\,\boldsymbol{B}\wedge\boldsymbol{v} = B_{12}\,v_{2}\,\boldsymbol{\hat{e}}_1 -\,B_{12}\,v_{1}\,\boldsymbol{\hat{e}}_2 
\end{equation}
From relations~\eqref{inner product v B} and~\eqref{inner product B v} we conclude that the inner product of a vector $\boldsymbol{v}$ and a bivector $\boldsymbol{B}$ is antisymmetric,
\begin{equation}\label{inner product v B antisymmetric}
\boldsymbol{v}\cdot\boldsymbol{B} = -\,\boldsymbol{B}\cdot\boldsymbol{v}
\end{equation}
In view of relations~\eqref{geometric product v B} and~\eqref{geometric product B v},~\eqref{outer product v B symmetric} and~\eqref{inner product v B antisymmetric},
\begin{align}
\label{inner product antisymmetry v B}
&\boldsymbol{v}\cdot\boldsymbol{B} = \frac{1}{2}\left(\boldsymbol{v}\,\boldsymbol{B} -\,\boldsymbol{B}\,\boldsymbol{v}\right)\\
\label{outer product symmetry v B}
&\boldsymbol{v}\wedge\boldsymbol{B} = \frac{1}{2}\left(\boldsymbol{v}\,\boldsymbol{B} + \boldsymbol{B}\,\boldsymbol{v}\right)
\end{align}
To determine the geometric product of two bivectors $\boldsymbol{A}$ and $\boldsymbol{B}$, we write them as $\boldsymbol{A} = A_{12}\,\boldsymbol{\hat{e}}_1\,\boldsymbol{\hat{e}}_2 + A_{23}\,\boldsymbol{\hat{e}}_2\,\boldsymbol{\hat{e}}_3 + A_{31}\,\boldsymbol{\hat{e}}_3\,\boldsymbol{\hat{e}}_1$ and $\boldsymbol{B} = B_{12}\,\boldsymbol{\hat{e}}_1\,\boldsymbol{\hat{e}}_2 + B_{23}\,\boldsymbol{\hat{e}}_2\,\boldsymbol{\hat{e}}_3 + B_{31}\,\boldsymbol{\hat{e}}_3\,\boldsymbol{\hat{e}}_1$. The geometric product of the bivectors $\boldsymbol{A}$ and $\boldsymbol{B}$ reads,
\begin{align}\label{geometric product A V 0}
&\boldsymbol{A}\,\boldsymbol{B} = \left(A_{12}\,\boldsymbol{\hat{e}}_1\,\boldsymbol{\hat{e}}_2 + A_{23}\,\boldsymbol{\hat{e}}_2\,\boldsymbol{\hat{e}}_3 + A_{31}\,\boldsymbol{\hat{e}}_3\,\boldsymbol{\hat{e}}_1\right)\left(B_{12}\,\boldsymbol{\hat{e}}_1\,\boldsymbol{\hat{e}}_2 + B_{23}\,\boldsymbol{\hat{e}}_2\,\boldsymbol{\hat{e}}_3 + B_{31}\,\boldsymbol{\hat{e}}_3\,\boldsymbol{\hat{e}}_1\right)\nonumber\\
&\phantom{\boldsymbol{A}\,\boldsymbol{B}} = A_{12}\,B_{12}\,\boldsymbol{\hat{e}}_1\,\boldsymbol{\hat{e}}_2\,\boldsymbol{\hat{e}}_1\,\boldsymbol{\hat{e}}_2 + A_{12}\,B_{23}\,\boldsymbol{\hat{e}}_1\,\boldsymbol{\hat{e}}_2\,\boldsymbol{\hat{e}}_2\,\boldsymbol{\hat{e}}_3 + A_{12}\,B_{31}\,\boldsymbol{\hat{e}}_1\,\boldsymbol{\hat{e}}_2\,\boldsymbol{\hat{e}}_3\,\boldsymbol{\hat{e}}_1\\
&\phantom{\boldsymbol{A}\,\boldsymbol{B} =} + A_{23}\,B_{12}\,\boldsymbol{\hat{e}}_2\,\boldsymbol{\hat{e}}_3\,\boldsymbol{\hat{e}}_1\,\boldsymbol{\hat{e}}_2 + A_{23}\,B_{23}\,\boldsymbol{\hat{e}}_2\,\boldsymbol{\hat{e}}_3\,\boldsymbol{\hat{e}}_2\,\boldsymbol{\hat{e}}_3 + A_{23}\,B_{31}\,\boldsymbol{\hat{e}}_2\,\boldsymbol{\hat{e}}_3\,\boldsymbol{\hat{e}}_3\,\boldsymbol{\hat{e}}_1\nonumber\\
&\phantom{\boldsymbol{A}\,\boldsymbol{B} =} + A_{31}\,B_{12}\,\boldsymbol{\hat{e}}_3\,\boldsymbol{\hat{e}}_1\,\boldsymbol{\hat{e}}_1\,\boldsymbol{\hat{e}}_2 + A_{31}\,B_{23}\,\boldsymbol{\hat{e}}_3\,\boldsymbol{\hat{e}}_1\,\boldsymbol{\hat{e}}_2\,\boldsymbol{\hat{e}}_3 + A_{31}\,B_{31}\,\boldsymbol{\hat{e}}_3\,\boldsymbol{\hat{e}}_1\,\boldsymbol{\hat{e}}_3\,\boldsymbol{\hat{e}}_1\nonumber
\end{align}
which reduces to,
\begin{equation}\label{geometric product A V}
\begin{split}
&\boldsymbol{A}\,\boldsymbol{B} = -\,\left(A_{12}\,B_{12} + A_{23}\,B_{23} + A_{31}\,B_{31}\right) + \left(A_{31}\,B_{23} -\,A_{23}\,B_{31}\right)\boldsymbol{\hat{e}}_1\,\boldsymbol{\hat{e}}_2\\
&\phantom{\boldsymbol{A}\,\boldsymbol{B} =} + \left(A_{12}\,B_{31} -\,A_{31}\,B_{12}\right)\boldsymbol{\hat{e}}_2\,\boldsymbol{\hat{e}}_3 + \left(A_{23}\,B_{12} -\,A_{12}\,B_{23}\right)\boldsymbol{\hat{e}}_3\,\boldsymbol{\hat{e}}_1 
\end{split}
\end{equation}
The geometric product of the bivectors $\boldsymbol{B}$ and $\boldsymbol{A}$ reads,
\begin{align}\label{geometric product B V 0}
&\boldsymbol{B}\,\boldsymbol{A} = \left(B_{12}\,\boldsymbol{\hat{e}}_1\,\boldsymbol{\hat{e}}_2 + B_{23}\,\boldsymbol{\hat{e}}_2\,\boldsymbol{\hat{e}}_3 + B_{31}\,\boldsymbol{\hat{e}}_3\,\boldsymbol{\hat{e}}_1\right)\left(A_{12}\,\boldsymbol{\hat{e}}_1\,\boldsymbol{\hat{e}}_2 + A_{23}\,\boldsymbol{\hat{e}}_2\,\boldsymbol{\hat{e}}_3 + A_{31}\,\boldsymbol{\hat{e}}_3\,\boldsymbol{\hat{e}}_1\right)\nonumber\\
&\phantom{\boldsymbol{B}\,\boldsymbol{A}} = A_{12}\,B_{12}\,\boldsymbol{\hat{e}}_1\,\boldsymbol{\hat{e}}_2\,\boldsymbol{\hat{e}}_1\,\boldsymbol{\hat{e}}_2 + A_{23}\,B_{12}\,\boldsymbol{\hat{e}}_1\,\boldsymbol{\hat{e}}_2\,\boldsymbol{\hat{e}}_2\,\boldsymbol{\hat{e}}_3 + A_{12}\,B_{31}\,\boldsymbol{\hat{e}}_1\,\boldsymbol{\hat{e}}_2\,\boldsymbol{\hat{e}}_3\,\boldsymbol{\hat{e}}_1\\
&\phantom{\boldsymbol{B}\,\boldsymbol{A} =} + A_{12}\,B_{23}\,\boldsymbol{\hat{e}}_2\,\boldsymbol{\hat{e}}_3\,\boldsymbol{\hat{e}}_1\,\boldsymbol{\hat{e}}_2 + A_{23}\,B_{23}\,\boldsymbol{\hat{e}}_2\,\boldsymbol{\hat{e}}_3\,\boldsymbol{\hat{e}}_2\,\boldsymbol{\hat{e}}_3 + A_{31}\,B_{23}\,\boldsymbol{\hat{e}}_2\,\boldsymbol{\hat{e}}_3\,\boldsymbol{\hat{e}}_3\,\boldsymbol{\hat{e}}_1\nonumber\\
&\phantom{\boldsymbol{B}\,\boldsymbol{A} =} + A_{12}\,B_{31}\,\boldsymbol{\hat{e}}_3\,\boldsymbol{\hat{e}}_1\,\boldsymbol{\hat{e}}_1\,\boldsymbol{\hat{e}}_2 + A_{23}\,B_{31}\,\boldsymbol{\hat{e}}_3\,\boldsymbol{\hat{e}}_1\,\boldsymbol{\hat{e}}_2\,\boldsymbol{\hat{e}}_3 + A_{31}\,B_{31}\,\boldsymbol{\hat{e}}_3\,\boldsymbol{\hat{e}}_1\,\boldsymbol{\hat{e}}_3\,\boldsymbol{\hat{e}}_1\nonumber
\end{align}
which reduces to,
\begin{equation}\label{geometric product B V}
\begin{split}
&\boldsymbol{B}\,\boldsymbol{A} = -\,\left(A_{12}\,B_{12} + A_{23}\,B_{23} + A_{31}\,B_{31}\right) + \left(A_{23}\,B_{31} -\,A_{31}\,B_{23}\right)\boldsymbol{\hat{e}}_1\,\boldsymbol{\hat{e}}_2\\
&\phantom{\boldsymbol{B}\,\boldsymbol{A} =} + \left(A_{31}\,B_{12} -\,A_{12}\,B_{31}\right)\boldsymbol{\hat{e}}_2\,\boldsymbol{\hat{e}}_3 + \left(A_{12}\,B_{23} -\,A_{23}\,B_{12}\right)\boldsymbol{\hat{e}}_3\,\boldsymbol{\hat{e}}_1 
\end{split}
\end{equation}
The geometric product~\eqref{geometric product A V} of two bivectors $\boldsymbol{A}$ and $\boldsymbol{B}$ is the sum symmetric product and an antisymmetric product,
\begin{equation}\label{geometric product bivectors}
\boldsymbol{A}\,\boldsymbol{B} = \boldsymbol{A}\cdot\boldsymbol{B} + \boldsymbol{A}\times\boldsymbol{B}
\end{equation}
According to relations~\eqref{geometric product A V} and~\eqref{geometric product B V}, the inner product, or anticommutator, of two bivectors $\boldsymbol{A}$ and $\boldsymbol{B}$ yields a scalar,
\begin{equation}\label{geometric product bivectors inner}
\boldsymbol{A}\cdot\boldsymbol{B} = \boldsymbol{B}\cdot\boldsymbol{A} = \frac{1}{2}\left(\boldsymbol{A}\,\boldsymbol{B} + \boldsymbol{B}\,\boldsymbol{A}\right) =  -\,\left(A_{12}\,B_{12} + A_{23}\,B_{23} + A_{31}\,B_{31}\right)
\end{equation}
According to relations~\eqref{geometric product A V} and~\eqref{geometric product B V}, the cross product, or commutator, of two bivectors $\boldsymbol{A}$ and $\boldsymbol{B}$ yields a bivector,
\begin{align}\label{geometric product bivectors outer}
&\boldsymbol{A}\times\boldsymbol{B} = -\,\boldsymbol{B}\times\boldsymbol{A} = \frac{1}{2}\left(\boldsymbol{A}\,\boldsymbol{B} -\,\boldsymbol{B}\,\boldsymbol{A}\right)\nonumber\\
&= \left(A_{31}\,B_{23} -\,A_{23}\,B_{31}\right)\boldsymbol{\hat{e}}_1\,\boldsymbol{\hat{e}}_2 + \left(A_{12}\,B_{31} -\,A_{31}\,B_{12}\right)\boldsymbol{\hat{e}}_2\,\boldsymbol{\hat{e}}_3\\
&\phantom{=} + \left(A_{23}\,B_{12} -\,A_{12}\,B_{23}\right)\boldsymbol{\hat{e}}_3\,\boldsymbol{\hat{e}}_1\nonumber
\end{align}
The fact that the anticommutator of two bivectors yields a scalar and the commutator of two bivectors yields a bivector is specific to the geometric algebra $\mathbb{G}^{3}$. To determine the geometric product of a vector $\boldsymbol{v}$ and the pseudoscalar $I$, we write them as $\boldsymbol{B} = B_{12}\,\boldsymbol{\hat{e}}_1\,\boldsymbol{\hat{e}}_2 + B_{23}\,\boldsymbol{\hat{e}}_2\,\boldsymbol{\hat{e}}_3 + B_{31}\,\boldsymbol{\hat{e}}_3\,\boldsymbol{\hat{e}}_1$ and $I = \boldsymbol{\hat{e}}_1\,\boldsymbol{\hat{e}}_2\,\boldsymbol{\hat{e}}_3$. The geometric product of the vector $\boldsymbol{v}$ and the pseudoscalar $I$ reads,
\begin{align}
\label{v I}
&\boldsymbol{v}\,I = \left(v_1\,\boldsymbol{\hat{e}}_1 + v_2\,\boldsymbol{\hat{e}}_2 + v_3\,\boldsymbol{\hat{e}}_3\right)\left(\boldsymbol{\hat{e}}_1\,\boldsymbol{\hat{e}}_2\,\boldsymbol{\hat{e}}_3\right) = v_1\,\boldsymbol{\hat{e}}_2\,\boldsymbol{\hat{e}}_3 + v_2\,\boldsymbol{\hat{e}}_3\,\boldsymbol{\hat{e}}_1 + v_3\,\boldsymbol{\hat{e}}_1\,\boldsymbol{\hat{e}}_2\\
\label{I v}
&I\,\boldsymbol{v} = \left(\boldsymbol{\hat{e}}_1\,\boldsymbol{\hat{e}}_2\,\boldsymbol{\hat{e}}_3\right)\left(v_1\,\boldsymbol{\hat{e}}_1 + v_2\,\boldsymbol{\hat{e}}_2 + v_3\,\boldsymbol{\hat{e}}_3\right) = v_1\,\boldsymbol{\hat{e}}_2\,\boldsymbol{\hat{e}}_3 + v_2\,\boldsymbol{\hat{e}}_3\,\boldsymbol{\hat{e}}_1 + v_3\,\boldsymbol{\hat{e}}_1\,\boldsymbol{\hat{e}}_2
\end{align}
According to relation~\eqref{v I} and~\eqref{I v}, the vector $\boldsymbol{v}$ commutes with the pseudoscalar $I$,
\begin{equation}\label{commutation v I}
\boldsymbol{v}\,I = I\,\boldsymbol{v}
\end{equation}
To determine the geometric product of a vector $\boldsymbol{v}$ and the pseudoscalar $I$, we write them as $\boldsymbol{v} = v_1\,\boldsymbol{\hat{e}}_1 + v_2\,\boldsymbol{\hat{e}}_2 + v_3\,\boldsymbol{\hat{e}}_3$ and $I = \boldsymbol{\hat{e}}_1\,\boldsymbol{\hat{e}}_2\,\boldsymbol{\hat{e}}_3$. The geometric product of the bivector $\boldsymbol{B}$ and the pseudoscalar $I$ reads,
\begin{align}
\label{B I}
&\boldsymbol{B}\,I = \left(B_{12}\,\boldsymbol{\hat{e}}_1\,\boldsymbol{\hat{e}}_2 + B_{23}\,\boldsymbol{\hat{e}}_2\,\boldsymbol{\hat{e}}_3 + B_{31}\,\boldsymbol{\hat{e}}_3\,\boldsymbol{\hat{e}}_1\right)\left(\boldsymbol{\hat{e}}_1\,\boldsymbol{\hat{e}}_2\,\boldsymbol{\hat{e}}_3\right)\nonumber\\
&\phantom{\boldsymbol{B}\,I} = -\left(B_{23}\,\boldsymbol{\hat{e}}_1 + B_{31}\,\boldsymbol{\hat{e}}_2 + B_{12}\,\boldsymbol{\hat{e}}_3\right)\\
\label{I B}
&I\,\boldsymbol{B} = \left(\boldsymbol{\hat{e}}_1\,\boldsymbol{\hat{e}}_2\,\boldsymbol{\hat{e}}_3\right)\left(B_{12}\,\boldsymbol{\hat{e}}_1\,\boldsymbol{\hat{e}}_2 + B_{23}\,\boldsymbol{\hat{e}}_2\,\boldsymbol{\hat{e}}_3 + B_{31}\,\boldsymbol{\hat{e}}_3\,\boldsymbol{\hat{e}}_1\right) = -\,B_{12}\,\boldsymbol{\hat{e}}_3\nonumber\\
&\phantom{I\,\boldsymbol{B}} = -\left(B_{23}\,\boldsymbol{\hat{e}}_1 + B_{31}\,\boldsymbol{\hat{e}}_2 + B_{12}\,\boldsymbol{\hat{e}}_3\right)
\end{align}
According to relation~\eqref{B I} and~\eqref{I B}, the bivector $\boldsymbol{B}$ commutes with the pseudoscalar $I$,
\begin{equation}\label{commutation B I}
\boldsymbol{B}\,I = I\,\boldsymbol{B}
\end{equation}

\section{Duality in geometric algebra}
\label{Duality in geometric algebra}

\noindent The reverse of the scalar $s$, the vector $\boldsymbol{v} = v_3\,\boldsymbol{\hat{e}}_3$, of the bivector $\boldsymbol{B} = B_{12}\,\boldsymbol{\hat{e}}_1\,\boldsymbol{\hat{e}}_2$ and of the pseudoscalar $I$ is obtained by reversing the order of the basis vectors,~\cite{Macdonald:2011}
\begin{align}
\label{reverse s}
&\boldsymbol{s}^{\dag} = s\\
\label{reverse v}
&\boldsymbol{v}^{\dag} = v_3\,\boldsymbol{\hat{e}}_3 = \boldsymbol{v}\\
\label{reverse B}
&\boldsymbol{B}^{\dag} = B_{12}\,\boldsymbol{\hat{e}}_2\,\boldsymbol{\hat{e}}_1 = -\,B_{12}\,\boldsymbol{\hat{e}}_1\,\boldsymbol{\hat{e}}_2 = -\,\boldsymbol{B}\\
\label{reverse I}
&I^{\dag} = \boldsymbol{\hat{e}}_3\,\boldsymbol{\hat{e}}_2\,\boldsymbol{\hat{e}}_1 = -\,\boldsymbol{\hat{e}}_1\,\boldsymbol{\hat{e}}_2\,\boldsymbol{\hat{e}}_3 = -\,I
\end{align}
Thus, the reverse of the multivector~\eqref{multivector} is given by,
\begin{equation}\label{multivector reverse}
M^{\dag} = s^{\dag} + \boldsymbol{v}^{\dag} + \boldsymbol{B}^{\dag} + s^{\prime}I^{\dag} = s + \boldsymbol{v} -\,\boldsymbol{B} -\,s^{\prime}I
\end{equation}
The square of the modulus of a vector $\boldsymbol{v}$, a bivector $\boldsymbol{B}$ and a pseudovector $I$ are defined as,
\begin{align}
\label{modulus v}
&\left|\boldsymbol{v}\right|^2 = \boldsymbol{v}^{\dag}\cdot\boldsymbol{v} = \boldsymbol{v}\cdot\boldsymbol{v} = \boldsymbol{v}^2\\
\label{modulus B}
&\left|\boldsymbol{B}\right|^2 = \boldsymbol{B}^{\dag}\cdot\boldsymbol{B} = -\,\boldsymbol{B}\cdot\boldsymbol{B} = -\,\boldsymbol{B}^2\\
\label{modulus I}
&\left|I\right|^2 = I^{\dag}\cdot I = -\,I\cdot I = -\,I^2 = 1
\end{align}
The geometric interpretation of the modulus is clear : the modulus of a vector $\left|\boldsymbol{v}\right|$ is the length of a line element, the modulus of a bivector $\left|\boldsymbol{B}\right|$ is the surface of a plane element and the modulus of a pseudoscalar $\left|s^{\prime}I\right|$ is the volume of a space element. Thus, the modulus of the bivector obtained by taking the outer product of a vector $\boldsymbol{u}$ and a vector $\boldsymbol{v}$ is the surface of the parallelogram spanned by these vectors,
\begin{equation}\label{area bivector}
\left|\boldsymbol{u}\wedge\boldsymbol{v}\right| = \left|\boldsymbol{u}\right|\,\left|\boldsymbol{v}\right|\,\sin\theta
\end{equation}
where $\theta$ is the acute angle between these vectors. This modulus is the same as the modulus of the cross product of these vectors,
\begin{equation}\label{cross product length}
\left|\boldsymbol{u}\times\boldsymbol{v}\right| = \left|\boldsymbol{u}\right|\,\left|\boldsymbol{v}\right|\,\sin\theta
\end{equation}
Thus,
\begin{equation}\label{modulus duality}
\left|\boldsymbol{u}\wedge\boldsymbol{v}\right| = \left|\boldsymbol{u}\times\boldsymbol{v}\right|
\end{equation}
The inverse of a vector $\boldsymbol{v}$, a bivector $\boldsymbol{B}$ and a pseudovector $I$ are defined as,
\begin{align}
\label{inverse v}
&\boldsymbol{v}^{-1} = \frac{\boldsymbol{v}}{\boldsymbol{v}^2} = \frac{\boldsymbol{v}^{\dag}}{\left|\boldsymbol{v}\right|^2}\\
\label{inverse B}
&\boldsymbol{B}^{-1} = \frac{\boldsymbol{B}}{\boldsymbol{B}^2} = \frac{\boldsymbol{B}^{\dag}}{\left|\boldsymbol{B}\right|^2}\\
\label{inverse I}
&I^{-1} = \frac{I}{I^2} = \frac{I^{\dag}}{\left|I\right|^2} = -\,I
\end{align}
The dual of a vector $\boldsymbol{v}$, a bivector $\boldsymbol{B}$ and a pseudovector $I$ are defined as,~\cite{Macdonald:2011}
\begin{align}
\label{dual v}
&\boldsymbol{v}^{\ast} = \frac{\boldsymbol{v}}{I} = \boldsymbol{v}\,I^{-1} = -\,\boldsymbol{v}\,I\\
\label{dual B}
&\boldsymbol{B}^{\ast} = \frac{\boldsymbol{B}}{I} = \boldsymbol{B}\,I^{-1} = -\,\boldsymbol{B}\,I\\
\label{dual I}
&I^{\ast} = \frac{I}{I} = I\,I^{-1} = 1
\end{align}
This duality is the transformation as the Hodge duality in differential forms. The dual of the dual of a vector $\boldsymbol{v}$, a bivector $\boldsymbol{B}$ and a pseudovector $I$ are their opposite,
\begin{align}
\label{dual dual v}
&\left(\boldsymbol{v}^{\ast}\right)^{\ast} = -\,\boldsymbol{v}^{\ast}\,I = \boldsymbol{v}\,I^2 = -\,\boldsymbol{v}\\
\label{dual dual B}
&\left(\boldsymbol{B}^{\ast}\right)^{\ast} = -\,\boldsymbol{B}^{\ast}\,I = \boldsymbol{B}\,I^2 = -\,\boldsymbol{B}\\
\label{dual dual I}
&\left(I^{\ast}\right)^{\ast} = -\,I^{\ast}\,I = I\,I^2 = -\,I
\end{align}
The dual of the vector $\boldsymbol{v} = v_3\,\boldsymbol{\hat{e}}_3$ is,
\begin{equation}\label{dual vector v 0}
\boldsymbol{v}^{\ast} = -\,\boldsymbol{v}\,I = -\,\left(v_3\,\boldsymbol{\hat{e}}_3\right)\left(\boldsymbol{\hat{e}}_1\,\boldsymbol{\hat{e}}_2\,\boldsymbol{\hat{e}}_3\right) = -\,v_3\,\boldsymbol{\hat{e}}_1\,\boldsymbol{\hat{e}}_2
\end{equation}
Defining the bivector $\boldsymbol{V}$ as,
\begin{equation}\label{dual vector v 1}
\boldsymbol{V} = v_{12}\,\boldsymbol{\hat{e}}_1\,\boldsymbol{\hat{e}}_2 \qquad\text{where}\qquad \left|\boldsymbol{V}\right| = \left|\boldsymbol{v}\right| \qquad\text{and thus}\qquad v_{12} = v_3
\end{equation}
it satisfies the duality,
\begin{equation}\label{dual vector v}
\boldsymbol{v}^{\ast} = -\,\boldsymbol{V} = \left(\boldsymbol{V}^{\ast}\right)^{\ast} \qquad\text{and thus}\qquad \boldsymbol{V}^{\ast} = \boldsymbol{v}
\end{equation}
The dual of the bivector $\boldsymbol{B} = B_{12}\,\boldsymbol{\hat{e}}_1\,\boldsymbol{\hat{e}}_2$ is,
\begin{equation}\label{dual bivector B 0}
\boldsymbol{B}^{\ast} = -\,\boldsymbol{B}\,I = -\,\left(B_{12}\,\boldsymbol{\hat{e}}_1\,\boldsymbol{\hat{e}}_2\right)\left(\boldsymbol{\hat{e}}_1\,\boldsymbol{\hat{e}}_2\,\boldsymbol{\hat{e}}_3\right) = B_{12}\,\boldsymbol{\hat{e}}_3
\end{equation}
Defining the vector $\boldsymbol{b}$ as,
\begin{equation}\label{dual bivector B 1}
\boldsymbol{b} = b_{3}\,\boldsymbol{\hat{e}}_3 \qquad\text{where}\qquad \left|\boldsymbol{b}\right| = \left|\boldsymbol{B}\right| \qquad\text{and thus}\qquad b_3 = B_{12}
\end{equation}
it satisfies the duality,
\begin{equation}\label{dual bivector B}
\boldsymbol{B}^{\ast} = \boldsymbol{b} = -\,\left(\boldsymbol{b}^{\ast}\right)^{\ast} \qquad\text{and thus}\qquad \boldsymbol{b}^{\ast} = -\,\boldsymbol{B}
\end{equation}
There is a duality between a vector and a bivector of same modulus and there is a duality between a scalar and a pseudoscalar of same modulus. In view of relations~\eqref{dual v} and~\eqref{dual bivector B},
\begin{equation}\label{dual bivector B bis}
\boldsymbol{B} = -\,\boldsymbol{b}^{\ast} = \boldsymbol{b}\,I
\end{equation}
which implies that by duality the multivector~\eqref{multivector} is recast as,~\cite{Hestenes:2015}
\begin{equation}\label{multivector bis}
M = \left(s + s^{\prime}I\right) + \left(\boldsymbol{v} + \boldsymbol{b}\,I\right)
\end{equation}
and the reverse of the multivector~\eqref{multivector reverse} is recast as,
\begin{equation}\label{multivector reverse bis}
M^{\dag} = \left(s -\,s^{\prime}I\right) + \left(\boldsymbol{v} -\,\boldsymbol{b}\,I\right)
\end{equation}
which means that the reverse of a multivector is like a complex conjugate where the pseudoscalar $I$ is like the imaginary number $i$. The bivector $\boldsymbol{W}$ and the vector $\boldsymbol{w}$ are defined as the wedge product and the cross product of two vectors $\boldsymbol{u}$ and $\boldsymbol{v}$ respectively,
\begin{equation}\label{wedge and cross products}
\boldsymbol{W} = \boldsymbol{u}\wedge\boldsymbol{v} \qquad\text{and}\qquad \boldsymbol{w} = \boldsymbol{u}\times\boldsymbol{v}
\end{equation}
According to relation~\eqref{modulus duality} the modulus of the bivector $\boldsymbol{W}$ and vector $\boldsymbol{w}$ are equal, which means that the vector $\boldsymbol{w}$ is the dual of the bivector $\boldsymbol{W}$,
\begin{equation}\label{wedge and cross products duality 0}
\left|\boldsymbol{W}\right| = \left|\boldsymbol{w}\right| \qquad\text{and thus}\qquad \boldsymbol{W}^{\ast} = \boldsymbol{w} \qquad\text{and}\qquad \boldsymbol{w}^{\ast} = -\,\boldsymbol{W}
\end{equation}
Thus, the cross product of the vectors $\boldsymbol{u}$ and $\boldsymbol{v}$ is the dual of the wedge product of these vectors,
\begin{equation}\label{wedge and cross products duality}
\left(\boldsymbol{u}\wedge\boldsymbol{v}\right)^{\ast} = \boldsymbol{u}\times\boldsymbol{v} \qquad\text{and}\qquad \left(\boldsymbol{u}\times\boldsymbol{v}\right)^{\ast} = -\,\boldsymbol{u}\wedge\boldsymbol{v} 
\end{equation}
To establish the duality between the inner and outer product of two vectors, we choose a vector $\boldsymbol{u} = u_1\,\boldsymbol{\hat{e}}_1 + u_2\,\boldsymbol{\hat{e}}_2$ and a vector $\boldsymbol{v} = v_1\,\boldsymbol{\hat{e}}_1 + v_2\,\boldsymbol{\hat{e}}_2$ in the same plane. The dual of the outer product of the vectors $\boldsymbol{u}$ and $\boldsymbol{v}$ yields,
\begin{equation}\label{duality vectors outer}
\begin{split}
&\left(\boldsymbol{u}\wedge\boldsymbol{v}\right)^{\ast} = -\,\Big(\left(u_1\,\boldsymbol{\hat{e}}_1 + u_2\,\boldsymbol{\hat{e}}_2\right)\wedge\left(v_1\,\boldsymbol{\hat{e}}_1 + v_2\,\boldsymbol{\hat{e}}_2\right)\Big)\left(\boldsymbol{\hat{e}}_1\,\boldsymbol{\hat{e}}_2\,\boldsymbol{\hat{e}}_3\right)\\
&\phantom{\left(\boldsymbol{u}\wedge\boldsymbol{v}\right)^{\ast}} = -\,\left(u_1\,v_2\,\boldsymbol{\hat{e}}_1\,\boldsymbol{\hat{e}}_2 + u_2\,v_1\,\boldsymbol{\hat{e}}_2\,\boldsymbol{\hat{e}}_1\right)\left(\boldsymbol{\hat{e}}_1\,\boldsymbol{\hat{e}}_2\,\boldsymbol{\hat{e}}_3\right)\\
&\phantom{\left(\boldsymbol{u}\wedge\boldsymbol{v}\right)^{\ast}} = \left(u_1\,v_2 -\,u_2\,v_1\right)\boldsymbol{\hat{e}}_3
\end{split}
\end{equation}
The inner product of the vector $\boldsymbol{u}$ and $\boldsymbol{v}^{\ast}$ yields,
\begin{equation}\label{duality vectors outer bis}
\begin{split}
&\boldsymbol{u}\cdot\boldsymbol{v}^{\ast} = -\,\left(u_1\,\boldsymbol{\hat{e}}_1 + u_2\,\boldsymbol{\hat{e}}_2\right)\cdot\Big(\left(v_1\,\boldsymbol{\hat{e}}_1 + v_2\,\boldsymbol{\hat{e}}_2\right)\left(\boldsymbol{\hat{e}}_1\,\boldsymbol{\hat{e}}_2\,\boldsymbol{\hat{e}}_3\right)\Big)\\
&\phantom{\boldsymbol{u}\cdot\boldsymbol{v}^{\ast}} = -\,\left(u_1\,\boldsymbol{\hat{e}}_1 + u_2\,\boldsymbol{\hat{e}}_2\right)\cdot\left(v_1\,\boldsymbol{\hat{e}}_2\,\boldsymbol{\hat{e}}_3 -\,v_2\,\boldsymbol{\hat{e}}_1\,\boldsymbol{\hat{e}}_3\right)\\
&\phantom{\boldsymbol{u}\cdot\boldsymbol{v}^{\ast}} = \left(u_1\,v_2 -\,u_2\,v_1\right)\boldsymbol{\hat{e}}_3
\end{split}
\end{equation}
The identification of relations~\eqref{duality vectors outer} and~\eqref{duality vectors outer bis} yields the vectorial duality,
\begin{equation}\label{vector duality}
\left(\boldsymbol{u}\wedge\boldsymbol{v}\right)^{\ast} = \boldsymbol{u}\cdot\boldsymbol{v}^{\ast}
\end{equation}
which is recast in terms of the dual bivector $\boldsymbol{V} = -\,\boldsymbol{v}^{\ast}$ as,
\begin{equation}\label{vector duality bis 0}
\left(\boldsymbol{u}\wedge\boldsymbol{v}\right)^{\ast} = -\,\boldsymbol{u}\cdot\boldsymbol{V} = \boldsymbol{V}\cdot\boldsymbol{u}
\end{equation}
In view of relations~\eqref{wedge and cross products duality} and~\eqref{vector duality}, we obtain,
\begin{equation}\label{wedge and cross products duality bis}
\boldsymbol{u}\times\boldsymbol{v} = \boldsymbol{u}\cdot\boldsymbol{v}^{\ast}
\end{equation}
which is recast in terms of the dual bivector $\boldsymbol{V} = -\,\boldsymbol{v}^{\ast}$ as,
\begin{equation}\label{wedge and cross products duality ter}
\boldsymbol{u}\times\boldsymbol{v} = -\,\boldsymbol{u}\cdot\boldsymbol{V} = \boldsymbol{V}\cdot\boldsymbol{u}
\end{equation}
The dual of the inner product of the vectors $\boldsymbol{u}$ and $\boldsymbol{v}$ yields,
\begin{equation}\label{duality vectors inner}
\begin{split}
&\left(\boldsymbol{u}\cdot\boldsymbol{v}\right)^{\ast} = -\,\Big(\left(u_1\,\boldsymbol{\hat{e}}_1 + u_2\,\boldsymbol{\hat{e}}_2\right)\cdot\left(v_1\,\boldsymbol{\hat{e}}_1 + v_2\,\boldsymbol{\hat{e}}_2\right)\Big)\left(\boldsymbol{\hat{e}}_1\,\boldsymbol{\hat{e}}_2\,\boldsymbol{\hat{e}}_3\right)\\
&\phantom{\left(\boldsymbol{u}\cdot\boldsymbol{v}\right)^{\ast}} = -\,\left(u_1\,v_1 + u_2\,v_2\right)\boldsymbol{\hat{e}}_1\,\boldsymbol{\hat{e}}_2\,\boldsymbol{\hat{e}}_3
\end{split}
\end{equation}
The outer product of the vectors $\boldsymbol{u}$ and $\boldsymbol{v}^{\ast}$ yields,
\begin{equation}\label{duality vectors inner bis}
\begin{split}
&\boldsymbol{u}\wedge\boldsymbol{v}^{\ast} = -\,\left(u_1\,\boldsymbol{\hat{e}}_1 + u_2\,\boldsymbol{\hat{e}}_2\right)\wedge\Big(\left(v_1\,\boldsymbol{\hat{e}}_1 + v_2\,\boldsymbol{\hat{e}}_2\right)\left(\boldsymbol{\hat{e}}_1\,\boldsymbol{\hat{e}}_2\,\boldsymbol{\hat{e}}_3\right)\Big)\\
&\phantom{\boldsymbol{u}\wedge\boldsymbol{v}^{\ast}} = -\,\left(u_1\,\boldsymbol{\hat{e}}_1 + u_2\,\boldsymbol{\hat{e}}_2\right)\wedge\left(v_1\,\boldsymbol{\hat{e}}_2\,\boldsymbol{\hat{e}}_3 -\,v_2\,\boldsymbol{\hat{e}}_1\,\boldsymbol{\hat{e}}_3\right)\\
&\phantom{\boldsymbol{u}\wedge\boldsymbol{v}^{\ast}} = \left(u_1\,v_1 + u_2\,v_2\right)\boldsymbol{\hat{e}}_1\,\boldsymbol{\hat{e}}_2\,\boldsymbol{\hat{e}}_3
\end{split}
\end{equation}
The identification of relations~\eqref{duality vectors inner} and~\eqref{duality vectors inner bis} yields the pseudoscalar duality,
\begin{equation}\label{vector duality bis}
\left(\boldsymbol{u}\cdot\boldsymbol{v}\right)^{\ast} = \boldsymbol{u}\wedge\boldsymbol{v}^{\ast}
\end{equation}
which is recast in terms of the dual bivector $\boldsymbol{V} = -\,\boldsymbol{v}^{\ast}$ as,
\begin{equation}\label{vector duality ter}
\left(\boldsymbol{u}\cdot\boldsymbol{v}\right)^{\ast} = -\,\boldsymbol{u}\wedge\boldsymbol{V}
\end{equation}
To establish the duality between the inner and outer product of a bivector and a vector, the spatial frame is oriented such that the bivector is given by $\boldsymbol{B} = B_{12}\,\boldsymbol{\hat{e}}_1\,\boldsymbol{\hat{e}}_2$ and the vector is written as $\boldsymbol{v} = v_1\,\boldsymbol{\hat{e}}_1 + v_2\,\boldsymbol{\hat{e}}_2 + v_3\,\boldsymbol{\hat{e}}_3$. The dual of the outer product of the vector $\boldsymbol{u}$ and the bivector $\boldsymbol{B}$ yields,
\begin{equation}\label{duality vector bivector outer}
\begin{split}
&\left(\boldsymbol{u}\wedge\boldsymbol{B}\right)^{\ast} = -\,\Big(\left(u_1\,\boldsymbol{\hat{e}}_1 + u_2\,\boldsymbol{\hat{e}}_2 + u_3\,\boldsymbol{\hat{e}}_3\right)\wedge\left(B_{12}\,\boldsymbol{\hat{e}}_1\,\boldsymbol{\hat{e}}_2\right)\Big)\left(\boldsymbol{\hat{e}}_1\,\boldsymbol{\hat{e}}_2\,\boldsymbol{\hat{e}}_3\right)\\
&\phantom{\left(\boldsymbol{u}\wedge\boldsymbol{B}\right)^{\ast}} = -\,\left(u_3\,B_{12}\,\boldsymbol{\hat{e}}_1\,\boldsymbol{\hat{e}}_2\,\boldsymbol{\hat{e}}_3\right)\left(\boldsymbol{\hat{e}}_1\,\boldsymbol{\hat{e}}_2\,\boldsymbol{\hat{e}}_3\right)\\
&\phantom{\left(\boldsymbol{u}\wedge\boldsymbol{B}\right)^{\ast}} = u_3\,B_{12}
\end{split}
\end{equation}
The inner product of the vector $\boldsymbol{u}$ and the bivector $\boldsymbol{B}^{\ast}$ yields,
\begin{equation}\label{duality vector bivector outer bis}
\begin{split}
&\boldsymbol{u}\cdot\boldsymbol{B}^{\ast} = -\,\left(u_1\,\boldsymbol{\hat{e}}_1 + u_2\,\boldsymbol{\hat{e}}_2 + u_3\,\boldsymbol{\hat{e}}_3\right)\cdot\Big(\left(B_{12}\,\boldsymbol{\hat{e}}_1\,\boldsymbol{\hat{e}}_2\right)\Big)\left(\boldsymbol{\hat{e}}_1\,\boldsymbol{\hat{e}}_2\,\boldsymbol{\hat{e}}_3\right)\Big)\\
&\phantom{\boldsymbol{u}\cdot\boldsymbol{B}^{\ast}} = \left(u_1\,\boldsymbol{\hat{e}}_1 + u_2\,\boldsymbol{\hat{e}}_2 + u_3\,\boldsymbol{\hat{e}}_3\right)\cdot\left(B_{12}\,\boldsymbol{\hat{e}}_3\right)\\
&\phantom{\boldsymbol{u}\cdot\boldsymbol{B}^{\ast}} = u_3\,B_{12}
\end{split}
\end{equation}
The identification of relations~\eqref{duality vector bivector outer} and~\eqref{duality vector bivector outer bis} yields the scalar duality,
\begin{equation}\label{scalar duality}
\left(\boldsymbol{u}\wedge\boldsymbol{B}\right)^{\ast} = \boldsymbol{u}\cdot\boldsymbol{B}^{\ast}
\end{equation}
which is recast in terms of the dual vector $\boldsymbol{b} = \boldsymbol{B}^{\ast}$ as,
\begin{equation}\label{scalar duality bis}
\left(\boldsymbol{u}\wedge\boldsymbol{B}\right)^{\ast} = \boldsymbol{u}\cdot\boldsymbol{b}
\end{equation}
and is the dual of the pseudoscalar duality~\eqref{vector duality ter} for $\boldsymbol{b} = \boldsymbol{v}$ and $\boldsymbol{B} = \boldsymbol{V}$. The dual of the inner product of the vector $\boldsymbol{u}$ and the bivector $\boldsymbol{B}$ yields,
\begin{equation}\label{duality vector bivector inner}
\begin{split}
&\left(\boldsymbol{u}\cdot\boldsymbol{B}\right)^{\ast} = -\,\Big(\left(u_1\,\boldsymbol{\hat{e}}_1 + u_2\,\boldsymbol{\hat{e}}_2 + u_3\,\boldsymbol{\hat{e}}_3\right)\cdot\left(B_{12}\,\boldsymbol{\hat{e}}_1\,\boldsymbol{\hat{e}}_2\right)\Big)\left(\boldsymbol{\hat{e}}_1\,\boldsymbol{\hat{e}}_2\,\boldsymbol{\hat{e}}_3\right)\\
&\phantom{\left(\boldsymbol{u}\cdot\boldsymbol{B}\right)^{\ast}} = -\,\left(u_1\,B_{12}\,\boldsymbol{\hat{e}}_2 -\,u_2\,B_{12}\,\boldsymbol{\hat{e}}_1\right)\left(\boldsymbol{\hat{e}}_1\,\boldsymbol{\hat{e}}_2\,\boldsymbol{\hat{e}}_3\right)\\
&\phantom{\left(\boldsymbol{u}\cdot\boldsymbol{B}\right)^{\ast}} = u_1\,B_{12}\,\boldsymbol{\hat{e}}_1\,\boldsymbol{\hat{e}}_3 + u_2\,B_{12}\,\boldsymbol{\hat{e}}_2\,\boldsymbol{\hat{e}}_3
\end{split}
\end{equation}
The outer product of the vector $\boldsymbol{u}$ and the dual of the bivector $\boldsymbol{B}^{\ast}$ yields,
\begin{equation}\label{duality vector bivector inner bis}
\begin{split}
&\boldsymbol{u}\wedge\boldsymbol{B}^{\ast} = -\,\left(u_1\,\boldsymbol{\hat{e}}_1 + u_2\,\boldsymbol{\hat{e}}_2 + u_3\,\boldsymbol{\hat{e}}_3\right)\wedge\Big(\left(B_{12}\,\boldsymbol{\hat{e}}_1\,\boldsymbol{\hat{e}}_2\right)\left(\boldsymbol{\hat{e}}_1\,\boldsymbol{\hat{e}}_2\,\boldsymbol{\hat{e}}_3\right)\Big)\\
&\phantom{\boldsymbol{u}\wedge\boldsymbol{B}^{\ast}} = \left(u_1\,\boldsymbol{\hat{e}}_1 + u_2\,\boldsymbol{\hat{e}}_2 + u_3\,\boldsymbol{\hat{e}}_3\right)\wedge\left(B_{12}\,\boldsymbol{\hat{e}}_3\right)\\
&\phantom{\boldsymbol{u}\wedge\boldsymbol{B}^{\ast}} = u_1\,B_{12}\,\boldsymbol{\hat{e}}_1\,\boldsymbol{\hat{e}}_3 + u_2\,B_{12}\,\boldsymbol{\hat{e}}_2\,\boldsymbol{\hat{e}}_3
\end{split}
\end{equation}
The identification of relations~\eqref{duality vector bivector inner} and~\eqref{duality vector bivector inner bis} yields the bivectorial duality,
\begin{equation}\label{bivectorial duality}
\left(\boldsymbol{u}\cdot\boldsymbol{B}\right)^{\ast} = \boldsymbol{u}\wedge\boldsymbol{B}^{\ast}
\end{equation}
which is recast in terms of the dual vector $\boldsymbol{b} = \boldsymbol{B}^{\ast}$ as,
\begin{equation}\label{bivectorial duality bis}
\left(\boldsymbol{u}\cdot\boldsymbol{B}\right)^{\ast} = \boldsymbol{u}\wedge\boldsymbol{b}
\end{equation}
and is the dual of the vectorial duality~\eqref{vector duality bis 0} for $\boldsymbol{b} = \boldsymbol{v}$ and $\boldsymbol{B} = \boldsymbol{V}$. To establish the duality between the inner and outer product of bivectors in the same plane, the spatial frame is oriented such that the bivectors $\boldsymbol{A}$ and $\boldsymbol{B}$ are written as $\boldsymbol{A} = A_{12}\,\boldsymbol{\hat{e}}_1\,\boldsymbol{\hat{e}}_2$ and $\boldsymbol{B} = B_{12}\,\boldsymbol{\hat{e}}_2\,\boldsymbol{\hat{e}}_3$. The dual of the inner product of the bivectors $\boldsymbol{A}$ and $\boldsymbol{B}$ yields,
\begin{equation}\label{duality bivectors inner}
\begin{split}
&\left(\boldsymbol{A}\cdot\boldsymbol{B}\right)^{\ast} = -\,\Big(\left(A_{12}\,\boldsymbol{\hat{e}}_1\,\boldsymbol{\hat{e}}_2\right)\cdot\left(B_{12}\,\boldsymbol{\hat{e}}_1\,\boldsymbol{\hat{e}}_2\right)\Big)\left(\boldsymbol{\hat{e}}_1\,\boldsymbol{\hat{e}}_2\,\boldsymbol{\hat{e}}_3\right)\\
&\phantom{\left(\boldsymbol{A}\cdot\boldsymbol{B}\right)^{\ast}} = A_{12}\,B_{12}\,\boldsymbol{\hat{e}}_1\,\boldsymbol{\hat{e}}_2\,\boldsymbol{\hat{e}}_3
\end{split}
\end{equation}
The outer product of the bivectors $\boldsymbol{A}$ and $\boldsymbol{B}^{\ast}$ yields,
\begin{equation}\label{duality bivectors inner bis}
\begin{split}
&\boldsymbol{A}\wedge\boldsymbol{B}^{\ast} = -\,\left(A_{12}\,\boldsymbol{\hat{e}}_1\,\boldsymbol{\hat{e}}_2\right)\wedge\Big(\left(B_{12}\,\boldsymbol{\hat{e}}_1\,\boldsymbol{\hat{e}}_2\right)\left(\boldsymbol{\hat{e}}_1\,\boldsymbol{\hat{e}}_2\,\boldsymbol{\hat{e}}_3\right)\Big)\\
&\phantom{\boldsymbol{A}\wedge\boldsymbol{B}^{\ast}} = \left(A_{12}\,\boldsymbol{\hat{e}}_1\,\boldsymbol{\hat{e}}_2\right)\wedge\left(B_{12}\,\boldsymbol{\hat{e}}_3\right)\\
&\phantom{\boldsymbol{A}\wedge\boldsymbol{B}^{\ast}} = A_{12}\,B_{12}\,\boldsymbol{\hat{e}}_1\,\boldsymbol{\hat{e}}_2\,\boldsymbol{\hat{e}}_3
\end{split}
\end{equation}
The identification of relations~\eqref{duality bivectors inner} and~\eqref{duality bivectors inner bis} yields the pseudoscalar duality,
\begin{equation}\label{bivectorial duality bi}
\left(\boldsymbol{A}\cdot\boldsymbol{B}\right)^{\ast} = \boldsymbol{A}\wedge\boldsymbol{B}^{\ast}
\end{equation}
The pseudoscalar duality~\eqref{bivectorial duality bi} is expressed in terms of the dual vector $\boldsymbol{b} = \boldsymbol{B}^{\ast}$,
\begin{equation}\label{bivectorial duality bi bis}
\left(\boldsymbol{A}\cdot\boldsymbol{B}\right)^{\ast} = \boldsymbol{A}\wedge\boldsymbol{b}
\end{equation}
In view of the identities~\eqref{outer product v B symmetric},~\eqref{bivectorial duality}, the inner product of two bivectors $\boldsymbol{A}$ and $\boldsymbol{B}$ is expressed in terms of the dual vectors $\boldsymbol{a} = \boldsymbol{A}^{\ast}$ and $\boldsymbol{b} = \boldsymbol{B}^{\ast}$ as,
\begin{equation}\label{grad bivec dual}
\begin{split}
&\left(\boldsymbol{A}\cdot\boldsymbol{B}\right)^{\ast} = \boldsymbol{A}\wedge\boldsymbol{b} = \boldsymbol{b}\wedge\boldsymbol{A} = -\,\boldsymbol{b}\wedge\boldsymbol{a}^{\ast}\\
&\phantom{\left(\boldsymbol{A}\cdot\boldsymbol{B}\right)^{\ast}} = -\,\left(\boldsymbol{b}\cdot\boldsymbol{a}\right)^{\ast} = -\,\left(\boldsymbol{a}\cdot\boldsymbol{b}\right)^{\ast}
\end{split}
\end{equation}
The dual of identity~\eqref{grad bivec dual} yields,
\begin{equation}\label{grad bivec}
\boldsymbol{A}\cdot\boldsymbol{B} = -\,\boldsymbol{a}\cdot\boldsymbol{b}
\end{equation}
To establish the duality of the inner product of a vector $\boldsymbol{v}$ and a trivector $T = t\,I$, we write that trivector as $T = t\,I$, where $t$ is a scalar. Thus, in view of identity~\eqref{dual I},
\begin{equation}\label{dual vector trivector}
T^{\ast} = \left(t\,I\right)^{\ast} = -\,\left(t\,I\right)I = t \qquad\text{and}\qquad t^{\ast} = -\,T
\end{equation}
Use the dualities~\eqref{dual B} and~\eqref{dual vector trivector}, and the commutation~\eqref{commutation v I} between the vector $\boldsymbol{v}$ and the pseudoscalar $I$ or the trivector $T = t\,I$, the dual of bivector $\boldsymbol{v}\cdot T$ is written as,
\begin{equation}\label{dual vector trivector inner}
\left(\boldsymbol{v}\cdot T\right)^{\ast} = \left(\boldsymbol{v}\,T\right)^{\ast} = -\,\left(\boldsymbol{v}\,t\,I\right)I = \boldsymbol{v}\,t = \boldsymbol{v}\,T^{\ast}
\end{equation}

\section{Algebraic identities in geometric algebra}
\label{Algebraic identities in geometric algebra}

\noindent The double cross product of the three vectors $\boldsymbol{u}$, $\boldsymbol{v}$ and $\boldsymbol{w}$ is written as,
\begin{equation}\label{double cross product vectors}
\boldsymbol{u}\times\left(\boldsymbol{v}\times\boldsymbol{w}\right) = \left(\boldsymbol{u}\cdot\boldsymbol{w}\right)\boldsymbol{v} -\,\left(\boldsymbol{u}\cdot\boldsymbol{v}\right)\boldsymbol{w}
\end{equation}
In view of the duality~\eqref{wedge and cross products duality} between the cross and wedge products, the vectorial duality~\eqref{wedge and cross products duality bis}, the double duality~\eqref{dual dual B} and the antisymmetry of the outer product~\eqref{outer product space u v}, the left-hand side of relation~\eqref{double cross product vectors} is recast as,
\begin{equation}\label{id cross}
\boldsymbol{u}\times\left(\boldsymbol{v}\times\boldsymbol{w}\right) = \boldsymbol{u}\times\left(\boldsymbol{v}\wedge\boldsymbol{w}\right)^{\ast} = \boldsymbol{u}\cdot\Big(\left(\boldsymbol{v}\wedge\boldsymbol{w}\right)^{\ast}\Big)^{\ast} = -\,\boldsymbol{u}\cdot\left(\boldsymbol{v}\wedge\boldsymbol{w}\right) = \boldsymbol{u}\cdot\left(\boldsymbol{w}\wedge\boldsymbol{v}\right)
\end{equation}
According to the identity~\eqref{id cross}, the double cross product~\eqref{double cross product vectors} yields the triple product,
\begin{equation}\label{triple product vectors}
\boldsymbol{u}\cdot\left(\boldsymbol{v}\wedge\boldsymbol{w}\right) = \left(\boldsymbol{u}\cdot\boldsymbol{v}\right)\boldsymbol{w} -\,\left(\boldsymbol{u}\cdot\boldsymbol{w}\right)\boldsymbol{v}
\end{equation}
This triple product~\eqref{triple product vectors} represents the projection of the vector $\boldsymbol{u}$ on the bivector $\boldsymbol{v}\wedge\boldsymbol{w}$. The vectors $\boldsymbol{u}$, $\boldsymbol{v}$ and $\boldsymbol{w}$ satisfy the identity,
\begin{equation}\label{id dot cross}
\boldsymbol{u}\cdot\left(\boldsymbol{v}\times\boldsymbol{w}\right) = \boldsymbol{v}\cdot\left(\boldsymbol{w}\times\boldsymbol{u}\right) = \boldsymbol{w}\cdot\left(\boldsymbol{u}\times\boldsymbol{v}\right)
\end{equation}
In view of the duality~\eqref{wedge and cross products duality} between cross product and wedge product and the scalar duality~\eqref{scalar duality}, the three terms of identity~\eqref{id dot cross} are recast as,
\begin{equation}\label{id dot cross wedge}
\begin{split}
&\boldsymbol{u}\cdot\left(\boldsymbol{v}\times\boldsymbol{w}\right) = \boldsymbol{u}\cdot\left(\boldsymbol{v}\wedge\boldsymbol{w}\right)^{\ast} = \left(\boldsymbol{u}\wedge\boldsymbol{v}\wedge\boldsymbol{w}\right)^{\ast}\\
&\boldsymbol{v}\cdot\left(\boldsymbol{w}\times\boldsymbol{u}\right) = \boldsymbol{v}\cdot\left(\boldsymbol{w}\wedge\boldsymbol{u}\right)^{\ast} = \left(\boldsymbol{v}\wedge\boldsymbol{w}\wedge\boldsymbol{u}\right)^{\ast}\\
&\boldsymbol{w}\cdot\left(\boldsymbol{u}\times\boldsymbol{v}\right) = \boldsymbol{w}\cdot\left(\boldsymbol{u}\wedge\boldsymbol{v}\right)^{\ast} = \left(\boldsymbol{w}\wedge\boldsymbol{u}\wedge\boldsymbol{v}\right)^{\ast}
\end{split}
\end{equation}
According to the relations~\eqref{id dot cross wedge}, the scalar identity~\eqref{id dot cross} becomes,
\begin{equation}\label{id dot cross bis}
\left(\boldsymbol{u}\wedge\boldsymbol{v}\wedge\boldsymbol{w}\right)^{\ast} = \left(\boldsymbol{v}\wedge\boldsymbol{w}\wedge\boldsymbol{u}\right)^{\ast} = \left(\boldsymbol{w}\wedge\boldsymbol{u}\wedge\boldsymbol{v}\right)^{\ast}
\end{equation}
The dual of the scalar identity~\eqref{id dot cross bis} yields the pseudoscalar identity,
\begin{equation}\label{id dot cross ter}
\boldsymbol{u}\wedge\boldsymbol{v}\wedge\boldsymbol{w} = \boldsymbol{v}\wedge\boldsymbol{w}\wedge\boldsymbol{u} = \boldsymbol{w}\wedge\boldsymbol{u}\wedge\boldsymbol{v}
\end{equation}
This identity represents the fact the oriented volume of the parallelepiped spanned by the vectors $\boldsymbol{u}$, $\boldsymbol{v}$, $\boldsymbol{w}$ is the same as that of the parallelepiped spanned by the vectors $\boldsymbol{v}$, $\boldsymbol{w}$, $\boldsymbol{u}$ and also the same as that of the parallelepiped spanned by the vectors $\boldsymbol{w}$, $\boldsymbol{u}$, $\boldsymbol{v}$ because these three parallelepipeds have the same orientation. According to relations~\eqref{inner product antisymmetry v B} and~\eqref{outer product symmetry v B} for $\boldsymbol{B} = \boldsymbol{v}\wedge\boldsymbol{w}$, the inner and outer products of the vector $\boldsymbol{u}$ and the bivector $\boldsymbol{v}\wedge\boldsymbol{w}$ are given by,
\begin{align}
\label{inner product vec bivec}
&\boldsymbol{u}\cdot\left(\boldsymbol{v}\wedge\boldsymbol{w}\right) = \frac{1}{2}\Big(\boldsymbol{u}\left(\boldsymbol{v}\wedge\boldsymbol{w}\right) -\,\left(\boldsymbol{v}\wedge\boldsymbol{w}\right)\boldsymbol{u}\Big)\\
\label{outer product vec bivec}
&\boldsymbol{u}\wedge\left(\boldsymbol{v}\wedge\boldsymbol{w}\right) = \frac{1}{2}\Big(\boldsymbol{u}\left(\boldsymbol{v}\wedge\boldsymbol{w}\right) + \left(\boldsymbol{v}\wedge\boldsymbol{w}\right)\boldsymbol{u}\Big)
\end{align}
To determine the mixed product of two vectors $\boldsymbol{u}$, $\boldsymbol{v}$ and a bivector $\boldsymbol{B}$, we use the dual vector defined as $\boldsymbol{b} = -\,\boldsymbol{B}\,I$. Using the identity~\eqref{inner product v B antisymmetric} and the triple product of vectors~\eqref{id dot cross}, we obtain,
\begin{equation}\label{triple product u v B}
\left(\boldsymbol{u}\wedge\boldsymbol{v}\right)\cdot\boldsymbol{B} = \left(\boldsymbol{u}\wedge\boldsymbol{v}\right)\cdot\boldsymbol{b}\,I = -\,\boldsymbol{b}\cdot\left(\boldsymbol{u}\wedge\boldsymbol{v}\right)\,I = -\,I\left(\boldsymbol{b}\cdot\boldsymbol{u}\right)\boldsymbol{v} + I\left(\boldsymbol{b}\cdot\boldsymbol{v}\right)\boldsymbol{u}
\end{equation}
which reduces to,
\begin{equation}\label{triple product u v B bis}
\left(\boldsymbol{u}\wedge\boldsymbol{v}\right)\cdot\boldsymbol{B} = \left(\boldsymbol{B}\cdot\boldsymbol{u}\right)\cdot\boldsymbol{v} -\,\left(\boldsymbol{B}\cdot\boldsymbol{v}\right)\cdot\boldsymbol{u} = \boldsymbol{u}\cdot\left(\boldsymbol{v}\cdot\boldsymbol{B}\right) -\,\boldsymbol{v}\cdot\left(\boldsymbol{u}\cdot\boldsymbol{B}\right)
\end{equation}
To determine the first kind of mixed product of two bivectors $\boldsymbol{A}$, $\boldsymbol{B}$ and a vector $\boldsymbol{v}$, we use the dual vectors defined as $\boldsymbol{a} = -\,\boldsymbol{A}\,I$ and $\boldsymbol{b} = -\,\boldsymbol{B}\,I$. Using the identities~\eqref{inner product v B antisymmetric} and $I^2 = -1$, we obtain,
\begin{equation}\label{triple product A v B 1st}
\left(\boldsymbol{A}\wedge\boldsymbol{v}\right)\cdot\boldsymbol{B} = \left(\boldsymbol{a}\,I\wedge\boldsymbol{v}\right)\cdot\boldsymbol{b}\,I = -\,\left(\boldsymbol{a}\wedge\boldsymbol{v}\right)\cdot\boldsymbol{b} = \boldsymbol{b}\cdot\left(\boldsymbol{a}\wedge\boldsymbol{v}\right)
\end{equation}
Using the triple product of vectors~\eqref{id dot cross}, the mixed product is recast as,
\begin{equation}\label{triple product A v B 1st bis}
\begin{split}
&\left(\boldsymbol{A}\wedge\boldsymbol{v}\right)\cdot\boldsymbol{B} = \boldsymbol{b}\cdot\left(\boldsymbol{a}\wedge\boldsymbol{v}\right) = \left(\boldsymbol{a}\cdot\boldsymbol{b}\right)\boldsymbol{v} -\,\left(\boldsymbol{b}\cdot\boldsymbol{v}\right)\boldsymbol{a}\\
&\phantom{\left(\boldsymbol{A}\wedge\boldsymbol{v}\right)\cdot\boldsymbol{B}} = -\,\left(\boldsymbol{A}\cdot\boldsymbol{B}\right)\boldsymbol{v} + \left(\boldsymbol{B}\cdot\boldsymbol{v}\right)\cdot\boldsymbol{A}
\end{split}
\end{equation}
To determine the first kind of mixed product of two bivectors $\boldsymbol{A}$, $\boldsymbol{B}$ and a vector $\boldsymbol{v}$, we use the dual vector defined as $\boldsymbol{a} = \boldsymbol{A}^{\ast} = -\,\boldsymbol{A}\,I$ and the duality $\left(\boldsymbol{v}\wedge\boldsymbol{a}\right)^{\ast} = -\left(\boldsymbol{v}\wedge\boldsymbol{a}\right)I$. Using the identities~\eqref{inner product v B antisymmetric},~\eqref{vector duality} and $I^2 = -1$, we obtain,
\begin{equation}\label{triple product A v B 2nd}
\begin{split}
&\left(\boldsymbol{A}\cdot\boldsymbol{v}\right)\wedge\boldsymbol{B} = -\,\left(\boldsymbol{v}\cdot\boldsymbol{A}\right)\wedge\boldsymbol{B} = \left(\boldsymbol{v}\cdot\boldsymbol{a}^{\ast}\right)\wedge\boldsymbol{B} = \left(\boldsymbol{v}\wedge\boldsymbol{a}\right)^{\ast}\wedge\boldsymbol{B}\\
&\phantom{\left(\boldsymbol{A}\cdot\boldsymbol{v}\right)\wedge\boldsymbol{B}} = -\,\left(\boldsymbol{v}\wedge\boldsymbol{a}\right)I\wedge\boldsymbol{B} = \boldsymbol{v}\wedge\left(\boldsymbol{A}\wedge\boldsymbol{B}\right)
\end{split}
\end{equation}
%


\bibliography{references}
\bibliographystyle{plainnat} 

\end{document}